\documentclass[aps,pra,twocolumn]{revtex4}
\usepackage{amsfonts}
\usepackage{amsmath}
\usepackage{amsthm}
\usepackage{bm}
\usepackage{graphics}
\usepackage{graphicx}
\usepackage{subfigure}
\usepackage{amssymb}
\usepackage{graphics}
\usepackage{graphicx}
\usepackage{color}
\usepackage{subfigure}
\usepackage{epstopdf}
\usepackage{tikz}
\usepackage{soul}

\def\kmax{k_{\rm max}}
\newcommand\bet{{g}}
 \newcommand\alp{\hbar/(2 m)}
\newcommand\alps{{\frac{\hbar^2}{2m}}}
\newcommand\mub{{\mu}}

\newcommand\dertt[1]{ \frac{\partial{ #1}}{\partial t} }
\newcommand\gd{\mbox{${\bf \nabla}^{2}$}}
\newcommand\psib{\overline{\psi}}

\newcommand{\dert}[1]{\frac{{\rm d}{#1}}{{\rm dt}}}

\begin{document}

\title{Quantitative estimation of effective viscosity in quantum turbulence}
\author{Vishwanath Shukla}
\email{research.vishwanath@gmail.com}
\affiliation{Institut de Physique de Nice, Universit\'e de C\^ote d'Azur,CNRS, Nice, France}
\author{Pablo D.~Mininni}
\email{mininni@df.uba.ar}
\affiliation{Departamento de F\'\i sica, Facultad de Ciencias
Exactas y Naturales, Universidad de Buenos Aires and IFIBA,
CONICET, Ciudad Universitaria, 1428 Buenos Aires, Argentina } 
\author{Giorgio Krstulovic}
\email{krstulovic@oca.eu}
\affiliation{Observatoire de la C\^ote d'Azur, Universit\'e C\^ote d'Azur, 
CNRS, Laboratoire Lagrange, Bd de l'Observatoire, CS 34229, 06304 Nice
cedex 4, France.}
\author{Patricio Clark di Leoni}
\email{patricio.clark@roma2.infn.it}
\affiliation{Department of Physics and INFN, University of
    Rome Tor Vergata, Via della Ricerca Scientifica 1, 00133 Rome, Italy.}
\author{Marc E. Brachet} 
\email{brachet@physique.ens.fr}
\affiliation{Laboratoire de Physique Statistique, 
\'{E}cole Normale Sup{\'e}rieure, 
PSL Research University;
UPMC Univ Paris 06, Sorbonne Universit\'{e}s;
Universit\'{e} Paris Diderot, Sorbonne Paris-Cit\'{e};
CNRS;
24 Rue Lhomond, 75005 Paris, France}
\date{\today}

\begin{abstract}
We study freely decaying quantum turbulence by performing high
resolution numerical simulations of the Gross-Pitaevskii equation
(GPE) in the Taylor-Green geometry. We use resolutions ranging from
$1024^3$ to $4096^3$ grid points.  The energy spectrum confirms the
presence of both a Kolmogorov scaling range for scales larger than the
intervortex scale $\ell$, and a second inertial range for scales
smaller than $\ell$. Vortex line visualizations show the existence of
substructures formed by a myriad of small-scale knotted
vortices. Next, we study finite temperature effects in the decay of
quantum turbulence by using the stochastic Ginzburg-Landau equation to
generate thermal states, and then by evolving a combination of these
thermal states with the Taylor-Green initial conditions using the
GPE. We extract the mean free path out of these simulations by
measuring the spectral broadening in the Bogoliubov dispersion
relation obtained from spatio-temporal spectra, and use it to quantify
the effective viscosity as a function of the temperature. Finally, in
order to compare the decay of high temperature quantum and that of
classical flows, and to further calibrate the estimations of
viscosity from the mean free path in the GPE simulations, we perform
low Reynolds number simulations of the Navier-Stokes equations.
\end{abstract}
\maketitle

\section{Introduction}

Turbulence in quantum fluids provides an exciting yet challenging
scenario to explore multi-scale and out of equilibrium dynamics. A
turbulent state in superfluid $^4$He was first envisioned by Feynman as
a random interacting tangle of quantum vortex lines \cite{Feynman55,
Donnelly}. More recently, quantum turbulence has been realized and
studied in experiments on a wide variety of superfluid systems and
Bose-Einstein condensates (BECs) \cite{Barenghi14}, such as bosonic
superfluid $^4$He~\cite{Vinen02}, its fermionic counterpart
$^3$He~\cite{Tsepelin2017}, and BECs in traps \cite{Henn09}. It must be
emphasized that the study of quantum turbulence in laboratory
experiments is a challenging task, which requires measurements at very
low temperatures and usually in small system sizes; therefore, any
experimental progress relies heavily on technical advancements. This is
where numerical and theoretical studies become important: firstly, by
providing explanations for the experimental observations; and
secondly, by probing regimes which are yet not directly accessible in
current experimental set-ups. However, these approaches have
limitations of their own.

From the experimental, numerical, and theoretical studies, turbulent
quantum fluids are known to display hydrodynamic behavior at 
the large scales, showcasing, for example, a Kolmogorov energy cascade
\cite{Nore97b, Maurer98}. At smaller scales, the dynamics are
dominated instead by nonlinear interactions of Kelvin waves
\cite{Lvov10}. The crossover scale between these two regimes is
determined by the mean intervortex length. Understanding how this
picture is affected by finite temperature effects in the experiments
is the matter of ongoing research.  Currently there is no single
theory  which covers all the systems and is capable of predicting
the known dynamic effects across all the involved length and time scales. Therefore, much
progress relies on phenomenological models, which at times are better
suited for one type of problem than for others. Three important
classes of phenomenological models are: (i) the two-fluid models, (ii)
the vortex filament model, and (iii) the Gross-Pitaevskii equation
(GPE), sometimes also referred to as the nonlinear Schr\"odinger
equation.

The phenomenological two-fluid model was proposed independently by
L.~Tisza and L.~Landau. In this model, originally developed for
superfluid helium, the fluid is regarded as a physically inseparable
mixture of two components: the superfluid and the normal fluid.  Many
of the flow properties of superfluid helium at low velocities can be
described within the framework of this model \cite{Landau}. One of its
great successes was the prediction of the propagation of second sound.
However, it does not account for the presence of quantum vortices, a
very important feature of quantum flows.

An extension of the two-fluid model is provided by the
Hall-Vinen-Bekharevich-Khalatnikov model \cite{Hall56,Bekarevich61}.
This model incorporates the effect of interactions between the quantized
vortices and the normal fluid by including a mutual friction term. In
this model the distinction between individual vortices is ignored, and
only length scales larger than the mean separation between quantum
vortices are considered. Therefore, it is an effective, coarse-grained
model, which provides a good description of superfluid turbulence at low
Mach numbers. The model has been used to study large scale flow
properties and intermittency in direct numerical simulations (DNSs) of
quantum turbulence \cite{Roche09,Shukla15,BiferaleHVBK}, and in reduced
dynamical systems based on shell models \cite{Wacks11,Boue13,Boue15,
Shukla16}.

The vortex filament model \cite{Schwarz85} overcomes some of the
limitations of the two-fluid and the HVBK models by regarding the
quantized vortices as filaments in three-dimensions (3D), and evolving
them under the Biot-Savart law plus a mutual friction term mimicking
the coupling between the normal and superfluid components. However, vortex
reconnection is taken care of on an \textit{ad hoc} basis. This model
is relevant in situations in which the core size is negligible in
comparison to the characteristic length scales in the hydrodynamic
description of the flow, e.g., the mean inter-vortex separation
$\ell$ or the radius of curvature of the vortex filaments $R$, and
has been used to study quantum turbulence at finite temperatures
\cite{TurbulencePolaBaggaleyPRL2012,Khomenko15}.

Finally, at zero or near-zero temperatures, and for weakly interacting
bosons, the GPE provides a good hydrodynamical description of a quantum
flow, that naturally includes quantum vortices as exact solutions which
can reconnect without the need for any extra \textit{ad hoc} assumptions
\cite{Koplik93}. The first 3D DNSs of decaying quantum turbulence, using
the GPE as a model of a zero-temperature quantum fluid, were performed
some $20$ years ago with linear resolutions up to $N=512$ grid points
in each spatial direction, in the geometry of the Taylor-Green (TG)
vortex flow \cite{Nore97a, Nore97b}. An important aspect of the above
studies was to introduce a preparation method to generate initial data
for vortex dynamics with minimal sound emission; also, for the
diagnostics, the total conserved energy was decomposed into the
incompressible kinetic energy and other energy components, each with
their corresponding spectra. The main achievement of these works was
showing that at the moment of maximum incompressible kinetic energy
dissipation, the incompressible kinetic energy spectrum displays a
power-law scaling which is compatible with Kolmogorov's $k^{-5/3}$
scaling. This scaling was later confirmed in both experimental
\cite{Maurer98} and in other numerical \cite{Kobayashi05} studies. The
GPE has also been used to study the small-scale Kelvin wave cascade
\cite{Clark15a, PRETangleVillois}, thus showing that both cascades can
be captured by its dynamics. More recently, high resolution
simulations resolving simultaneously the two inertial ranges (one for
scales larger than the intervortex length, the other for smaller
scales) were performed \cite{Clark17}. This study also showed that at
large scales the GPE can reproduce the dual cascade of energy and
helicity observed in classical turbulence \cite{Brissaud73}.

However, a problem with the GPE is that finite temperature effects can
be notoriously difficult to include \cite{Gardiner00,Gardiner02,
  Calzetta07}. One minimalistic approach, used in the present study,
is to use the so-called classical field models \cite{Proukakis08}, by
spectrally truncating the GPE \cite{Berloff14}.  It is well known that
the long time integration of the truncated system results in
microcanonical equilibrium states that can capture a condensation
transition \cite{Davis01}. This transition was later reproduced in
Ref.~\cite{Krstulovic11} using a grand-canonical method, where it was
shown to be a standard second-order $\lambda$-transition. Moreover,
dynamical counterflow effects on vortex motion, such as mutual
friction and thermalization dynamics, were also shown to be correctly
captured by this approach, and investigated in
Refs.~\cite{Krstulovic11, Krstulovic11a}. This scheme was also used 
to study the different regimes which appear during the relaxation dynamics of 
the turbulent two-dimensional (2D) GPE, where the complete thermalization 
states exhibit Berezinskii-Kosterlitz-Thouless transition
in the microcanonical ensemble framework~\cite{ShuklaNJP13,pandit2017pofoverview}; also, as
in the case of 3D this was method was extended to compute the mutual
friction coefficients in 2D~\cite{ShuklaFT14}.

Using this approach, finite temperature effects were recently studied
in helical quantum turbulence \cite{Clark18}, where it was observed
that close to the critical temperature the truncated system acts as a
classical viscous flow, with the decay of the incompressible kinetic
energy becoming exponential in time. From this observation, it was
proposed that a quantitative estimation of the effective viscosity can
be obtained by measuring the mean free path of the thermal excitations
directly on the spatio-temporal spectrum of the flow as a function of
the temperature, as this spectrum gives access to the spectrum of
phonons in the system \cite{Clark15a}. However, the calculation of
the spatio-temporal spectrum is computationally intensive, and
therefore, it is reasonable to perform it on a flow such as the TG
vortex, that (as a result of its symmetries) maximizes the scale
separation for given computational resources.

The purpose of the present paper is thus twofold.  First, we want to
extend the zero-temperature ($T=0$) TG vortex results at linear
resolution $N=512$, obtained $20$ years ago, to the resolutions
achievable with current computing resources. Second, we want to
measure, at the highest possible spatial resolution, the
spatio-temporal spectrum of the flow, in order to estimate the mean
free path and its associated effective viscosity.

The rest of the paper is organized as follows. In Sec.~\ref{methods}
we present the details of the GPE model which we use, and of its
numerical implementation. In particular, in Sec.~\ref{sec:basic} we
discuss the basic zero-temperature GPE theory and our diagnostics in
terms of different energies and associated spectra.
Section~\ref{sec:prepar}  contains the details of our zero-temperature
initial data preparation. Our methods of incorporating finite
temperature effects are reviewed in Sec.~\ref{FTGPtheory}.  We
describe the numerical implementation of the problem in
Sec.~\ref{sec:app3}, and in Sec.~\ref{units} we discuss the choice of
units. Section~\ref{Sec:results} contains our results. First, in 
Sec.~\ref{sec:hirest0} we present our results for zero-temperature GPE
dynamics with linear spatial resolutions up to $N=4096$. In
Sec.~\ref{sec:scans} we present the characterization of the
finite-temperature GPE states, including the condensation
transition. We give our results on the finite-temperature decaying GPE
runs in Sec.~\ref{sec:combined}.  In Sec.~\ref{sec:spatiotemp} we
compute and discuss the truncated GPE spatio-temporal correlation and
spectra. We evaluate the mean-free path in Sec.~\ref{sec:mfp} and
Sec.~\ref{sec:NS} is devoted to the comparison of the
finite-temperature freely decaying GPE runs with Navier-Stokes freely
decaying runs. Finally, we present our conclusions in
Sec.~\ref{sec:Conclusion}.

\section{Model, initial conditions, and numerical methods}
\label{methods}

\subsection{Gross-Pitaevskii theory}\label{sec:basic}

The GPE is a partial differential equation for a complex field $\psi$
that describes the dynamics of a zero-temperature and dilute
superfluid Bose-Einstein condensate. It reads
\begin{equation} 
    i\hbar\dertt{\psi}  =- \alps \gd \psi + \bet|\psi|^2\psi ,
    \label{Eq:GPE}
\end{equation}
where $|\psi|^2$ is the number of particles per unit volume, $m$ is
the mass of the bosons, $g=4 \pi  \tilde{a} \hbar^2/m$, and
$\tilde{a}$ is the $s$-wave scattering length. This equation conserves
the total energy $E$, the total number of particles $\mathcal{N}_p$, and
the momentum ${\bf P}$, defined in a volume $V$ respectively as
\begin{eqnarray}
  E&=&\int_{V} d^3 x \left( \alps |\nabla \psi |^2
    +\frac{g}{2}|\psi|^4 \right) , \label{Eq:defH}\\
  \mathcal{N}_p&=&\int_V  |\psi|^2 \,d^3x ,\label{Eq:defN}\\
  {\bf P}&=&\int_V  \frac{i\hbar}{2}\left(
    \psi {\bf \nabla}\psib - \psib {\bf
    \nabla}\psi\right)\,d^3x\label{Eq:defP},
\end{eqnarray}
where the overline denotes the complex conjugate.

Equation~(\ref{Eq:GPE}) can be mapped into hydrodynamic
equations of motion for a compressible irrotational fluid using the
Madelung transformation given by 
\begin{equation}
    \psi({\bf x},t)=\sqrt{\frac{\rho({\bf x},t)}{m}}\exp{[i
    \frac{m}{\hbar}\phi({\bf x},t)]},\label{Eq:defMadelung}
\end{equation}
where $\rho({\bf x},t)$ is the fluid density, and $\phi({\bf x},t)$ is
the velocity potential such that the fluid velocity is 
${\bf v}={\bf \nabla} \phi$. The Madelung transformation is singular on
the zeros of $\psi$. As two conditions are required in the singular
points (both real and imaginary parts of $\psi$ must vanish), these
singularities must take place on points in two-dimensions (2D) and on
curves in 3D. The Onsager-Feynman quantum of velocity circulation around
vortex lines with $\psi=0$ is given by $h/m$.

When Eq.~\eqref{Eq:GPE} is linearized around a constant state
$\psi=A_0$, one obtains the Bogoliubov dispersion relation
\begin{equation} 
    \omega_B(k)=\sqrt{\frac{g k^2 |A_0|^2}{m}+\frac{\hbar^2 k^4}{4
        m^2}}.
    \label{eq:Bog}
\end{equation} 
The sound velocity is thus given by $c=\sqrt{g|A_{0}|^2/m}$, with
dispersive effects taking place for length scales smaller than the 
coherence length defined by
\begin{equation}
    \xi=\sqrt{\hbar^2/(2gm|A_{\bf 0}|^2) } ; \label{Eq:defxi}
\end{equation}
$\xi$ is also proportional to the radius of the vortex cores
\cite{Nore97a,Nore97b}.

\subsubsection{Energy decomposition and associated spectra}
\label{sec:energies}

Following references \cite{Nore97a,Nore97b}, we define the total
energy per unit volume as $e_{\rm tot}=(E-\mu
\mathcal{N}_p)/V-\mu^2/2g$, where $\mu$ is the chemical potential. 
Using the hydrodynamic fields, $e_{\rm tot}$ can we written as the
sum of three components: the kinetic energy $e_{\rm kin}$, the
internal energy $e_{\rm int}$, and the quantum energy $e_{\rm q}$ (all
per unit volume), defined respectively as
\begin{eqnarray}
    e_{\rm kin}&=&\frac{1}{V}\int d^3x \frac{1}{2}(\sqrt \rho {\bf
                   v})^2 , \label{Eq:defEkin} \\ 
    e_{\rm int}&=&\frac{1}{V}\int d^3x
                   \frac{g}{2m^2}\left(\rho-\rho_0\right)^2
                   , \label{Eq:defEint}\\
   e_{\rm q}&=&\frac{1}{V}\int d^3x
                \frac{\hbar^2}{2m^2}\left(\nabla\sqrt{\rho}\right)^2
                ,\label{Eq:defEq} 
\end{eqnarray}
where $\rho_0=m |A_{\bf 0}|^2$ is the mean density of the fluid.  
The kinetic energy ${e_{\rm kin}}$ can be further decomposed into a
compressible component $e_{\rm kin}^{\rm c}$ and an incompressible
component $e_{\rm kin}^{\rm i}$, by making use of the relation 
$\sqrt \rho {\bf v}=(\sqrt {\rho} {\bf v})^{\rm c}+(\sqrt{\rho }{\bf
  v})^{\rm i}$ with 
${\bf \nabla}\cdot(\sqrt{\rho}{\bf v})^{\rm i}=0$ (see
\cite{Nore97a,Nore97b} for details).

Using Parseval's theorem one can also construct corresponding
energy spectra for each of these energies: e.g., the kinetic energy
spectrum $e_{\rm kin}(k)$ is defined as
\begin{equation}
    e_{\rm kin}(k)=\frac{1}{2} \int \left|\frac{1}{V} \int d^3 r e^{i
        {\bf r}\cdot{\bf  k}}\sqrt \rho {\bf v} \right|^2
        k^2d\Omega_k,
\end{equation}
where $d\Omega_k$ is the solid angle element on the sphere in Fourier
space.

% Vortex line length estimation
\subsubsection{Vortex line length estimation} 
\label{length_estim}

Working in a similar fashion as with the energy, we can define the
incompressible momentum power spectrum
\begin{equation}
    P^i (k) = \frac{1}{2} \int \left| \frac1V \int d^3r e^{i {\bf
          r}\cdot{\bf k}} (\rho {\bf v})^i \right|^2 k^2 d\Omega_k .
    \label{mom_spec}
\end{equation}
As was checked empirically in Refs.~\cite{Nore97a,Nore97b}, the high
wavenumber components of this spectrum can be approximated as the 
sum of the momentum of all the vortices present in the flow counted
individually. This fact provides an easy way to estimate the total line
length of the vortices in the flow. One simply has to calculate the
total incompressible momentum omitting the first wavenumbers, and
compare it to the momentum of a system where only one straight vortex
line spanning the whole box length is present. As a result, the  total
vortex length $L_V$ is
\begin{equation}
    \frac{L_V}{2\pi} = \frac{ \int_3^{k_{\rm max}} P^i(k) dk }{
      \int_3^{k_{\rm max}} P^i_{\mathrm{single}} (k) dk },
    \label{vortex_line_length}
\end{equation}
where $k_{\rm max}$ is the maximum resolved wavenumber in the
simulation, $P^i_{\mathrm{single}}(k)$ is the incompressible
momentum power spectrum of a single vortex core (which can be
calculated numerically by preparing the adequate initial conditions,
or semi-analytically by using an axisymmetric solution of the GPE, see 
\cite{Nore97a}), and the factor $2\pi$ is the length of the
computational domain. The average intervortex distance, $\ell$, can
finally be estimated from the total vortex length by looking at the
vortex line density, $L_V/V$, in the following way
\begin{equation}
    \ell^{-2} = \frac{L_V}{V}.
    \label{iv-distance}
\end{equation}

\subsection{Zero-temperature initial data
  preparation}\label{sec:prepar}

The TG initial condition $\psi_{\rm TG}$ is such that its nodal lines
correspond to vortex lines of the so-called Taylor-Green flow.  In
dimensionless units, the TG velocity flow ${\bf u}_{\rm TG}$ is
defined as
\begin{eqnarray}
    u^{\rm TG}_x(x,y,z) &=& \sin(x) \cos(y) \cos(z) , \nonumber \\
    u^{\rm TG}_y(x,y,z) &=&-\cos(x) \sin(y) \cos(z) , \nonumber \\
    u^{\rm TG}_z(x,y,z)  &=& 0 .
\label{tg3d}
\end{eqnarray}

\subsubsection{Taylor-Green symmetries}\label{sec:app1}

The symmetries of the TG velocity field are rotational symmetries of 
angle $\pi$ around the axes $x = z = \pi/2$, 
$y = z = \pi/2$, and $x = y = \pi/2$, and mirror symmetries with
respect to the planes $x=0$ \& $\pi$, $y=0$ \& $\pi$, and $z=0$ \&
$\pi$.  The TG velocity field is parallel to these planes, that form
the sides of an {\it impermeable box} which confines the flow. It is
demonstrated in Ref.~\cite{Brachet1983} that when using 
${\bf u}_{\rm TG}$ as initial data for the Navier-Stokes equations,
these symmetries are preserved by the dynamics, and that its solutions
admit the following Fourier expansion:
\begin{eqnarray}
    u_{x}&=&\sum_{m=0}^{\infty} \sum_{n=0}^{\infty} \sum_{p=0}^{\infty}
    \hat{u}_{x}(m,n,p) \sin mx \cos ny \cos pz,  \nonumber\\
    u_{y}&=&\sum_{m=0}^{\infty} \sum_{n=0}^{\infty} \sum_{p=0}^{\infty}
    \hat{u}_{y}(m,n,p) \cos mx \sin ny \cos pz,  \nonumber\\
    u_{z}&=&\sum_{m=0}^{\infty} \sum_{n=0}^{\infty} \sum_{p=0}^{\infty}
    \hat{u}_{z}(m,n,p) \cos mx \cos ny \sin pz, \label{tgv}
\end{eqnarray}
where $\hat {\bf u}(m,n,p)$ vanishes unless $m,n,p$ are either all
even or all odd integers. The expansion coefficients should also
satisfy:
\begin{eqnarray}
    \hat{u}_{x}(m,n,p) &=& (-1)^{r+1} \hat{u}_{y}(n,m,p), \nonumber \\
    \hat{u}_{z}(m,n,p) &=& (-1)^{r+1} \hat{u}_{z}(n,m,p),
    \label{coeftg}
\end{eqnarray}
where $r=1$ when $m,n,p$ are all even, and $r=2$ when $m,n,p$ are all
odd. These come from the fact that the TG flow has a rotational symmetry
of angle $\pi/2$ around the axis $x = y = \pi/2$.

%%%%%%%%%%%%%%%%%%%%%%%%%%%%%%%%%%%%%%%%%%%
\begin{table}
\begin{tabular}{p{1.2cm} p{2.3cm} p{2.3cm} p{2.3cm}}
 \hline \hline
 & \multicolumn{3}{c}{Resolution} \\
 \cline{2-4}
 $\xi k_{\rm max}$ & $N=128$ & $N=256$	&$N=512$ \\
 \hline \hline
 $1.5$           & --	       &$e_\lambda=4.17$     & -- \\
                     &                &$T_\lambda=3.25$    &      \\
 \hline
 $2.5$ &$e_\lambda=10.64$ &$e_\lambda=10.13$ &$e_\lambda=9.12$ \\
           &$T_\lambda=9.5$ &$T_\lambda=9.25$  &$T_\lambda=8.75$ \\
 \hline
 $4$              & --            &$e_\lambda=24.12$  & -- \\
                     &                 &$T_\lambda=23.5$    &      \\
 \hline \hline
\end{tabular}
\caption{Transition energy $e_\lambda=E_\lambda/V$, and temperature 
  $T_\lambda$ in the phase transition, for sets of runs at different
  spatial linear resolution $N$, and with different values of 
  $\xi k_{\rm max}$. Values of the energies and temperatures are in
  units of $MU^2L^{-3}$.}
\label{tab:1}
\end{table}
%%%%%%%%%%%%%%%%%%%%%%%%%%%%%%%%%%%%%%%%%%%

These symmetries can be extended to flows described by the
GPE in Eq.~\eqref{Eq:GPE}.  It is easy to show that the expressions in
Eq.~(\ref{tgv})  applied to $\rho v_j$, with $v_j=\partial_j \phi$,
(see Eq.~\ref{Eq:defMadelung}), correspond to the following
decomposition for the complex scalar $\psi(x,y,z,t)$ as a solution
of the GPE
\begin{equation} \label{expanpsi}
    \psi = \sum_{m=0}^{\infty} \sum_{n=0}^{\infty} \sum_{p=0}^{\infty}
    \hat{\psi}(m,n,p) \cos mx \cos ny \cos pz ,
\end{equation}
with $\hat{\psi}(m,n,p)=0$ unless $m,n,p$ are either all even or all odd
integers.  The additional conditions then become
\begin{equation} \label{coefpsi}
    \hat{\psi}(m,n,p) = (-1)^{r+1} \hat{\psi}(n,m,p)
\end{equation}
with the same convention as above.  Implementing these relations in a
numerical code yields savings of a factor $64$ in computational time
and memory size when compared to the general Fourier expansion.

\subsubsection{Taylor-Green initial data}\label{sec:app2}

In order to create the initial condition $\psi_{\rm TG}$ with zeros
along vortex lines of ${\bf u}_{\rm TG}$, we make use of the Clebsch
representation of the velocity field \cite{Nore97a,Nore97b}.  The
Clebsch potentials
\begin{eqnarray}
    \lambda(x,y,z) &=& \cos x \sqrt{2 \;|\cos z |} , \nonumber \\
    \mu(x,y,z) &=& \cos y  \sqrt{2 \; |\cos z |} \;
                   {\mbox{sgn}}(\cos z ) ,
    \label{clebsch3d}
\end{eqnarray}
(where sgn is the sign function) generate the TG flow in
Eq.~(\ref{tg3d}), in the sense that 
$\nabla \times {\bf u}_{\rm TG} = \nabla \lambda \times \nabla \mu$. 
Also, note that a zero in the $(\lambda,\mu)$ plane corresponds to a
vortex line of ${\bf u}_{\rm TG}$ (see \cite{Nore97a,Nore97b} for
details).

Defining the 2D complex field $\psi_e$ with a simple zero at the origin
of the ($\lambda,\mu$) plane,
\begin{equation}
    \psi_e(\lambda,\mu)=(\lambda + i \mu) 
    \frac{\tanh(\sqrt{\lambda^2 + \mu^2}/\sqrt{2} \xi)}
    {\sqrt{\lambda^2 + \mu^2}} ,
    \label{psie}
\end{equation}
we obtain a three dimensional field (as a function of $x$, $y$, and
$z$) with one nodal line. We can further define 
\begin{eqnarray}
    \psi_4 (\lambda,\mu) &=& \psi_e(\lambda -\frac{1}{\sqrt{2}}, \mu)
    \psi_e(\lambda, \mu-\frac{1}{\sqrt{2}}), \nonumber \\
    &\times& \psi_e(\lambda +\frac{1}{\sqrt{2}}, \mu)
    \psi_e(\lambda, \mu + \frac{1}{\sqrt{2}}),
    \label{psi4}
\end{eqnarray}
which contains four nodal lines. In order to match the circulation of 
${\bf u}_{\rm TG}$, we finally define a field which will be used below
as initial condition for an equation for data preparation as
\begin{equation}
    \psi_{\rm ARGLE}(x,y,z)=\psi_4
    (\lambda(x,y,z),\mu(x,y,z))^{[\gamma_d/4]}, 
    \label{psitot3d}
\end{equation}
where the ratio of the total circulation to the elementary defect's
circulation is $\gamma_d=(2/\pi) \, \alp$ with $\alp=c
\xi/\sqrt{2}$. Thus, initially each vortex line corresponds to a
multiple zero line.

The final step in the initial data preparation method consists in
running to convergence the Advective Real Ginzburg-Landau Equation
(ARGLE):
\begin{equation} \label{eqeglra}
    \dertt{\psi}=  \frac{\hbar}{2m} \gd \psi +\mu \psi -
    \frac{g}{\hbar}|\psi|^2 \psi -i {\bf u}_{\rm TG} \cdot \nabla
     \psi -\frac{|{\bf u}_{\rm TG}|^2}{2 \hbar/m} \psi ,
\end{equation}
with the initial condition $\psi_{\rm ARGLE}$. The ARGLE evolution
corresponds to the imaginary time propagation of the GPE with a local
Galilean transformation by the velocity field ${\bf u}_{\rm TG}$. Under
ARGLE dynamics the multiple zero lines in $\psi_{\rm ARGLE}$ will
spontaneously split into single zero lines, and the system will
finally converge to initial conditions for the GPE, compatible with
the TG flow, and with minimal sound emission. We denote the resulting
converged state as $\psi_{\rm TG}$.

%%%%%%%%%%%%%%%%%%%%%%%%%%%%%%%%%%%%%%%%%%%
\begin{table}
\begin{ruledtabular}
\begin{tabular}{ccc}
 $T/T_\lambda$ &$e_{\textrm{tot}} [M U^2 /L^3]$ & $n_0$ \\
 \hline \hline
 $0$&$0.129$& $1$\\
 $0.11$& $0.95$& $0.92$\\
 $0.22$&  $1.94$& $0.86$\\
 $0.33$&  $2.96$& $0.75$\\
 $0.44$& $4.02$& $0.65$\\
 $0.55$&  $5.14$& $0.55$\\
\end{tabular}
\end{ruledtabular}
\caption{\label{tab:2} Variation of the total energy
  $e_{\textrm{tot}}$ and of the condensed fraction $n_0$ in the
  thermal runs with $N=1024$ and $\xi k_{\rm max}=2.5$ (for these
  parameters $T_{\lambda}=8.58$, in units of  $M U^2 /L^3$).}
\end{table}
%%%%%%%%%%%%%%%%%%%%%%%%%%%%%%%%%%%%%%%%%%%

\subsection{A finite temperature model}
\label{FTGPtheory}

One of the different possible ways to include finite temperature effects
on the condensate dynamics is by imposing an ultra-violet cutoff on the
GPE. This amounts to performing a Galerkin truncation operation on the
GPE in Fourier space with a 
projection operator $\mathcal{P}_{\rm G}$ defined as
\begin{equation}
    \mathcal{P}_{\rm G}[\hat{\psi}({\bf k})] = 
    \Theta(\kmax-|\bf{k}|)\hat{\psi}(\bf k), 
\end{equation}
where $\hat{\psi}$ is the spatial Fourier transform of $\psi$, $\kmax$
is a suitably chosen ultraviolet cutoff (which, in practice, will be
the same as the maximum resolved wavenumber in the simulations), and
$\Theta$ is the Heaviside function. The resulting Galerkin truncated
GPE (TGPE) is
\begin{equation}
    i\hbar\dertt{\psi}  =\mathcal{P}_{\rm G} \left[- \alps \gd \psi +
      \bet\mathcal{P}_{\rm G} \left[ |\psi|^2 \right]\psi\right] . 
    \label{Eq:TGPEphys}
\end{equation}
The TGPE in Eq.~\eqref{Eq:TGPEphys} exactly conserves energy and 
mass; moreover, if we correctly de-alias it by using the 
$2/3-$dealiasing rule \cite{Got-Ors}, with $k_{\rm max}=2/3 \times
N/2$ (in dimensionless units), it also conserves momentum. We refer to
Ref.~\cite{Krstulovic11} for an explicit demonstration of the latter.
The Galerkin truncation operation also preserves the Hamiltonian
structure with the truncated Hamiltonian of the system given by
\begin{equation}
H=\int d^3 x \left[ \alps |\nabla \psi |^2
  +\frac{g}{2}\left( \mathcal{P}_{\rm G} |\psi|^2 \right)^2
  \right].
\label{Eq:HGalerkin}
\end{equation}

The grand canonical equilibrium states are given by the following
stationary probability distribution
\begin{equation}
	\mathbb{P}_{\rm st}[\psi]=\frac{1}{\mathcal{Z}}e^{-\beta
          [H-\mu \mathcal{N}_p]}, 
\label{Eq:StatProb}
\end{equation}
where $\mathcal{Z}$ is the grand partition function, 
$\beta=1/(k_{\rm B}T)$ is the inverse temperature, and $k_{\rm B}$ is
the Boltzmann constant. However, these states are difficult to compute
as the Hamiltonians in Eqs.~\eqref{Eq:defH} or \eqref{Eq:HGalerkin} are
not quadratic, and the resulting statistical distribution is
non-Gaussian. Nevertheless, it is possible to construct a stochastic
process that converges to a stationary solution with equilibrium
distribution given by Eq.~\eqref{Eq:StatProb}. This process is defined
by a Langevin equation consisting of a stochastic Ginzburg-Landau
equation (SGLE) that explicitly reads in physical space
\begin{eqnarray}
    \nonumber\hbar\dertt{\psi} &=&\mathcal{P}_{\rm G} \left[ \alps \gd
      \psi +\mub \psi- \bet\mathcal{P}_{\rm G} \left[|\psi|^2\right
      ]\psi \right]\\
    &&\hspace{1.5cm}+\sqrt{\frac{2 \hbar}{V\beta }}\mathcal{P}_{\rm G}
       \left[ \zeta({\bf x},t) \right] \label{Eq:SGLRphys},
\end{eqnarray}
where the Gaussian white noise $\zeta({\bf x},t)$ obeys
\begin{equation}
    \langle\zeta({\bf x},t) \overline{\zeta}({\bf
      x'},t')\rangle=\delta(t-t') \delta({\bf x}-{\bf x'}).
    \label{Eq:noise}
\end{equation}
We refer to Ref.~\cite{Krstulovic11} for more details on the
proof of the equivalence of this stationary probability
distribution to the grand canonical equilibrium state.

If one wants to control the number of particles $\mathcal{N}_p$
instead of the chemical potential $\mu$, then one must supplement the
SGLE with an \textit{ad-hoc} equation for the chemical potential
\begin{equation}
    \dert{\mu}=-\frac{\nu_N}{V}(\mathcal{N}_p-\mathcal{N}_p^*) ,
    \label{Eq:muAtRho}
\end{equation}
where $\mathcal{N}_p^*$ controls the mean number of particles and
$\nu_N$ governs the rate at which SGLE equilibrates.

We will call the thermal states generated by the SGLE
$\psi_{\mathrm{th}}$. These states can be used in the TGPE to compute 
their dynamical properties. Moreover, we can combine these thermal
states with an initial condition for a large-scale flow to simulate
quantum turbulence at finite temperature. For the TG flow, the
combined initial state in this case reads:
\begin{equation}
    \psi = \psi_{\mathrm{TG}} \times \psi_{\mathrm{th}}.
    \label{inipsi}
\end{equation}

In the present study, we perform several DNSs of the SGLE in
Eq.~\eqref{Eq:SGLRphys}) and of the TGPE in
Eq.~\eqref{Eq:TGPEphys}. For numerical purposes we rewrite the  
SGLE (omitting the Galerkin projector $\mathcal{P}_{\rm G}$) as
\begin{eqnarray}
    \nonumber\dertt{\psi} &=& \alpha_0\gd \psi +\Omega_{0} \psi
	- \beta_{0}|\psi|^2\psi+\sqrt{\frac{k_{B}T}{\alpha_0}}\zeta,
\end{eqnarray}
where $\alpha_0$, $\Omega_{0}$, and $\beta_{0}$ are parameters. We
can express physically relevant quantities, such as the coherence
length $\xi$ and the velocity of sound $c$, in terms of these new
parameters. These are related by
\begin{equation}
    \xi=\sqrt{\alpha_0/\Omega_{0}\,}, \hspace{.5cm}
    c=\sqrt{2\alpha_{0}\beta_{0}\rho_0} ,
\end{equation}
with $\rho_0=\Omega_{0}/\beta_{0}$. In all the DNS runs presented
below we set the density at $T=0$ to $\rho_0=1$ (in dimensionless
units as described below). In order to keep the value of intensive
variables constant in the thermodynamic limit, at constant volume $V$
and for $\kmax\to \infty$, the inverse temperature is expressed as
$\beta=1/(k_{\rm modes}T)$, where $k_{\rm modes}=V/\mathcal{N}_m$ with
$\mathcal{N}_m$ the number of Fourier modes in the system. With these
definitions the temperature $T$ has units of energy per volume, and
$4\pi\alpha_0$ is the quantum of circulation.

\subsection{Numerical implementation} \label{sec:app3}

The code, TYGRS (TaYlor-GReen Symmetric), is a pseudo-spectral code
that enforces the symmetries of the TG vortex in 3D for the GPE, the
Navier-Stokes equations, and the magnetohydrodynamic equations within
periodic cubes of length $2\pi$ (in dimensionless units). As a result
of the symmetries discussed in Sec.~\ref{sec:app1}, the
Fourier-transformed fields are non-zero only for wave vectors
$(k_x,k_y,k_z) = (m,n,p)$ with jointly even or jointly odd
components. Time integration of only these Fourier modes is performed
using a fourth-order Runge-Kutta method.

Pseudo-spectral codes are known to be optimal on  periodic domains
\cite{Got-Ors}. However, they require global spectral transforms, and
thus are hard to implement in distributed memory environments, a crucial
limitation until domain decomposition techniques (DDTs) arose
\cite{Calvin96, Dmitruk01} that allowed computation of serial Fast
Fourier Transforms (FFTs) in different directions in space (local in
memory) after performing transpositions. However, distributed
parallelization using the Message Passing Interface (MPI) in
pseudo-spectral codes is limited in the number of processors that can
be used, unless more transpositions are done per FFT (thus increasing
communication time). To overcome this limitation, the hybrid
(MPI-OpenMP) parallelization scheme we have implemented in TYGRS
builds upon a general purpose one-dimensional (slab-based) DDT that is
effective for parallel scaling using MPI alone \cite{gomez05},
extended with OpenMP to obtain an (in practice) 2D DDT without the
need of extra communication \cite{Mininni11}. In this scheme, each MPI
task creates multiple threads using OpenMP which operate over a
fraction of the available data. This method has been extended in TYGRS
to the sine/cosine with even/odd wavenumber FFTs needed to implement
the symmetries of TG flows, using loop-level OpenMP directives and
multi-threaded FFTs. The method was shown to scale with high parallel
efficiency to over 100,000 CPU cores \cite{Mininni11}.

The runs were performed on the IDRIS BlueGene/P machine. At resolution
$N=4096$ we used $512$ MPI processes, each process spawning $4$ 
OpenMP threads, needing a total of $2048$ CPU cores per simulation.

\subsection{Units}
\label{units}

In the following, all quantities are expressed in terms of a unit
length $L$, a unit speed $U$, and a unit mass $M$. These are related
to the simulation length $L'$, the characteristic speed $U'$, and the
actual mass $M'$ in the following way 
\begin{align}
    L &= \frac{L'}{2\pi}, \label{eq:unit1}
    \\
    U &= \frac{U'}{2}, \label{eq:unit2}
    \\
    M &= \frac{M'}{(2\pi)^3}. \label{eq:unit3}
\end{align}
With these choices the simulation box is $2\pi L$ long (in each
spatial direction), the speed of sound $c$ is $2 U$, and the mean
density $\rho_0$ is equal to $1 \, M/L^3$. The factors in
Eqs.~(\ref{eq:unit1}) to (\ref{eq:unit3}) result from the
dimensionless scheme used in the simulations (done in a periodic box
of dimensionless side $2\pi$).

In some of the simulations we present next, the healing length $\xi$
is such that $\xi k_{\max}=1.5$, or $\xi k_{\max}=2.5$, with a few
cases with $\xi k_{\max}=4$ (each case is appropriately indicated in
the text). As a result, in the simulation with the largest spatial
resolution in this work with $\xi k_{\max}=2.5$, the healing length is 
$\xi \approx 0.0018 L'$. While the resolution in this simulation is
state-of-the-art, the scale separation is not sufficient to be able to
compare with superfluid $^4$He experiments, where the characteristic
system size is  $L' \approx 10^{-2}$ m, the speed of sound is
$c'\approx 230$ m/s, the fluid density is $\approx 125$ kg/m$^3$ 
(thus $M' \approx 1.25 \times 10^{-4}$ kg), and the healing length is 
$\xi' \approx 10^{-8} \, \textrm{m} \approx 10^{-6} L'$
\cite{Barenghi14}. On the other hand, scale separation in BEC
experiments of quantum turbulence, where $L' \approx 10^{-4}$ m, 
$c'\approx 2\times10^{-3}$ m/s, and 
$\xi \approx 5 \times 10^{-7} \, \textrm{m} \approx 0.005 L'$
\cite{White14, Tsatsos16}, is within our reach.

Except when explicitly noted, temperatures will be expressed in
terms of the transition temperature $T_\lambda$. Finally, note that
the intensity of non-linear interactions is controlled by the inverse
of $\xi\kmax$ \cite{Krstulovic11}. Indeed, for $\xi\kmax$ very large,
most of the excitations correspond to free particles. More details on
how units can be handled in DNSs of the GPE and SGLE can be found in
\cite{Nore97a,Krstulovic11a,Krstulovic11,Clark18}.

\section{Results}
\label{Sec:results}

We start this section by discussing the temporal evolution of the TG
flow at zero temperature, using a resolution of $4096^3$ collocation
points as well as simulations at lower resolution. We then perform a
series of temperature scans to study the decay of the TG
initial conditions at finite temperature. Finally, by computing the
spatio-temporal spectra of these flows, we provide an estimation of 
the effective viscosity in flows evolved under the TGPE.

\subsection{High resolution GPE runs at $T=0$}
\label{sec:hirest0}

\begin{figure}
    \centering
    \includegraphics[width=8.5cm]{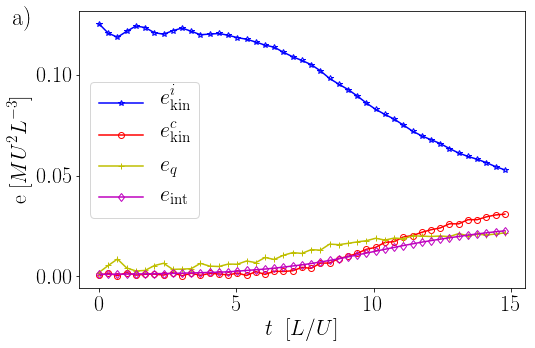}
    \includegraphics[width=8.5cm]{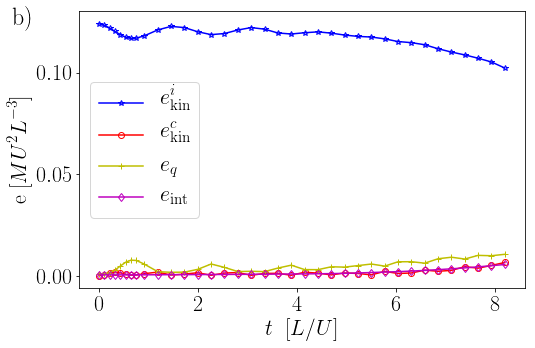}
    \includegraphics[width=8.5cm]{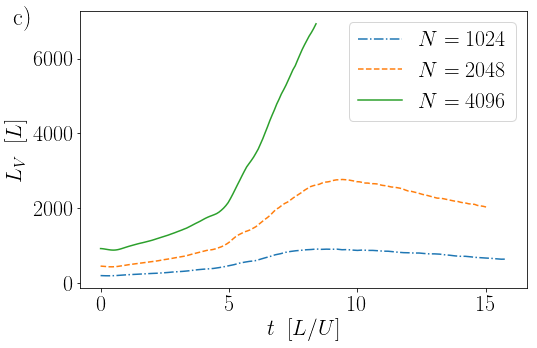}
    \caption{{\it (Color online)} Time evolution of the different
      energy components at zero temperature in (a) a 3D simulation
      with $N=2048$ grid points in each spatial direction, and (b)
      with $N=4096$ grid points in each direction. Both DNSs have 
      $\xi k_{\rm max}=2.5$. (c) Time evolution of the total vortex
      line length $L$ in simulations at different spatial
      resolutions.}
    \label{fig:energies}
\end{figure}

The TG initial data obtained following
Eqs.~(\ref{psi4})-(\ref{eqeglra}) was first produced with 
$\xi k_{\rm max}=2.5$ and with linear spatial resolutions of $N=1024$,
$2048$ and $4096$ grid points. The time evolution of the energies
defined in Eqs.~\eqref{Eq:defEkin}, \eqref{Eq:defEint}, and 
\eqref{Eq:defEq} under the GPE dynamics are shown in
Figs.~\ref{fig:energies}(a) and \ref{fig:energies}(b) for the
simulations with $N=2048$ and $N=4096$, respectively. In all cases,
the total energy was conserved within a $2\%$ error. The
incompressible kinetic energy per unit volume $e_{\rm kin}^i$ remains
approximately constant until $t \approx 4 \, L/U$, and afterwards it
starts decaying as the other energy components increase to keep the
total energy fixed. This indicates a transfer of energy from 
$e_{\rm kin}^i$ to the other energy components as turbulence develops
(most conspicuously at late times for the $N=2048$ run, to the
compressible component $e_{\rm kin}^c$). The vortex line length $L_V$,
as defined in Eq.~\eqref{vortex_line_length}, is shown in
Fig.~\ref{fig:energies}(c) for all three simulations. At around 
$t \approx 8 \, L/U$, $L_V$ peaks, and thus the maximum of
incompressible kinetic energy dissipation is reached.

%%%%%%%%%%%%%%%%%%%%%%%%%%%%%%%%%%%%%%%%%%%
%%%%%%%%%%%%%%%%%%%%%%%%%%%%%%%%%%%%%%%%%%%

\begin{figure}
    \centering
    \includegraphics[width=8.5cm]{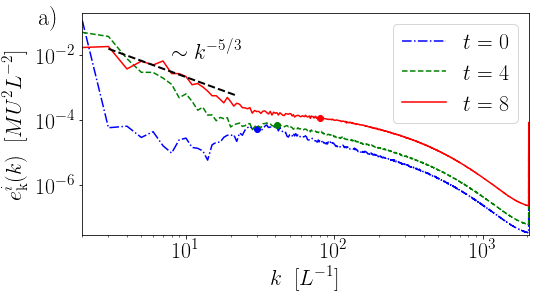}
    \includegraphics[width=8.5cm]{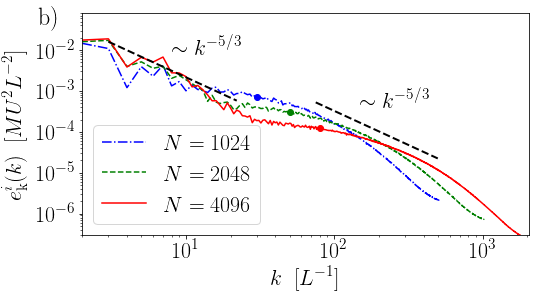}
    \includegraphics[width=8.5cm]{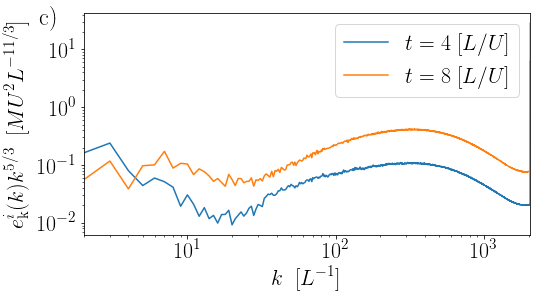}
    \includegraphics[width=8.5cm]{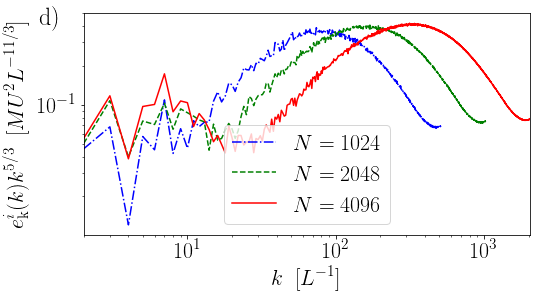}
    \caption{{\it (Color online)} a) Incompressible kinetic energy
      spectra $e^i_{\rm kin}(k)$ at zero temperature, at different
      times for the $N=4096$ simulation. b) Spectra at $t=8$ for three
      DNSs at different spatial resolution. In both panels, the
      circular marks indicate the mean intervortex wavenumber at the
      corresponding time, or for the corresponding spatial
      resolution. Power laws $\sim k^{-5/3}$ are shown as a reference,
      for scales larger and smaller than the intervortex scale. Note
      the emergence of a second intertial range, after a bottleneck,
      for scales smaler than the intervortex scale. 
      (c) and (d) show the
	spectra from panel (a) and (b), but compensated by Kolmogorov
        scaling.	}
    \label{fig:spec}
\end{figure}

In Fig.~\ref{fig:spec}(a) we show the incompressible kinetic energy
spectra of the $N=4096$ simulation at different times, while in
Fig.~\ref{fig:spec}(b) we present the same spectra at $t=8 \, L/U$ for
the $N=1024$, $N=2046$, and $N=4096$ simulations.  Round markers
indicate the mean intervortex wavenumber $k_\ell = 2\pi/\ell$, and the
dashed lines indicate $k^{-5/3}$ power laws as a reference. On the one
hand, at wavenumbers smaller than $k_\ell$, strong hydrodynamic
turbulence is known to be the principal mechanism to transfer energy
towards smaller scales. A Kolmogorov-like spectrum can thus be expected
in this range of scales. On other hand, at wavenumbers larger than
$k_\ell$, energy is expected to be carried towards even smaller scales
by the Kelvin wave cascade \cite{Lvov10}. This cascade, predicted with
weak-wave turbulence theory, also leads to a $k^{-5/3}$ scaling but with
a different origin from the one of Kolmogorov. Note that the Kelvin wave
cascade has been studied using the GPE before \cite{Krstulovic12}, and
it has been observed in GPE turbulence using spatio-temporal analysis
\cite{Clark15a} and by direct measurement of vortex line excitations
\cite{PRETangleVillois}. Also, the Kolmogorov and Kelvin wave cascades
transfer the energy towards smaller scales at different rates. It is
thus expected that energy should accumulate near the wavenumber
$k_\ell$, resulting in a bottleneck in the spectrum
\cite{LvovNazarenkoBottleneck}. We indeed observe the emergence of a
bottleneck in the vicinity of this wavenumber, although not as
pronounced as the thermalization scaling $\sim k^2$. This difference
might be due to the fact that the present simulations are freely
decaying, and have no force acting to sustain turbulence. The existence
of two simultaneous inertial ranges separated by a bottleneck was also
observed before in high resolution simulations using different initial
conditions \cite{Clark17}, but was not visible in the $N=512$ DNS of a
TG flow in \cite{Nore97a} possibly as a result of the limited spatial
resolution in that study. To further illustrate these ranges, and the
scale separation involved, in Fig.~\ref{fig:spec}(c) and (d) we show the
incompressible kinetic energy spectra compensated by Kolmogorov scaling
$\sim k^{-5/3}$.

%%%%%%%%%%%%%%%%%%%%%%%%%%%%%%%%%%%%%%%%%%%
%%%%%%%%%%%%%%%%%%%%%%%%%%%%%%%%%%%%%%%%%%%

\begin{figure}
    \centering
    \includegraphics[width=8.1cm]{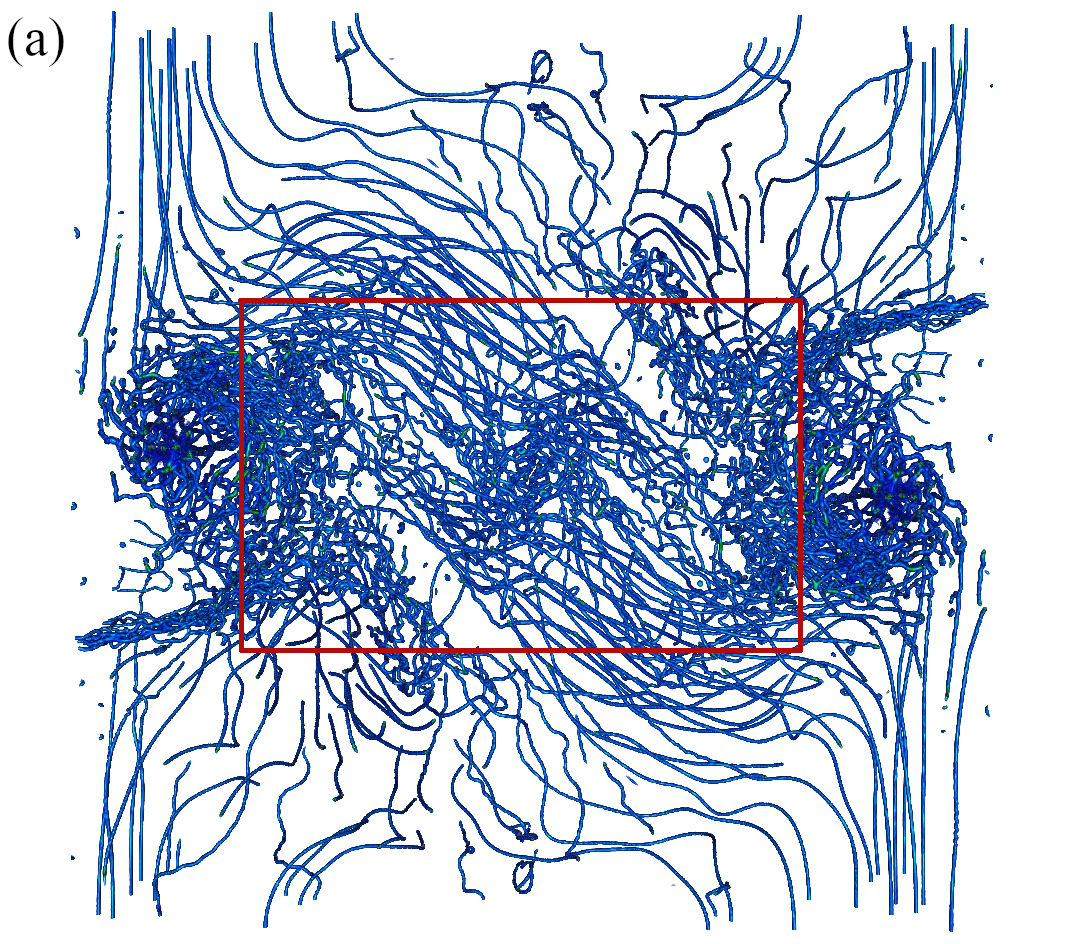}
    \includegraphics[width=8.1cm]{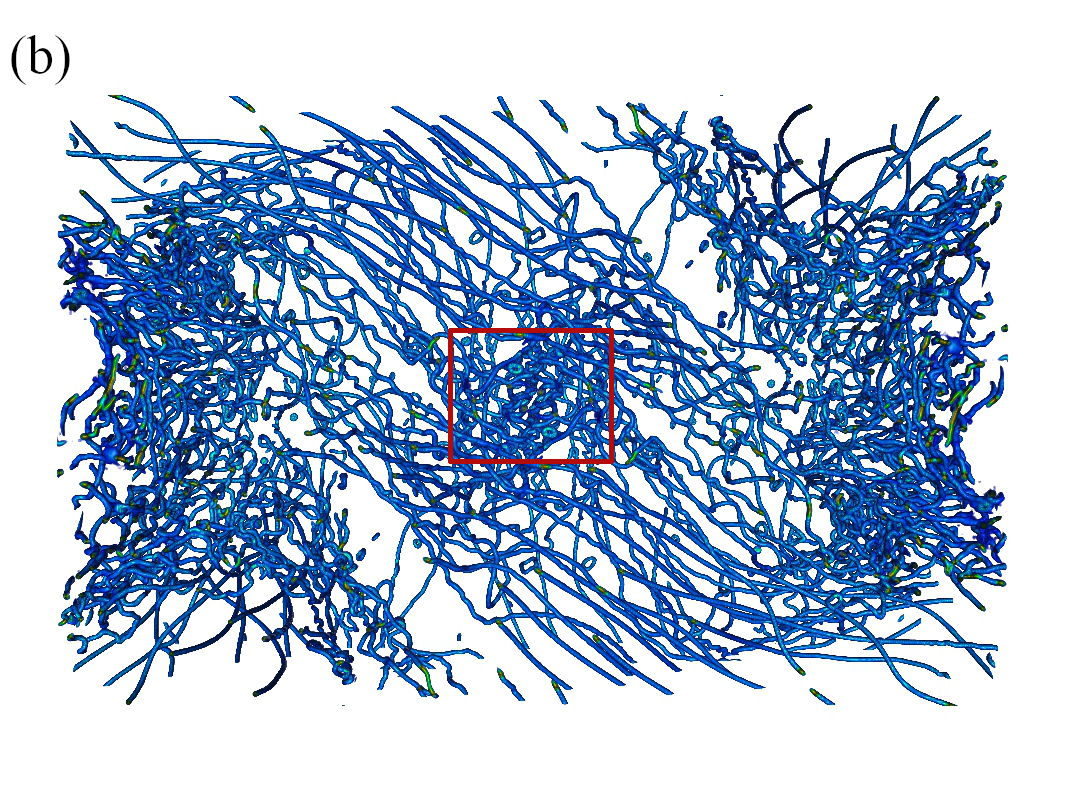}
    \includegraphics[width=8.1cm]{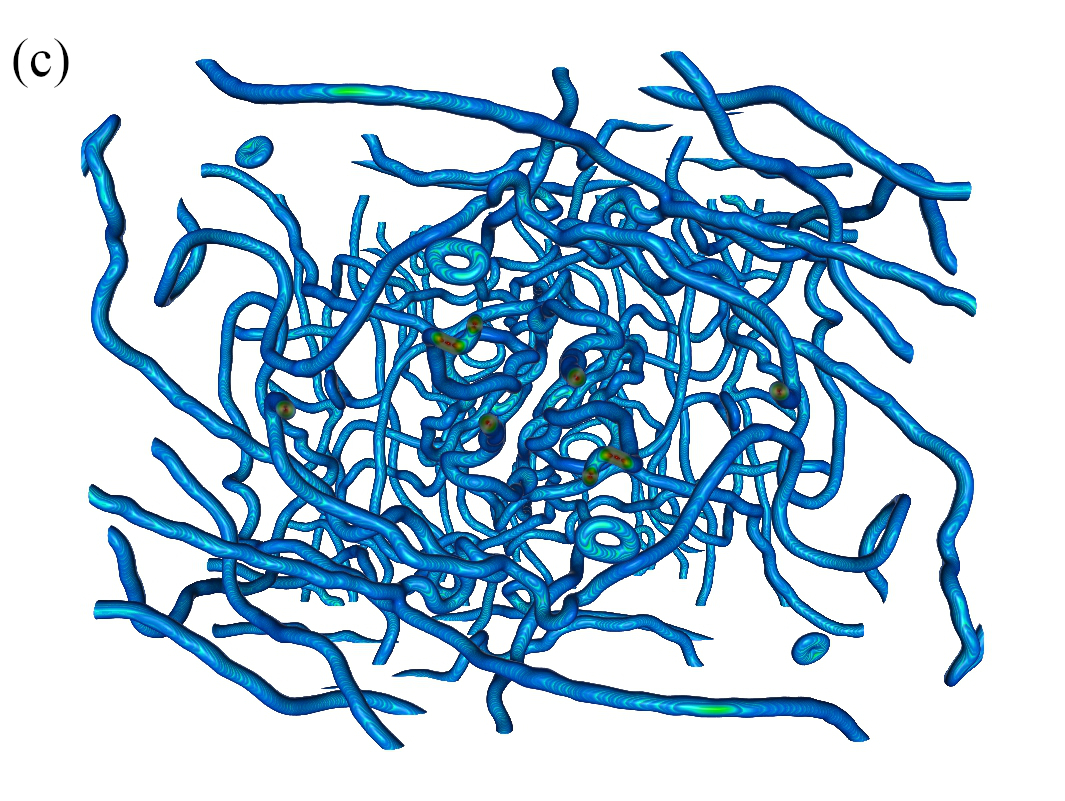}
    \caption{{\it (Color online)} Three-dimensional renderings of 
      vortex lines at the onset of the decay in the $4096^4$ GPE
      run. (a) The full impermeable box. The (red) box indicates the region shown in
      panel (b), which shows a zoom into a region of the domain with
      large shear of the vortex lines. Again, the (red) box in this
      panel indicates the region zoomed into in panel (c). Note in the
      latter panel the tangle of vortices, the many links between vortices,
      and the helical deformations of individual vortices along the
      vortex line. Visualizations were prepared using the software
      VAPOR \cite{Clyne07}.}
    \label{fig:vaporfb}
\end{figure}

Visualizations of the vortex lines in the $N=4096$ run close to the
time of maximum energy dissipation are shown in Fig.~\ref{fig:vaporfb}.
The intricate vortex line tangle in the entire computational domain
(for the TG {\it impermeable box}) is shown first. The large-scale
flow shows inhomogeneous regions with high density of vortices and
quiet regions with low density. Details into the central regions with
high density of vortices (and large shear) are also shown. It should
be noted that the tangle of vortices results from many reconnections
taking place after $t \approx 4 \, L/U$.  Comparing these $N=4096$
results with those obtained $20$ years ago at resolution $N=512$ and
presented in Fig.~18 of Ref.~\cite{Nore97a}, we can note the presence
of substructures made by a myriad of small-scale and knotted and
linked vortices that were not apparent at the lower resolution.

%%%%%%%%%%%%%%%%%%%%%%%%%%%%%%%%%%%%%%%%%%%
%%%%%%%%%%%%%%%%%%%%%%%%%%%%%%%%%%%%%%%%%%%

%%%%%%%%%%%%%%%%%%%%%%%%%%%%%%%%%%%%%%%%%%%
\begin{figure}
    \centering
    \includegraphics[width=0.98\linewidth]{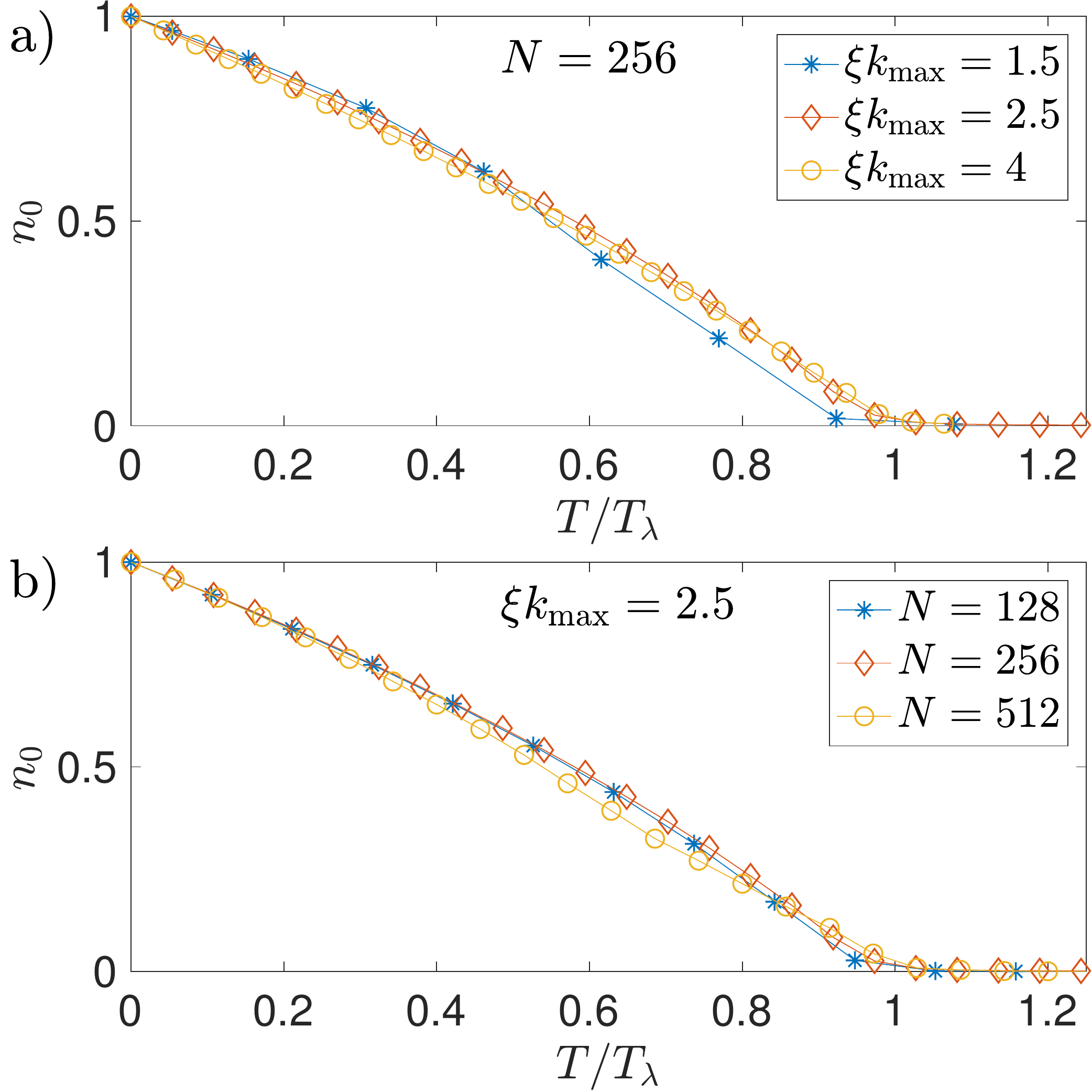}
    \caption{{\it (Color online)} (a) Condensed fraction as a function
    of $T/T_\lambda$ at linear resolution $N=256$ and for different
    values of $\xi\kmax$. (b) Condensed fraction as a function of
    $T/T_\lambda$ at $\xi\kmax=2.5$ and for different linear
    resolutions $N$. The condensed fraction is computed using
    Eq.~\eqref{Eq:DefCond} with $k_c=2$ for simulations with $N=128$,
    and with $k_c=4$ for all the other numerical simulations.}
    \label{Fig:CondFraction}
\end{figure}
%%%%%%%%%%%%%%%%%%%%%%%%%%%%%%%%%%%%%%%%%%%

\subsection{SGLE temperature scans} \label{sec:scans}

We now study only thermal states, in order to determine the
condensation temperature $T_\lambda$ in our system with symmetries. We
thus performed a series of SGLE temperature scans, with various values
for the linear resolution $N$ and $\xi k_{\rm max}$, as indicated in
Table \ref{tab:1}. Each box in the table indicates the transition
energy and temperature obtained, for a fixed value of $N$ and 
$\xi k_{\rm max}$, by performing 12 to 24 simulations in each set
varying the temperature. Boxes without data correspond to cases not
explored.

It is well known that the TGPE can capture the condensation transition
\cite{Proukakis08,Davis01,Krstulovic11}. The order parameter of this
phase transition is the condensed fraction, which is usually defined
as the fraction of atoms that are in the ground state. In terms of
Fourier modes, it is given by 
$|\hat{\psi}({\bf k}=0)|^2/\mathcal{N}_p$. However, for the TG flow the
symmetries cancel exactly the energy (and mass density) of some
Fourier modes, decreasing the availability of Fourier modes at low
wavenumbers, and thus affecting the dynamics of the condensed
fraction. As a result, we define the condensed fraction as
\begin{equation}
    n_0=\frac{1}{\mathcal{N}_p}
      \int_{{k=0}}^{k_c} |\widehat{\psi}_{\bf k}|^2 k^2 d\Omega_k ,
    \label{Eq:DefCond}
\end{equation}
where $k_c$ is a small wave-number (either 2 or 4, in dimensionless
units, depending on the spatial resolution $N$).

The condensed fraction as a function of the temperature is displayed
in Fig.~\ref{Fig:CondFraction}(a) for fixed spatial resolution $N$ and
different values of $\xi \kmax$, and in Fig.~\ref{Fig:CondFraction}(b)
for fixed $\xi \kmax$ and different values of $N$. Note the transition
in all cases at $T=T_\lambda$.  The corresponding values of the
transition temperatures $T_{\lambda}$ and energies $e_\lambda$ are
given in Table \ref{tab:1}. 

\begin{figure}
    \centering
    \includegraphics[width=0.98\linewidth]{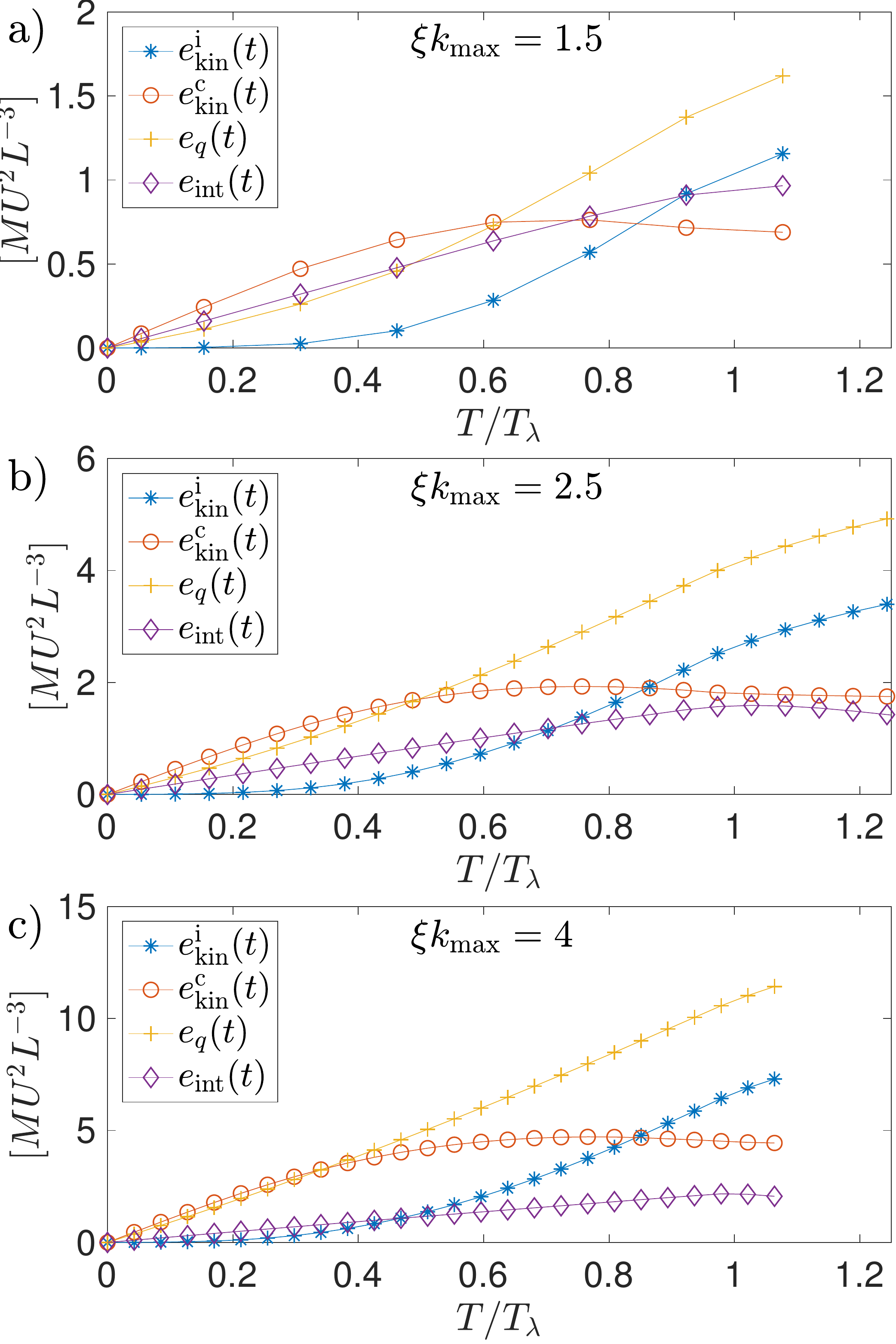}
    \caption{{\it (Color online)} Energy components as a function of
      the temperature, for different values of $\xi k_{\rm max}$, and
      for fixed linear spatial resolution ($N=256$). (a) Temperature
      scan with $\xi k_{\rm max} = 1.5$, (b) same with 
      $\xi k_{\rm max} = 2.5$, and (c) same with 
      $\xi k_{\rm max} = 4$.}
    \label{Fig:EnerguScan}
\end{figure}

The behavior of the different energy components as the temperature is
varied, at fixed linear resolution of $N=256$ and for different values
of $\xi k_{\rm max}$, is shown in Fig.~\ref{Fig:EnerguScan}. Note that
for fixed $\xi k_{\rm max}$, $e_{\rm q}$ increases in all cases with
$T$ up to $T_\lambda$, as also do $e_{\rm int}$ and $e_{\rm kin}^i$,
while $e_{\rm kin}^c$ displays a maximum at intermediate
temperatures. As expected, increasing the value of $\xi\kmax$
decreases the  non-linear interactions, that can be quantified by the
relative value of $e_{\rm int}(t)$. 

%%%%%%%%%%%%%%%%%%%%%%%%%%%%%%%%%%%%%%%%%%%
%%%%%%%%%%%%%%%%%%%%%%%%%%%%%%%%%%%%%%%%%%%

Comparing these TG symmetric results with those obtained for a
general periodic geometry as given in Fig.~2 of
Ref.~\cite{Krstulovic11}, it can be seen that the overall properties
of the condensation transition are nor significantly affected by the
nature of the geometry imposed by the symmetries at the largest
scales. As a result, we now consider the combination of the thermal
states with the TG flow.

\subsection{Thermal equilibria combined with the TG
  flow}\label{sec:combined}

To study finite temperature effects in the TG flow, we prepared
several high resolution thermal states (up to $T=0.55\,T_{\lambda}$) 
at linear resolution of $N=1024$, and with $\xi k_{\rm max}=2.5$ (see
Table~\ref{tab:2} for more details on the energy and condensed
fraction in these states as $T$ is varied). We then combined these
thermal states with the TG initial data prepared following 
Eqs.~(\ref{psi4})-(\ref{eqeglra}).

\begin{figure*}
\centering
\includegraphics[width=0.4\linewidth]{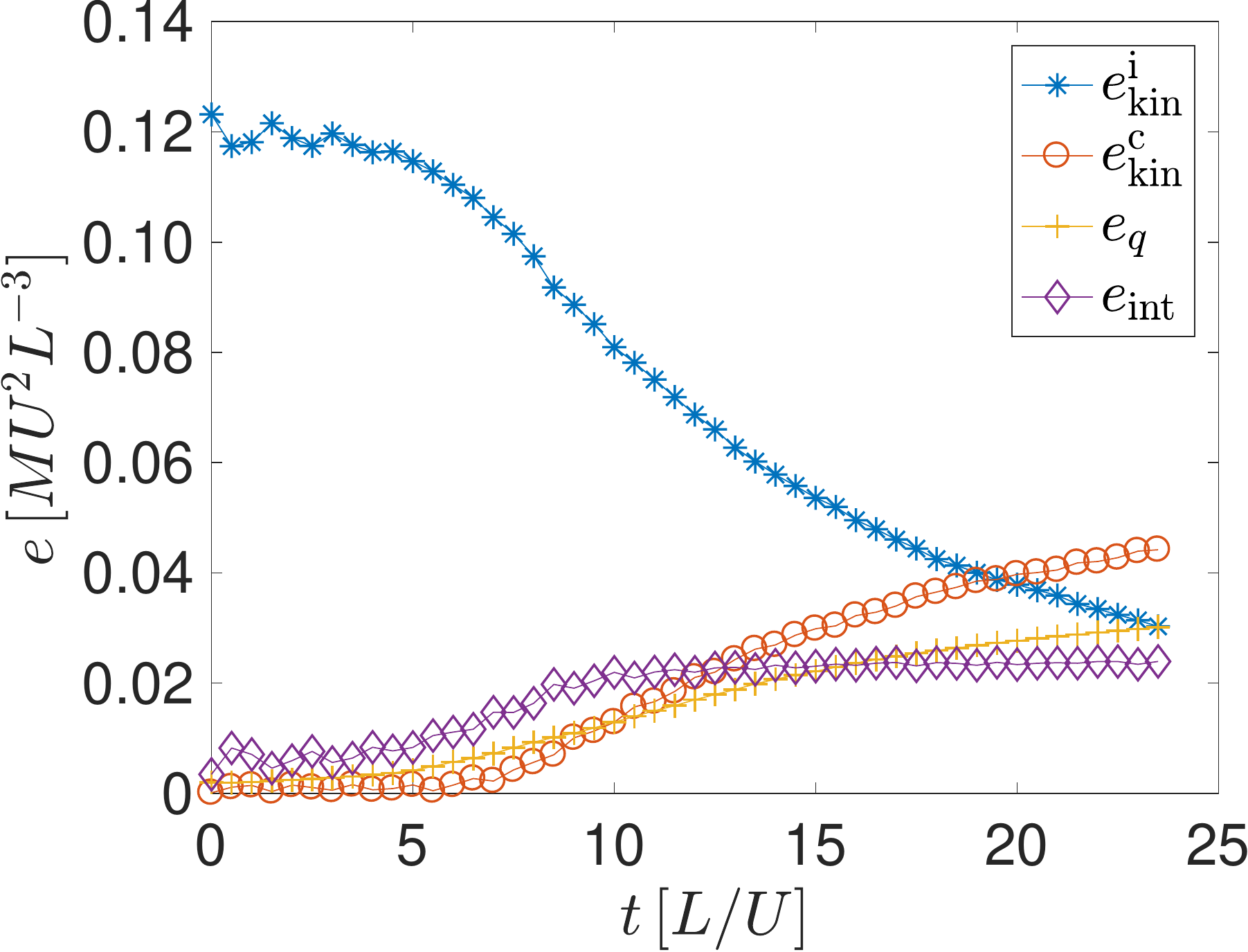}
\put(-25,87){\bf\large (a)}
\hskip .5cm
\includegraphics[width=0.4\linewidth]{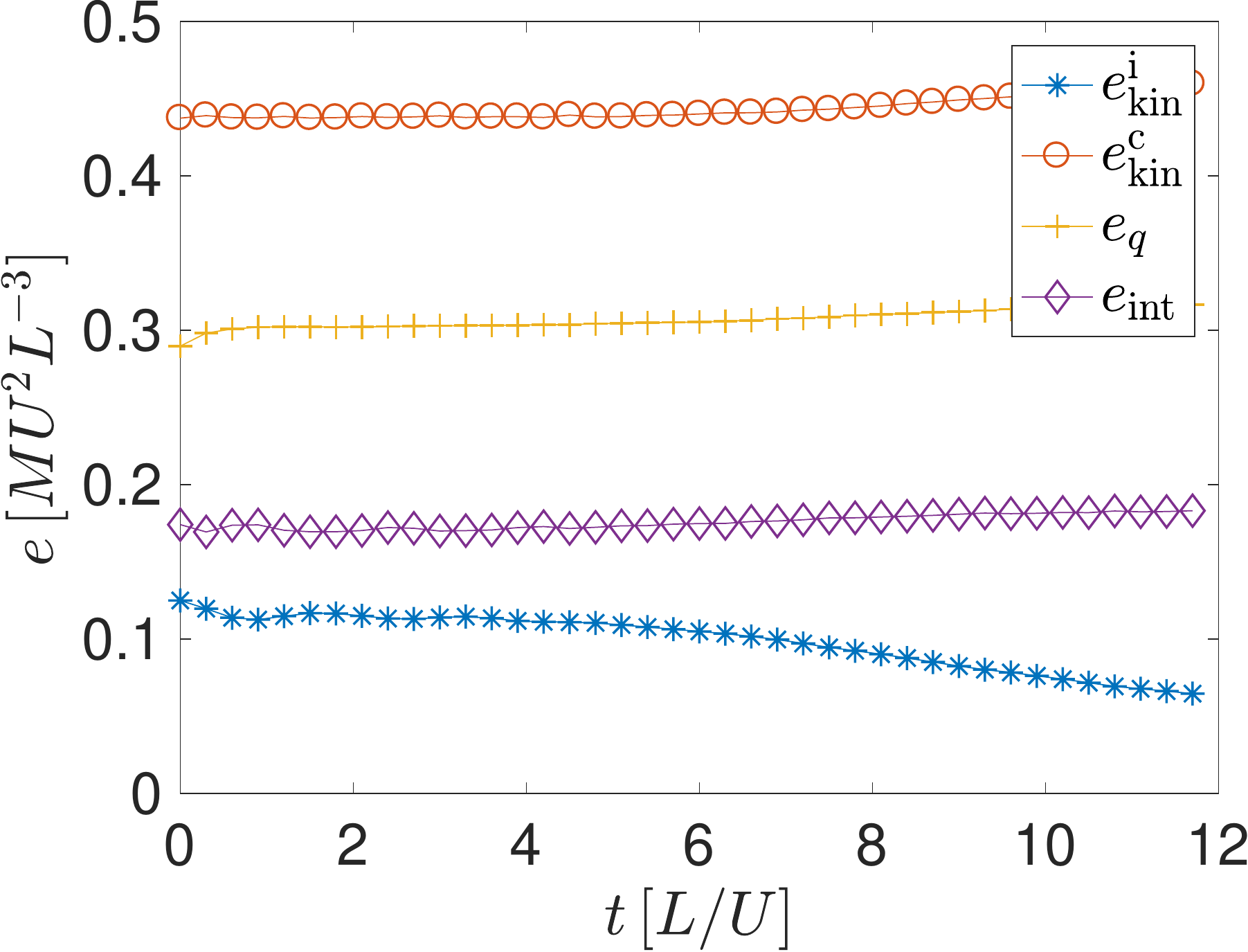}
\put(-25,87){\bf\large (b)}
\\
\includegraphics[width=0.4\linewidth]{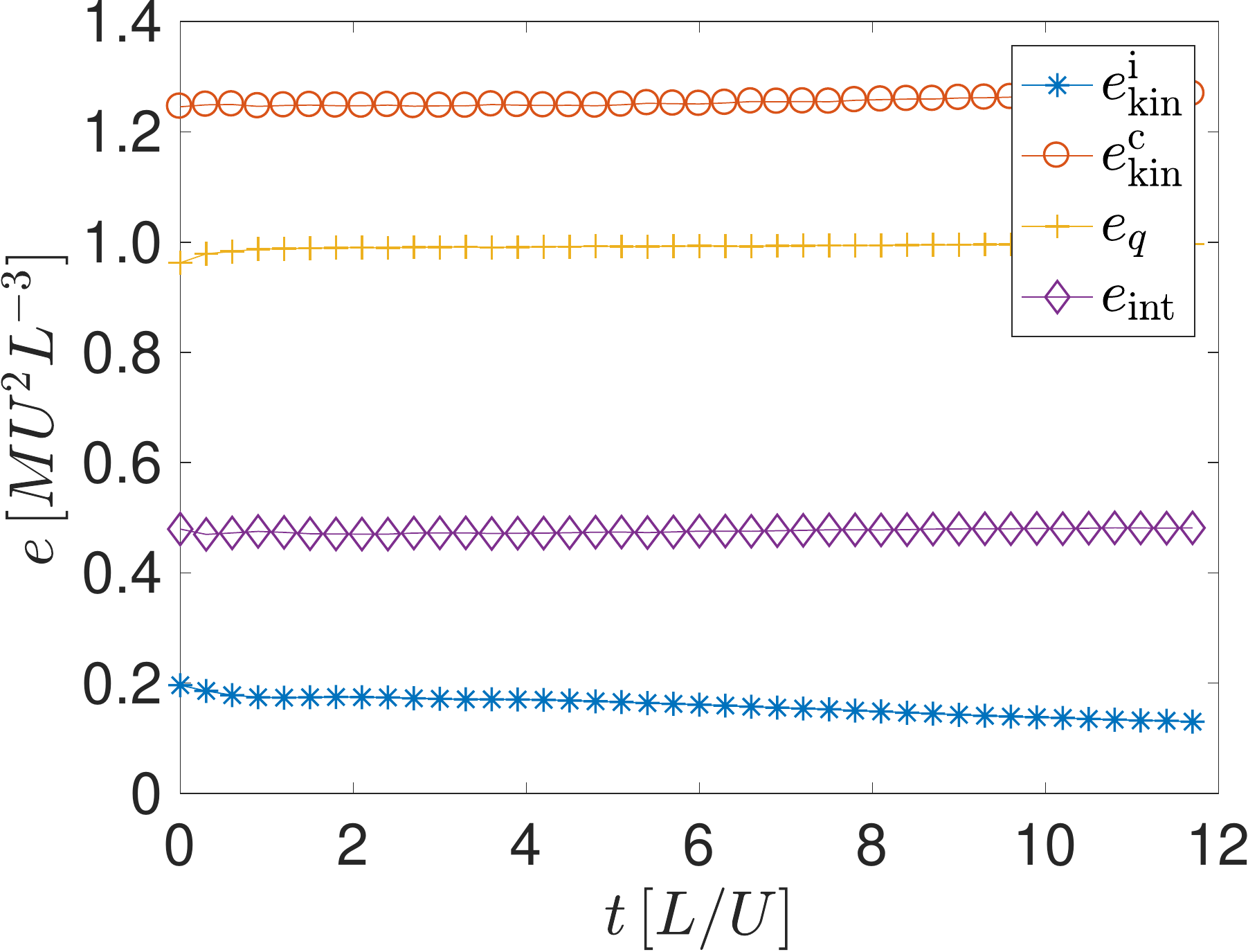}
\put(-25,87){\bf\large (c)}
\hskip .5cm
\includegraphics[width=0.4\linewidth]{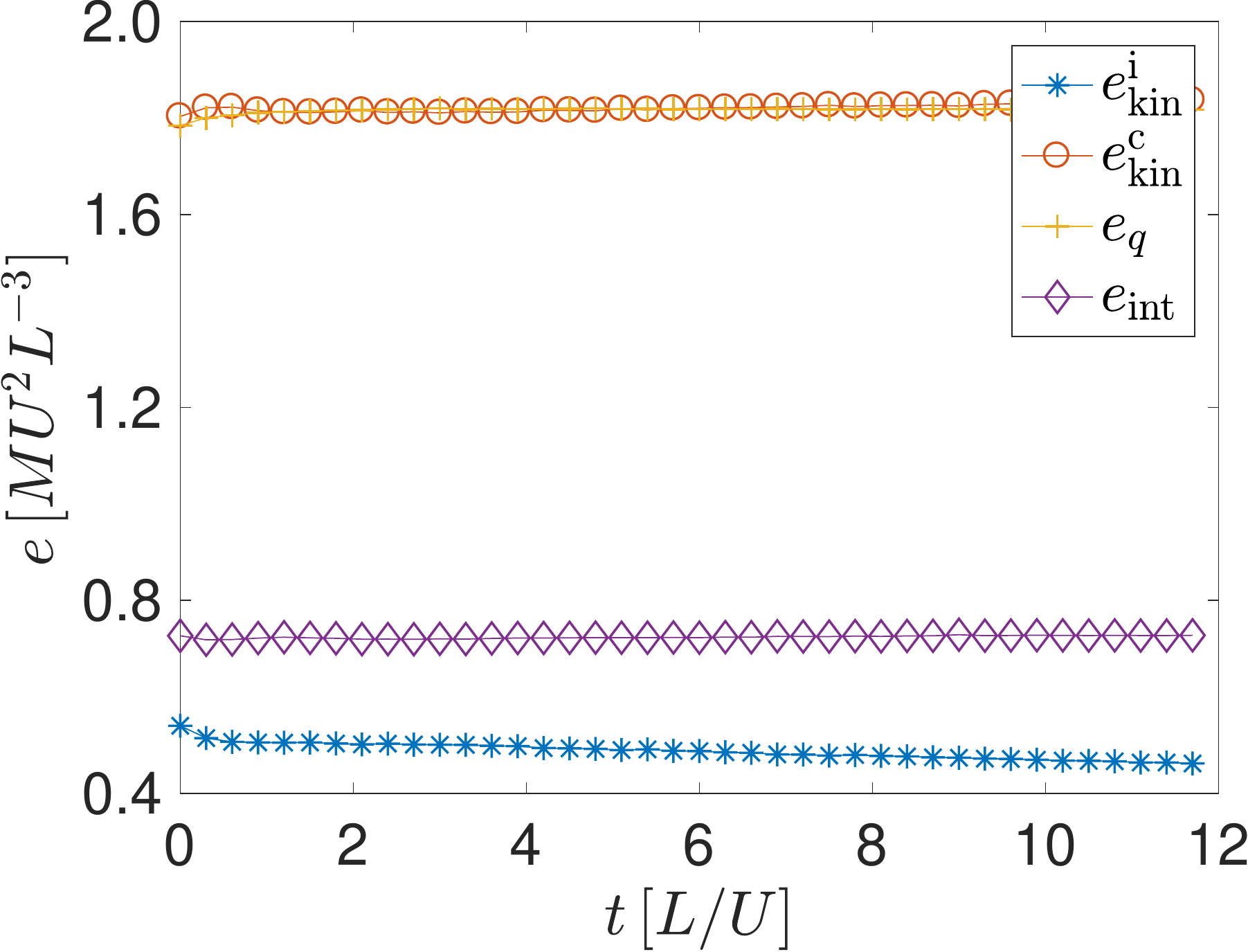}
\put(-25,87){\bf\large (d)}
\caption{{\it (Color online)}
	Time evolution of the energy components in the TGPE runs
	with linear spatial resolution $N=1024$ and 
        $\xi k_{\rm max}=2.5$ at four different temperatures: (a)
        $T=0\,T_{\lambda}$,  (b) $0.11\,T_{\lambda}$, (c)
        $0.33\,T_{\lambda}$, and (d) $0.55\,T_{\lambda}$. 
	%The labels
        %of the curves for the four energy components, the
        %incompressible kinetic energy $e^i_{\rm kin}$, the
        %compressible kinetic energy $e^c_{\rm kin}$, the quantum
        %energy $e_q$, and the internal energy $e_{\rm int}$, are
        %indicated in the insets.
	}
\label{fig:energetics1024}
\end{figure*}

In Figs.~\ref{fig:energetics1024}(a)-(d) we show the GPE temporal
evolution of the energy components $e^{i}_{\rm kin}$, $e^{c}_{\rm kin}$, 
$e_{\rm q}$, and $e_{\rm int}$ at four different temperatures
($T=0,\,0.11\,T_{\lambda}, 0.33\,T_{\lambda}$, and 
$0.55\,T_{\lambda}$). Similar to the case of high resolution 
runs at $N=2048$ and $4096$, the incompressible kinetic energy
$e^{i}_{\rm kin}$  at $T=0$ shown in
Fig.~\ref{fig:energetics1024}(a) stays roughly constant till
$t\approx 5\,L/U$, followed by a large decay of approximately $30\%$
of its initial value in a similar interval of time (up to 
$t\approx 10\,L/U$); thereafter, its decay slows down. During the
initial phase of the dynamical evolution of the TG flow, the decrease
in the incompressible kinetic energy is accompanied by an increase in
the other energy components, with $e_{\rm int}$ gaining the maximum
share until $t\approx 10\,L/U$, after which it saturates. At the later
stages, the compressible kinetic energy $e^c_{\rm kin}$ is the
dominant component, with the internal energy having magnitude similar
to $e_{\rm q}$ for the rest of the duration of the run.

%%%%%%%%%%%%%%%%%%%%%%%%%%%%%%%%%%%%%%%%%%%
%%%%%%%%%%%%%%%%%%%%%%%%%%%%%%%%%%%%%%%%%%%

At higher temperatures, the plots of the different energy components in
Figs.~\ref{fig:energetics1024}(c)-(d) show that $e^{c}_{\rm kin}$,
$e_{\rm q}$, and $e_{\rm int}$ start with higher values when compared
to the $T=0$ case, because of finite temperature effects included via
the thermal states, and in agreement with the discussion in
Sec.~\ref{sec:scans} (see Fig.~\ref{Fig:EnerguScan}). In particular,
in the runs with $T=0.33T_{\lambda}$ and with $0.55 T_{\lambda}$, the
incompressible kinetic energy has the lowest share of the total
energy, while the compressible component is the dominant form of the
energy (although at $T=0.55\,T_{\lambda}$, $e_{\rm q}$ also becomes
comparable to $e^c_{\rm kin}$).

\begin{figure}
\includegraphics[width=0.9\linewidth]{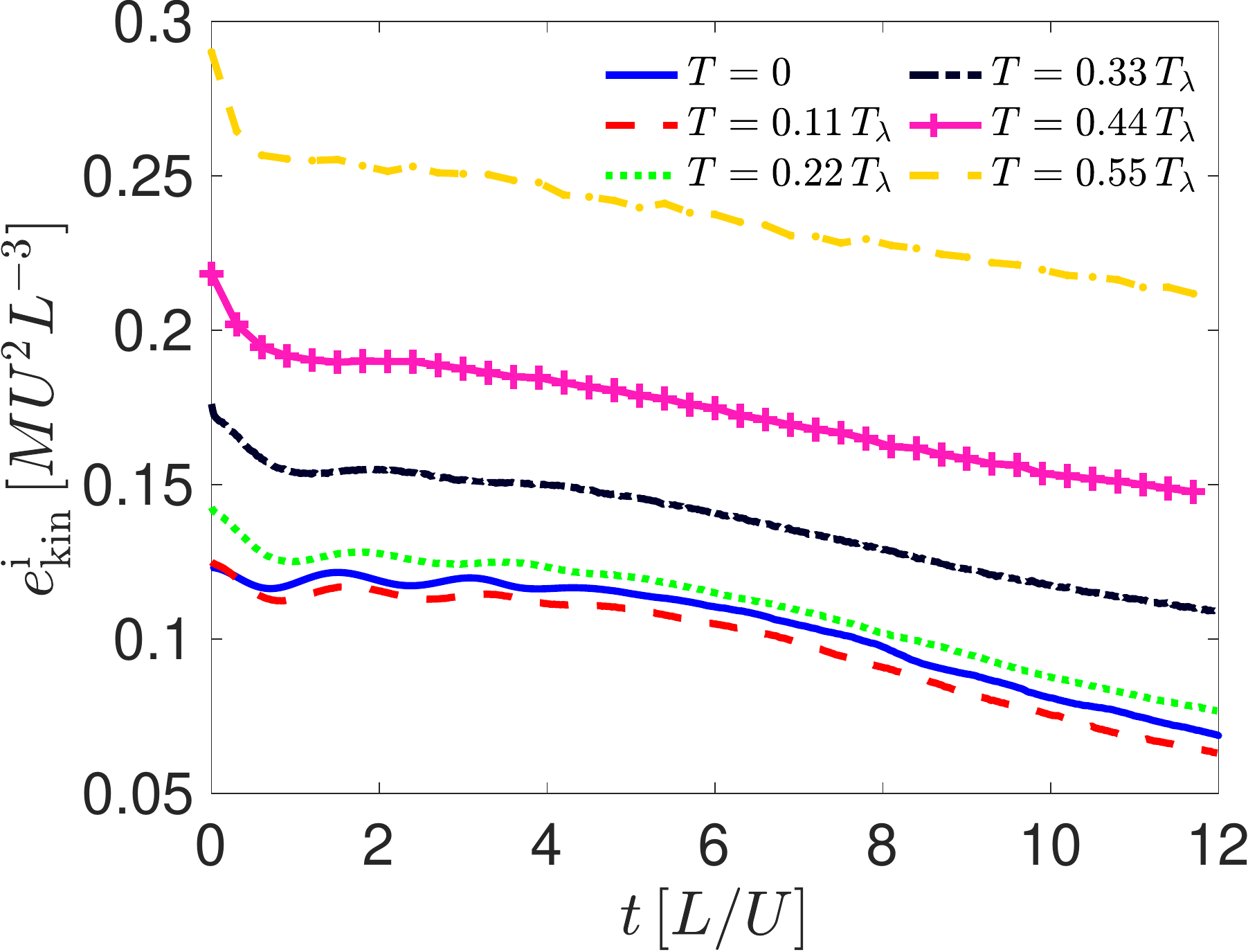}
\put(-175,36){\bf\large (a)} \\
\includegraphics[width=0.9\linewidth]{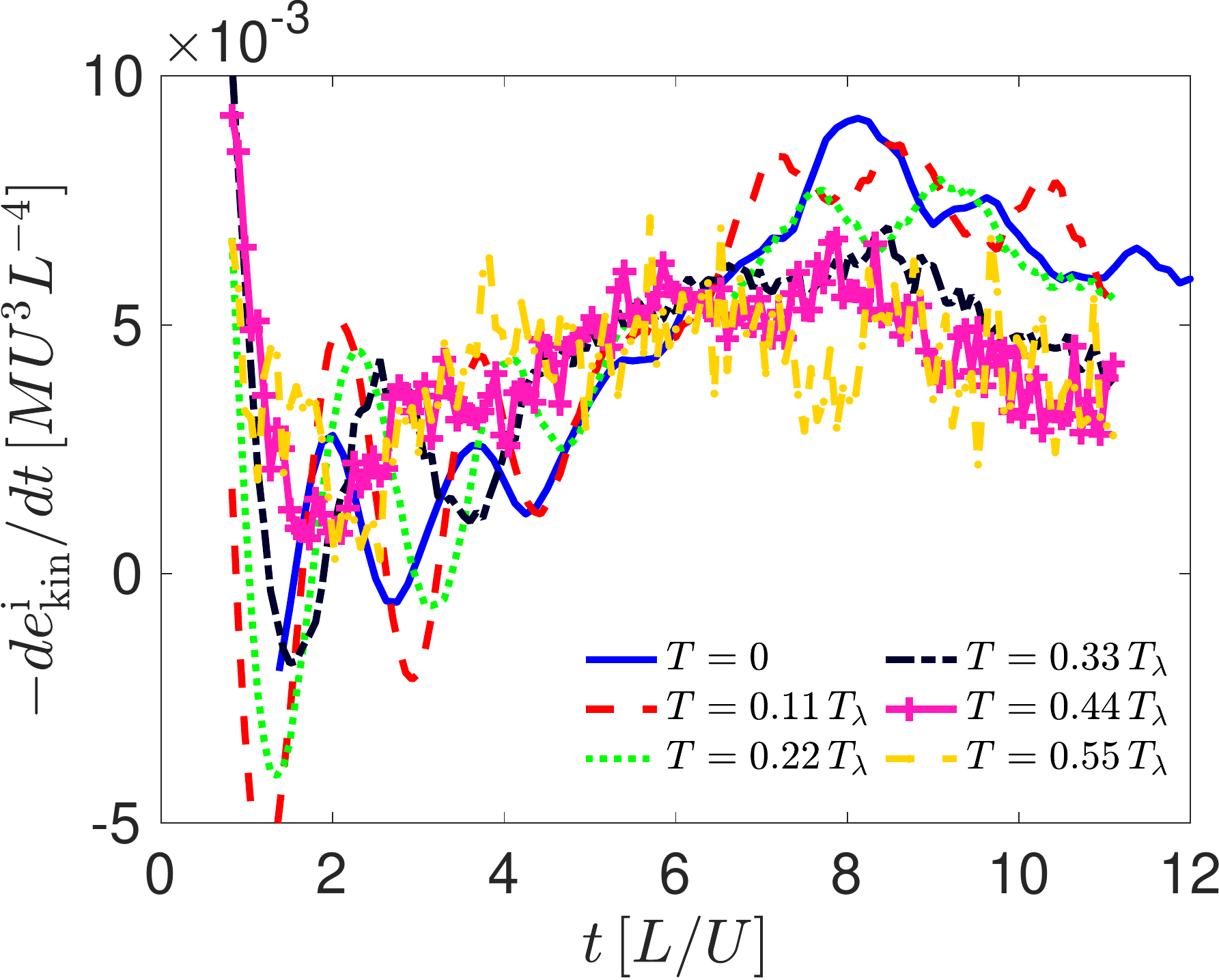}
\put(-175,36){\bf\large (b)}
\caption{{\it (Color online)} 
	Time evolution of: 
	(a) The incompressible kinetic energy $e^i_{\rm kin}$, and of
	(b) the incompressible kinetic energy dissipation rate 
	$-d e^i_{\rm kin}/dt$, at six different temperatures in TGPE
        runs with linear spatial resolution $N=1024$ and 
        $\xi k_{max}=2.5$. 
	%Labels for the curves corresponding to
        %simulations at different temperatures are indicated in the
        %figures. 
	}
\label{fig:epsilon1024}
\end{figure}

In spite of these differences at $t=0$ and at very early times, in all
the cases presented in Fig.~\ref{fig:energetics1024} we observe at
later times a qualitatively similar decrease in $e^i_{\rm kin}$ as the
TG flow evolves. For a better comparison, we plot the time evolution
of $e^i_{\rm kin}$ at five different temperatures in
Fig.~\ref{fig:epsilon1024}(a). We note once again that besides the
initial adaptation period which lasts roughly until $t\approx
1\,L/U$, the $e^i_{\rm kin}$ energy component decreases very slowly
during the time interval $t \approx 2\,L/U$ to $4\,L/U$, wherein
$e^i_{\rm kin}$ at  $T=0$ remains roughly constant. After this time
interval the decay of $e^i_{\rm kin}$ is faster. This behavior is
still better captured by computing the decay rate 
$-de^i_{\rm kin}/dt$,  a quantity frequently studied in freely
decaying classical fluid turbulence. In Fig.~\ref{fig:epsilon1024}(b)
we show the temporal evolution of $-de^i_{\rm kin}/dt$ for different
temperatures. If we discard the initial adaptation period, at least
the low temperature curves (up to $T=0.44\,T_{\lambda}$) exhibit a
peak at  $t\approx 8\,L/U$, and the peak value decreases, along with
an accompanying flattening, as we increase the temperature. At
$T=0.55\,T_{\lambda}$, $-de^i_{\rm kin}/dt$ fluctuates around the
value $5\times 10^{-3} MU^3L^{-4}$; however, because of the presence
of strong fluctuations (and the lack of enough statistics), we are
unable to make more precise statements for higher temperatures. Note
that given these fluctuations, to identify trends while varying
temperature, we use a filtering technique to smooth out the
curves (not shown). It is also interesting to note that at around
$t\approx 8\,L/U$ (when the peak in $-de^i_{\rm kin}/dt$ is observed),
the vortex line length $L_V$ is maximum in the $T=0$ runs, see
Sec.~\ref{sec:hirest0} and Fig.~\ref{fig:energies}(c).

%%%%%%%%%%%%%%%%%%%%%%%%%%%%%%%%%%%%%%%%%%%
%%%%%%%%%%%%%%%%%%%%%%%%%%%%%%%%%%%%%%%%%%%

\begin{figure*}
\centering
\includegraphics[width=0.4\linewidth]{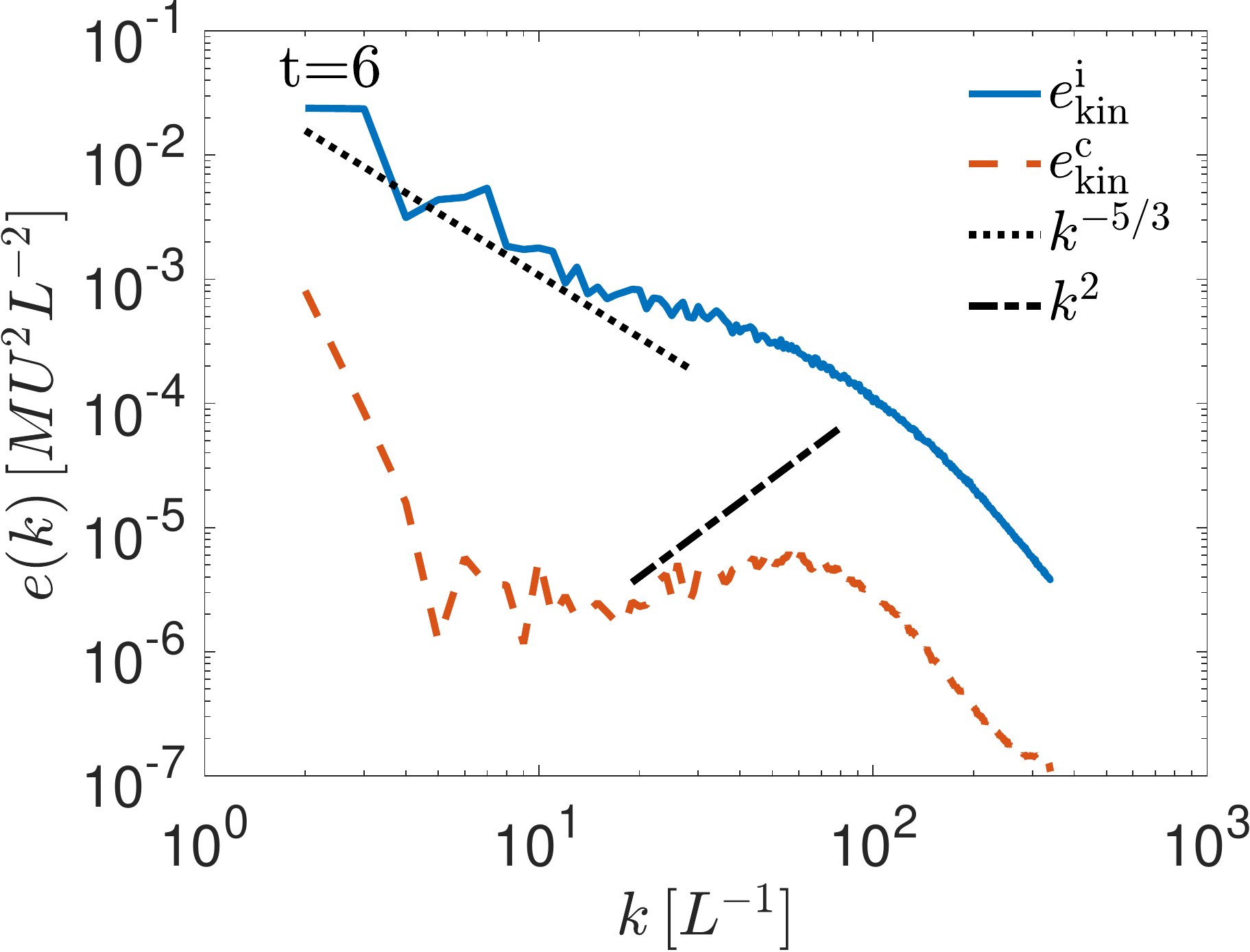}
\put(-164,38){\bf\large (a)}
\hskip .5cm
\includegraphics[width=0.4\linewidth]{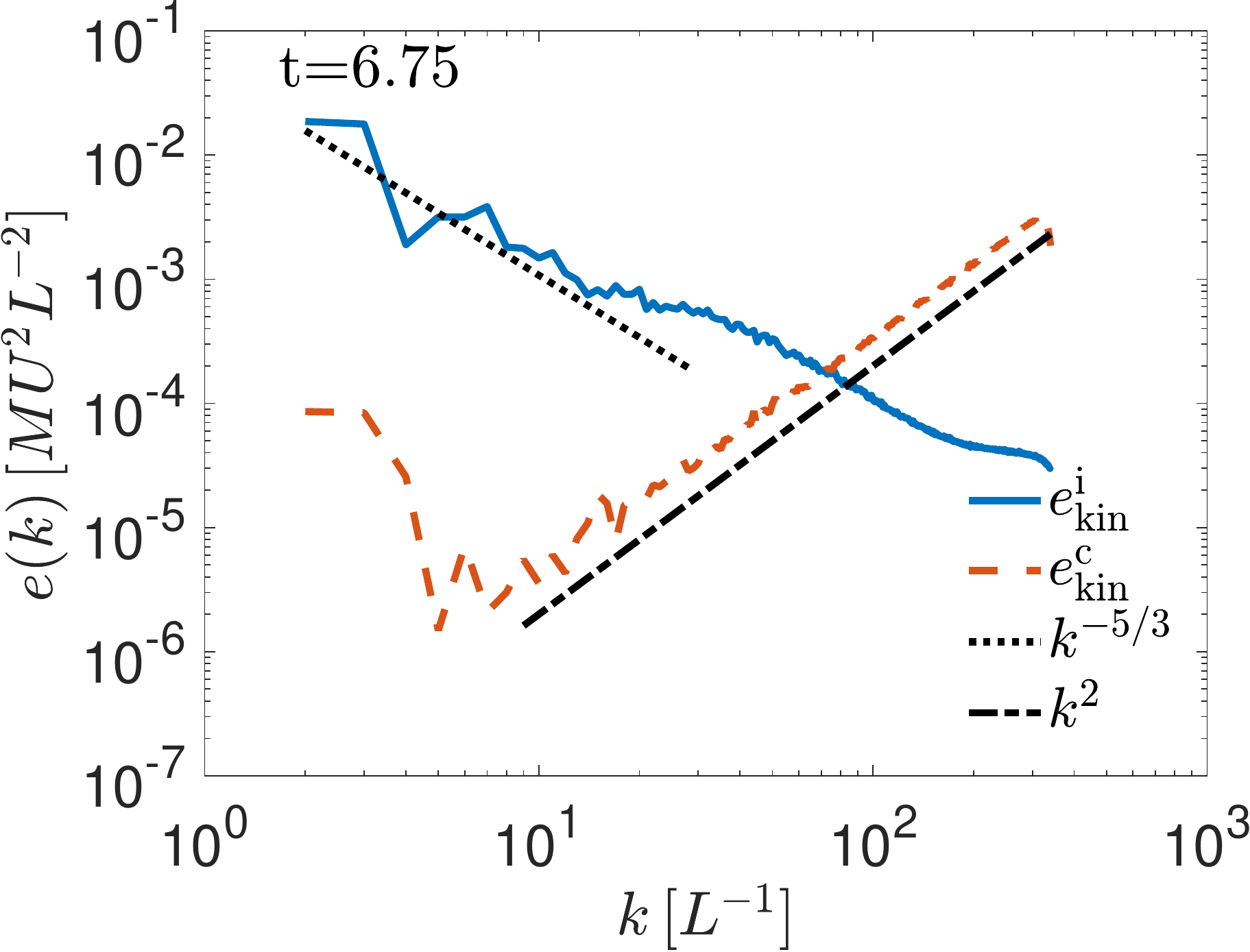}
\put(-164,38){\bf\large (b)}
\\
\includegraphics[width=0.4\linewidth]{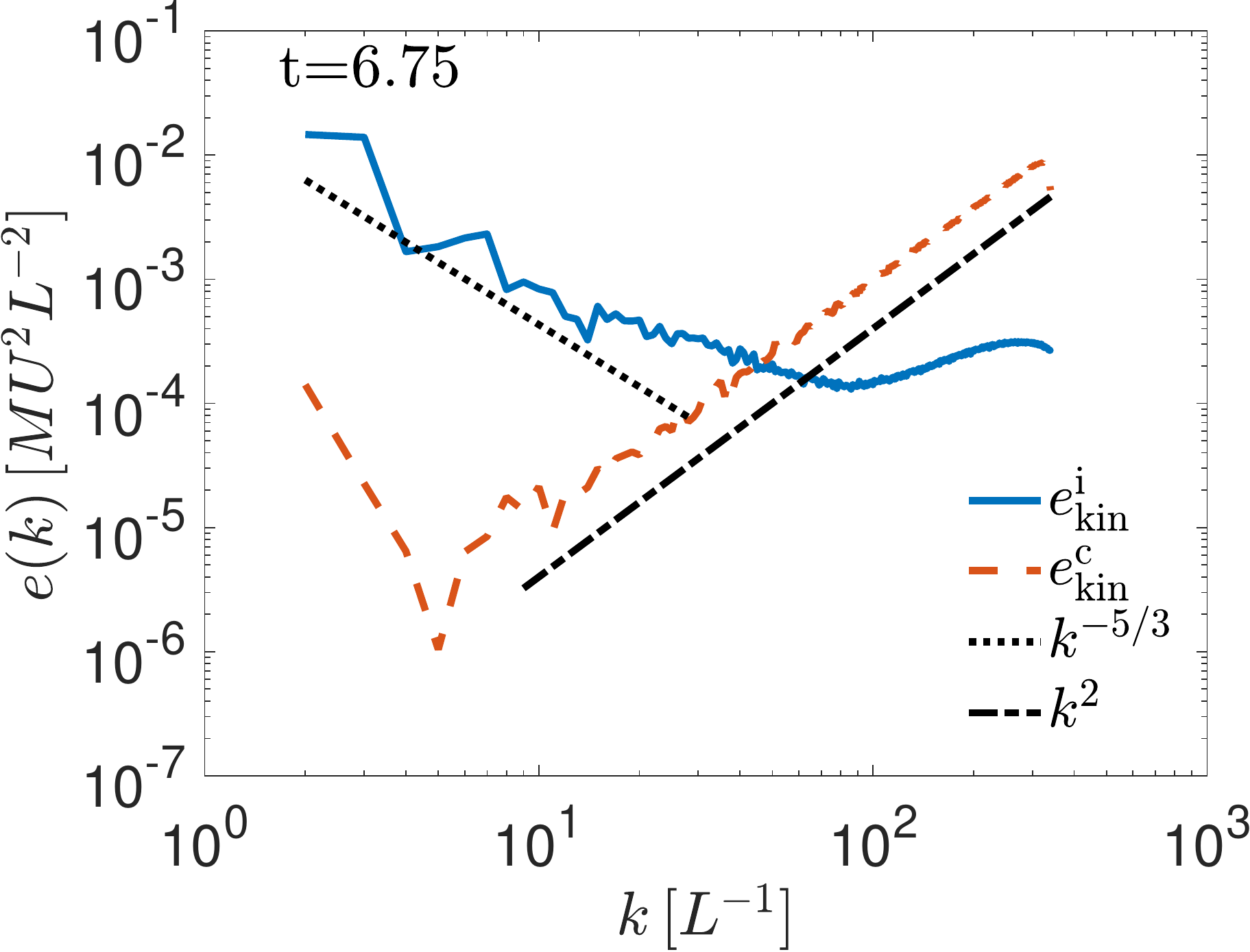}
\put(-164,38){\bf\large (c)}
\hskip .5cm
\includegraphics[width=0.4\linewidth]{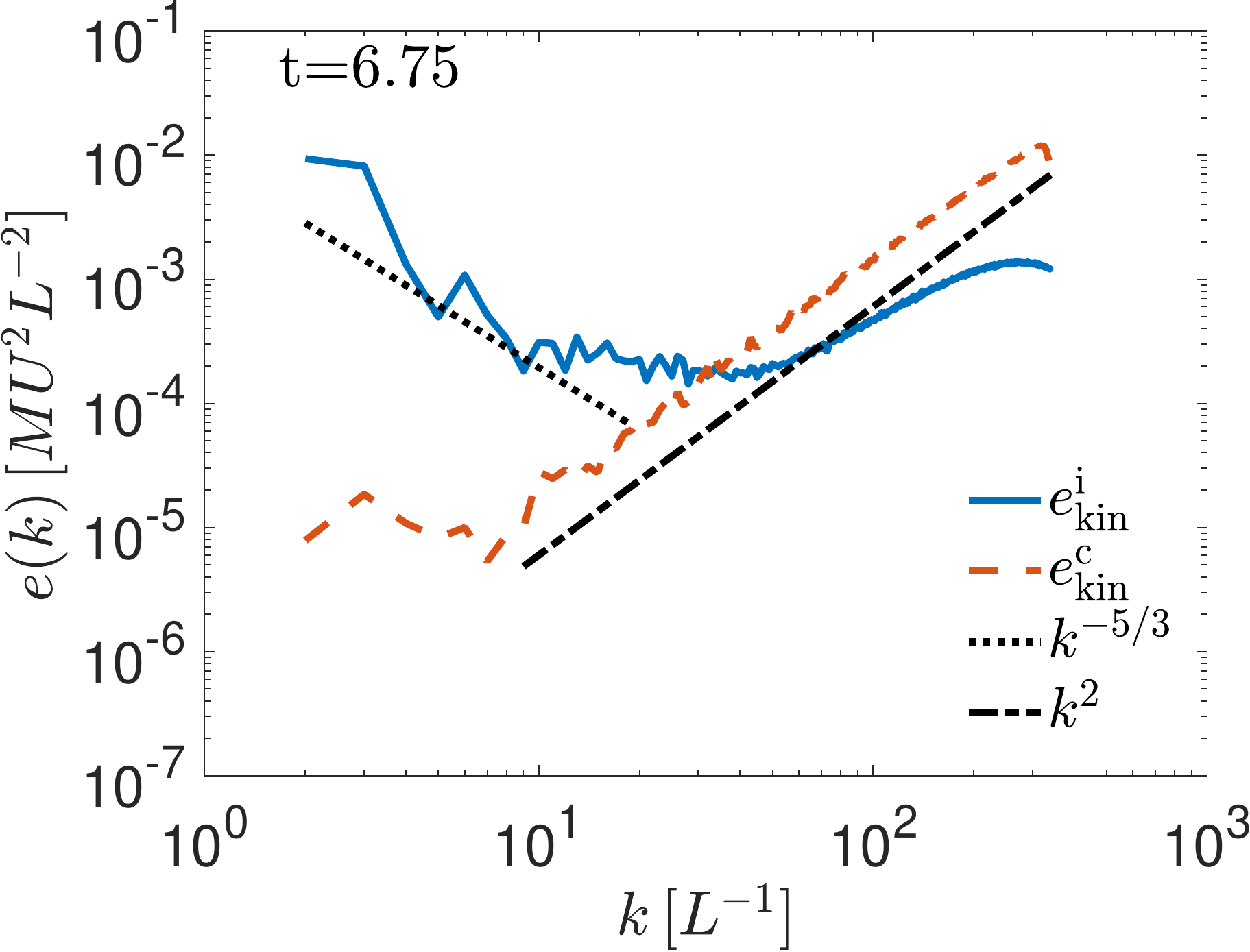}
\put(-164,38){\bf\large (d)}
	\caption{{\it (Color online)} 
	Kinetic energy spectra $e^i_{\rm kin}(k)$ and $e^c_{\rm kin}(k)$ 
	for four different temperatures: (a) $T=0\,T_{\lambda}$, 
	(b) $0.11\,T_{\lambda}$, (c) $0.33\,T_{\lambda}$, and (d)
        $0.55\,T_{\lambda}$, from TGPE runs with linear spatial
	resolution $N=1024$ and $\xi k_{\rm max}=2.5$. Videos M1 ($T=0$) 
	and M2 ($T>0$),
	see Supplemental Material~\cite{SMarXiv}, show the
        time evolution of the compensated incompressible kinetic
        energy  spectra at different temperatures.}
\label{fig:spectra1024}
\end{figure*}

To better understand how the thermal fluctuations act across the
different length scales during the evolution of the TG flow towards the
turbulent state, in Fig.~\ref{fig:spectra1024}(a)-(d) we show the
incompressible and compressible kinetic energy spectra for the $T=0$
case, and for three different temperatures ($T=0.11\,T_{\lambda}$,
$0.33\,T_{\lambda}$, and $0.55\,T_{\lambda}$), at a time at which we
observe a range of wavenumbers in the spectrum compatible with
self-similar scaling (i.e., with a possible inertial range). For
$T=0$, shown in Fig.~\ref{fig:spectra1024}(a), we observe that 
$e^i_{\rm kin}(k)\sim k^{-5/3}$ (at $t=6\,L/U$) roughly over a decade
at small wavenumbers, followed by a bottleneck around 
$k\approx 20\,L^{-1}$, and an exponential decrease for even larger 
wavenumbers. The amplitude of the spectrum of compressible energy is
amost negligible in this case and at this instant of time. For the
runs at larger temperatures, shown in
Figs.~\ref{fig:spectra1024}(b)-(d), we continue to observe a range of
wavenumbers compatible with $e^i_{\rm kin}(k)\sim k^{-5/3}$ scaling
(for small wavenumbers), but now at high wavenumbers we see an
accumulation of energy and the begining of a thermalized region
approaching $\sim k^2$ scaling, indicating small scale fluctuations
become more energetic as we increase the temperature. At the same
time, the amplitude of the compressible energy spectrum increases with
increasing $T$, with $e^c_{\rm kin}(k) \sim k^2$ at highwave numbers
as it is expected for a thermal state. In particular, for
$T=0.55\,T_{\lambda}$ in Fig.~\ref{fig:spectra1024}(d), $e^i_{\rm
  kin}(k)$ does not show anymore a significant range with Kolmogorov
scaling. This can be understood as the crossover region at which the
range of wavenumbers compatible with an inertial range finishes, at
$k\approx 20 L^{-1}$ in the $T=0$ case, is strongly modified in this
high temperature run by the thermal bath, which now affects and
extends to smaller wave numbers, thereby reducing the inertial range.

\begin{figure}[h!]
\includegraphics[width=0.9\linewidth]{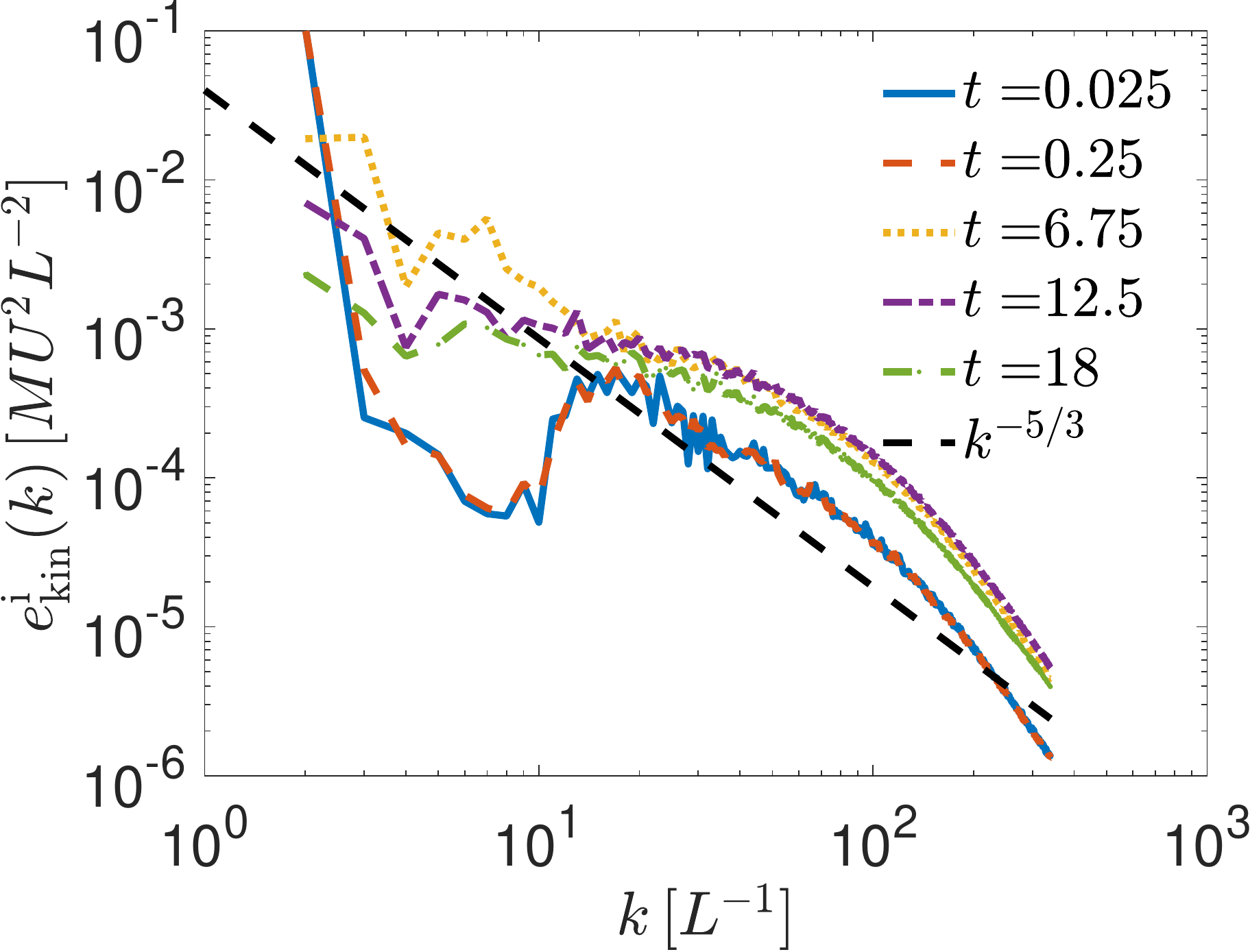}
\put(-170,39){\bf\large (a)} \\
\includegraphics[width=0.9\linewidth]{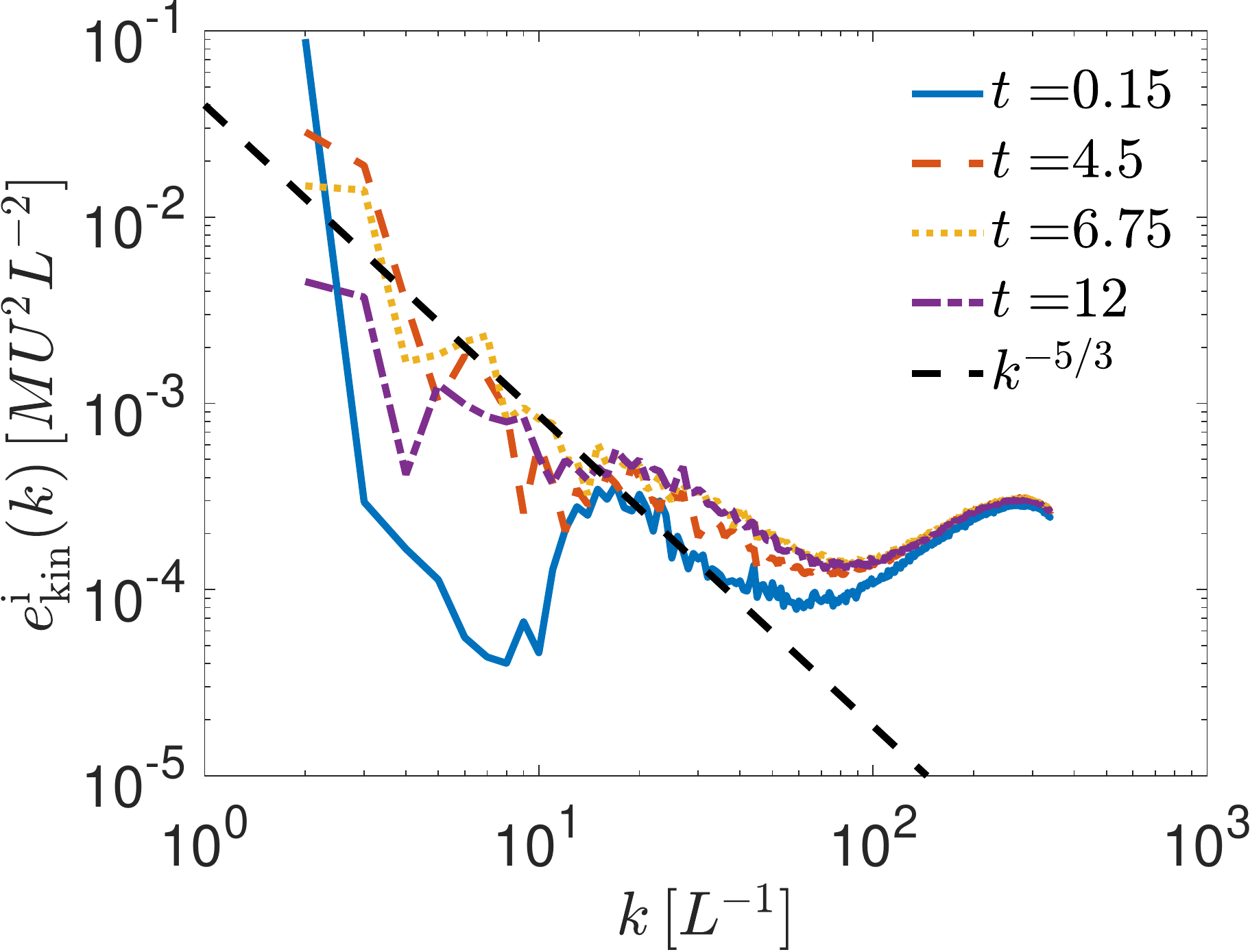}
\put(-170,39){\bf\large (b)} \\
\includegraphics[width=0.9\linewidth]{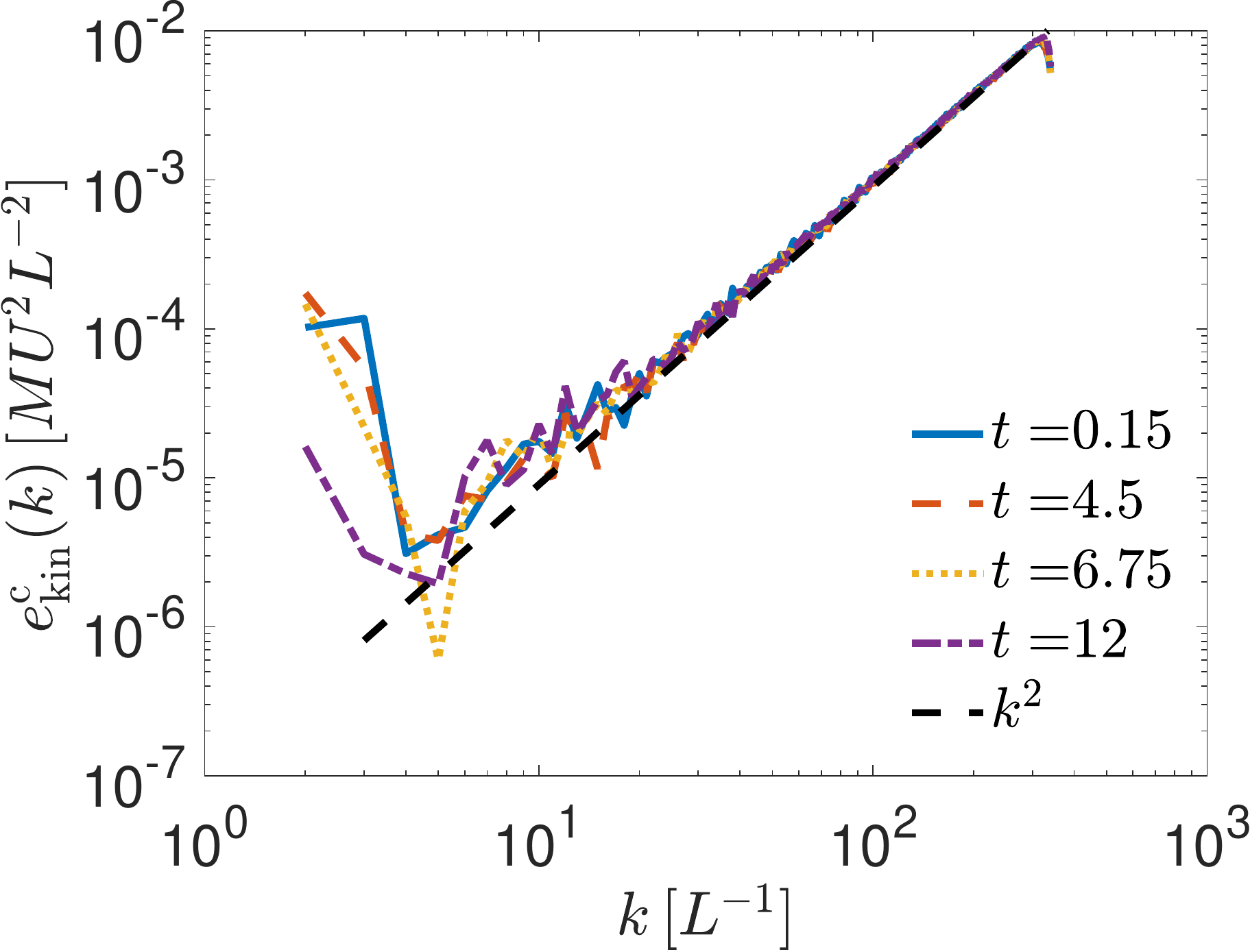}
\put(-170,39){\bf\large (c)} \\
\caption{{\it (Color online)} Temporal evolution of the kinetic energy
  spectra: (a) Incompressible component $e^i_{\rm kin}(k)$ at $T=0$; 
  (b) incompressible component $e^i_{\rm kin}(k)$,  and (c)
  compressible component $e^c_{\rm kin}(k)$, for the TGPE run at
  $T=0.33\,T_{\lambda}$ with linear spatial  resolution $N=1024$ and
  $\xi k_{\rm max}=2.5$.
  }
\label{fig:spectra1024-2}
\end{figure}

Finally, to illustrate the time evolution of the spectra, we show in
Fig.~\ref{fig:spectra1024-2} the incompressible and compressible
kinetic energy spectra in two runs at $T=0$ and at
$T=0.33\,T_{\lambda}$, at different instants in time. The development
in time of an inertial range at intermediate wavenumbers can be seen
in all cases in $e^i_{\textrm{kin}}(k)$ (Kolmogorov scaling is
indicated in Fig.~\ref{fig:spectra1024-2} as a reference, as well as the
thermalization scaling $\sim k^2$ for the spectrum of
$e^c_{\textrm{kin}}$). A more detailed time evolution of these spectra
can also be seen in the videos M1 ($T=0$) and M2 ($T>0$) in the supplemental material
\cite{SMarXiv}.

%%%%%%%%%%%%%%%%%%%%%%%%%%%%%%%%%%%%%%%%%%%
%%%%%%%%%%%%%%%%%%%%%%%%%%%%%%%%%%%%%%%%%%%

%%%%%%%%%%%%%%%%%%%%%%%%%%%%%%%%%%%%%%%%%%%
%%%%%%%%%%%%%%%%%%%%%%%%%%%%%%%%%%%%%%%%%%%

\subsection{TGPE spatio-temporal spectra in thermal equilibrium}
\label{sec:spatiotemp}

We now compute the spatio-temporal spectra (STS) of the thermal
equilibria, that will be later used to estimate the mean-free path and
the effective viscosity. To this end, some of the SGLE equilibria of
Table \ref{tab:1} were used as initial data for GPE runs (with the
same values of $N$ and $\xi k_{\rm max}$), and integrated in time.

The STS provides the power spectrum of a given quantity as a function
of the wavenumber and of the frequency \cite{Clark14, Clark15b,
  Clark15a}. To compute this spectrum, quantities of interest must be
stored with high time cadence, so a Fourier transform in time and
space can be computed resolving the relevant high frequencies involved
in the problem. The result is a spectrum that shows the amplitude of
the excitations as a function of $k$ and $\omega$, and that can be
used to extract the amplitude of waves in a disordered state, as wave
excitations should accumulate near the theoretical dispersion relation
in $k-\omega$ space. In the following we consider the spatio-temporal
power spectrum of $\psi$, which is the STS of mass fluctuations.

\begin{figure}
    \centering
    \includegraphics[width=0.49\linewidth]{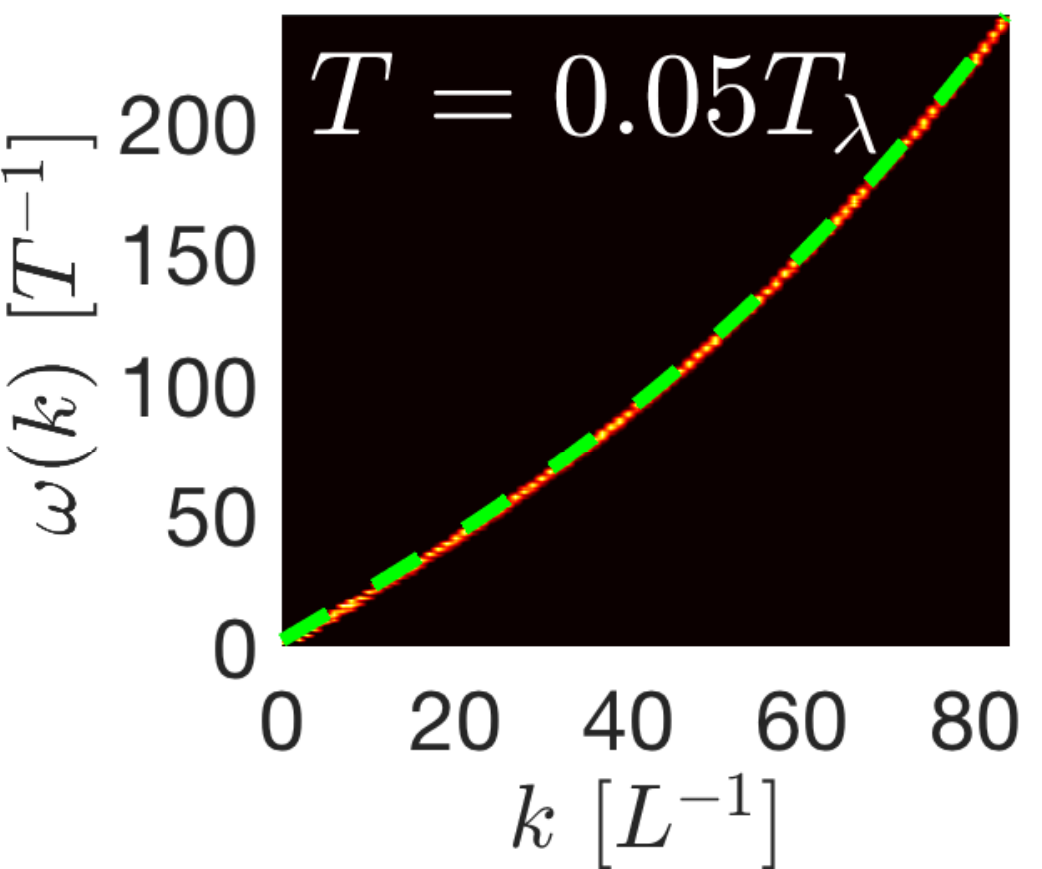}
    \includegraphics[width=0.49\linewidth]{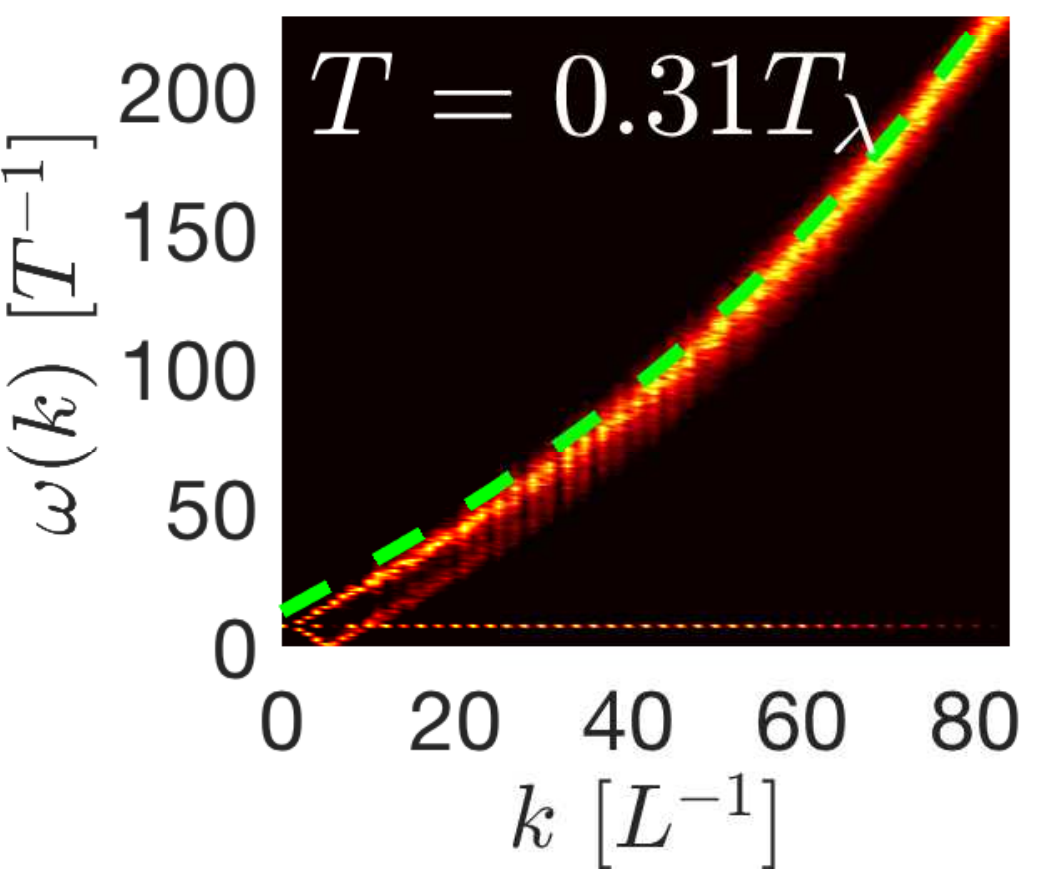}
    \includegraphics[width=0.49\linewidth]{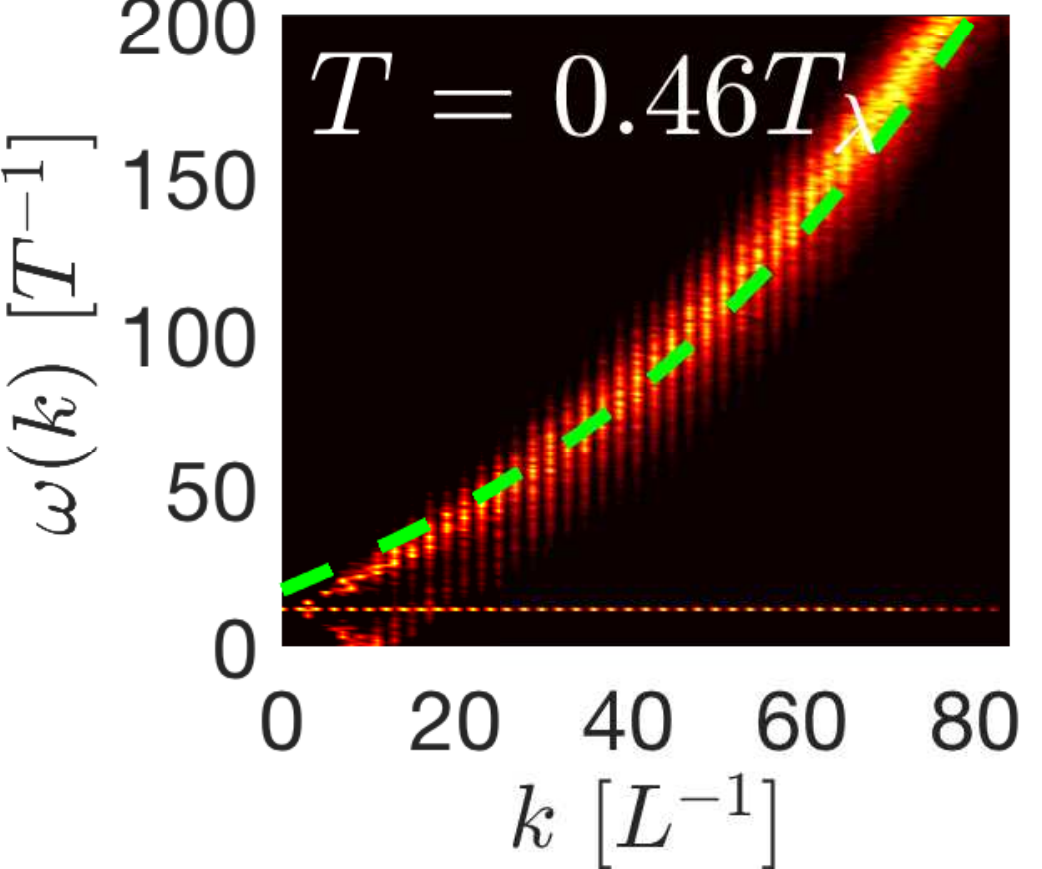}
    \includegraphics[width=0.49\linewidth]{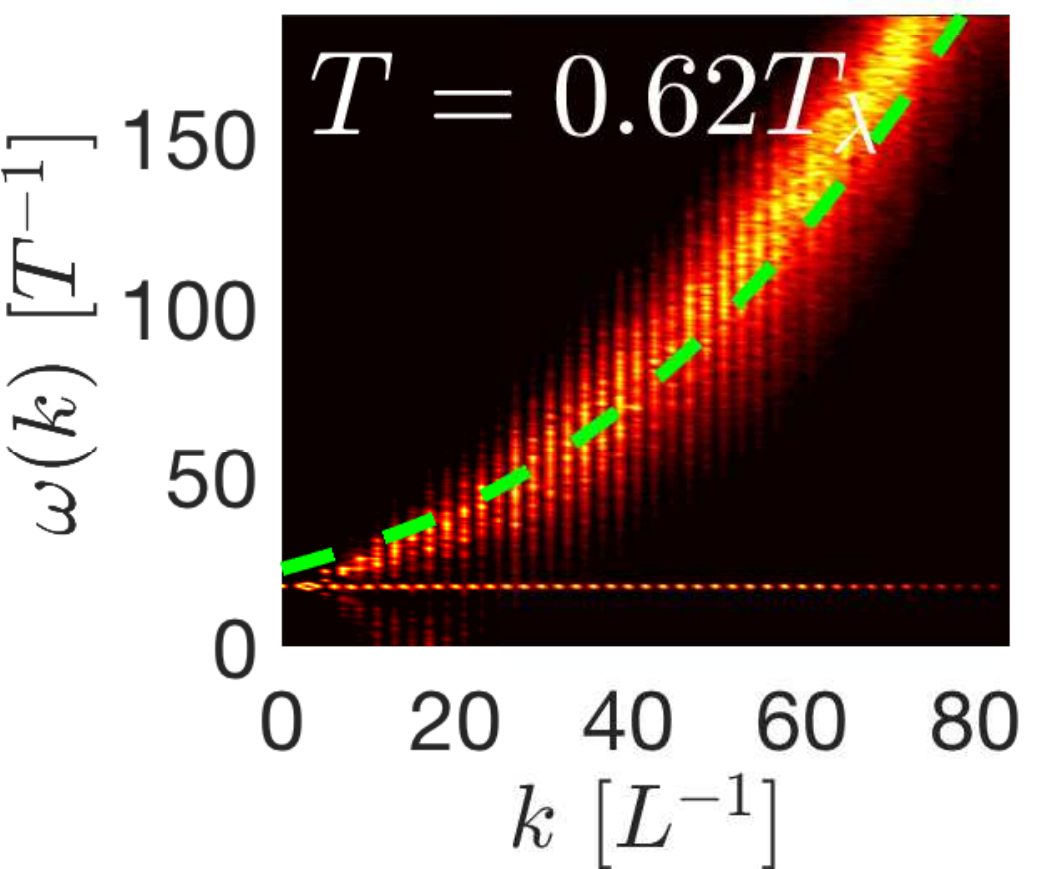}
    \includegraphics[width=0.49\linewidth]{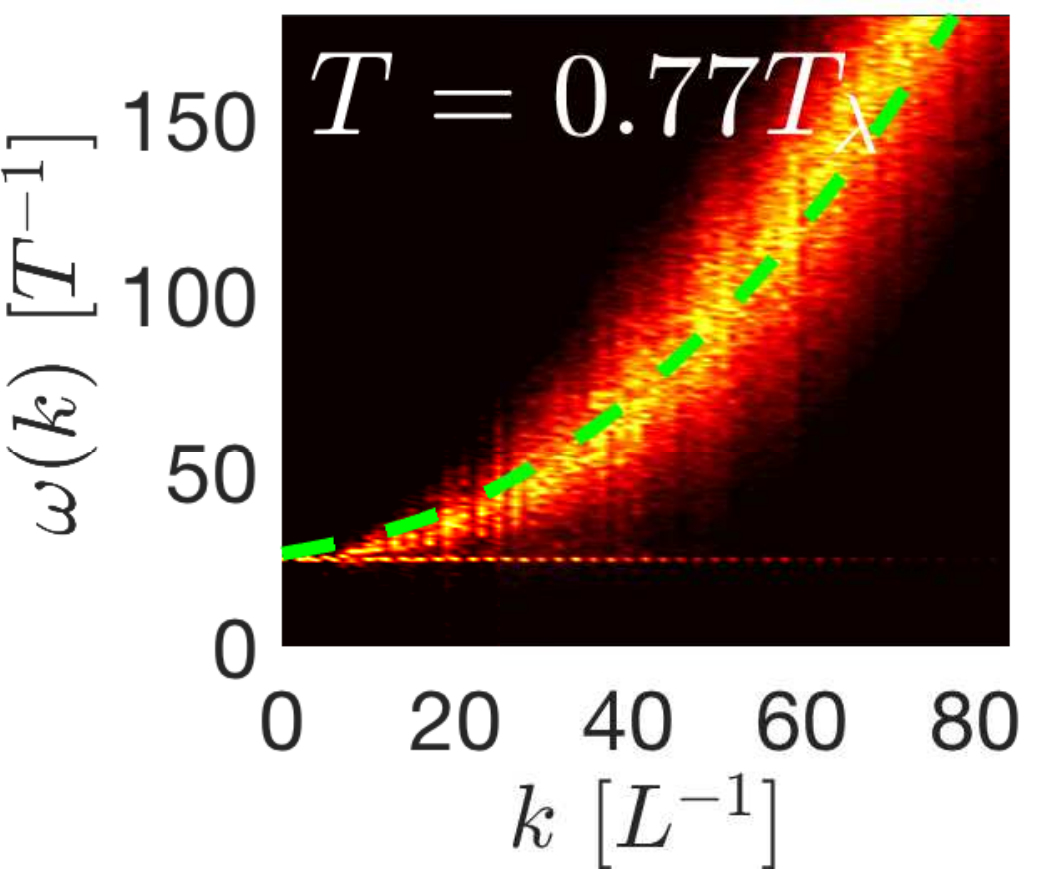}
    \includegraphics[width=0.49\linewidth]{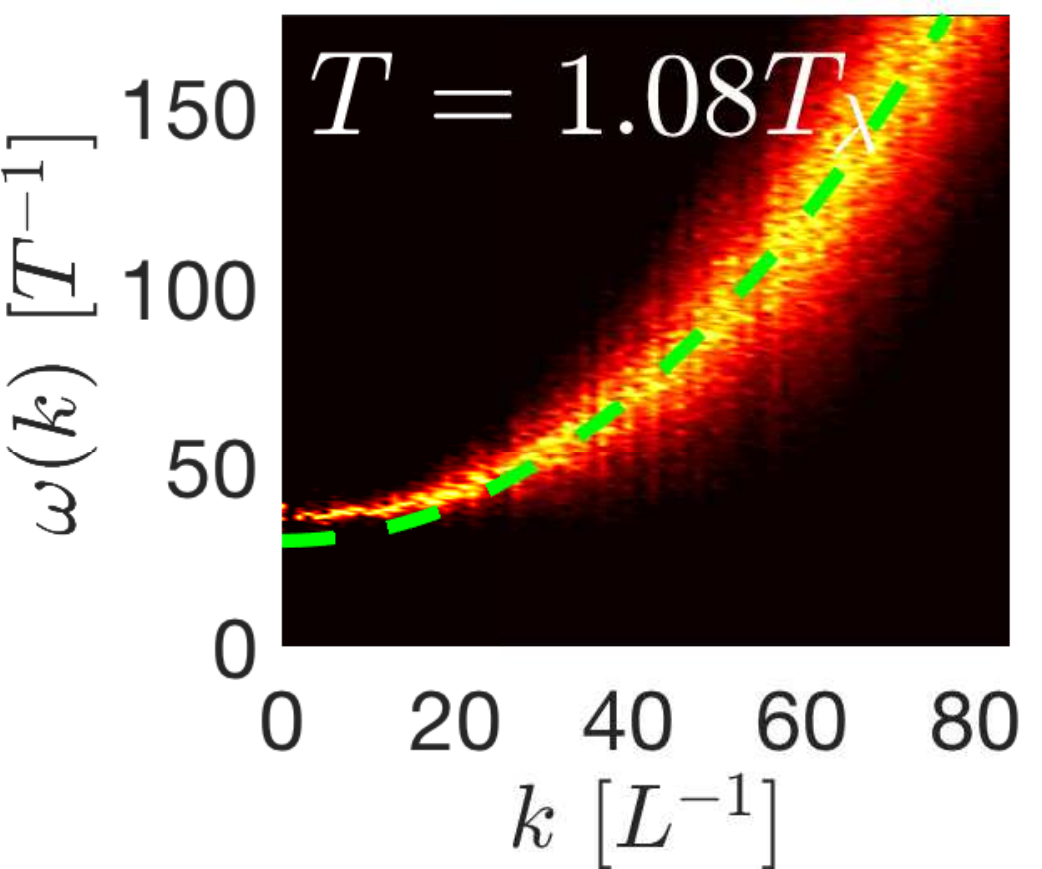}
    \caption{{\it (Color online)} Spatio-temporal spectra with $\xi
      k_{\rm max}=1.5$ and $N=256^3$, for different temperatures as
      indicated in each panel. The (green) dashed line indicates the
      theoretical Bogoliubov dispersion relation. Bright (red to
      yellow) areas indicate modes with large excitation.}
    \label{Fig:STS15}
\end{figure}

\begin{figure}
    \centering
    \includegraphics[width=0.49\linewidth]{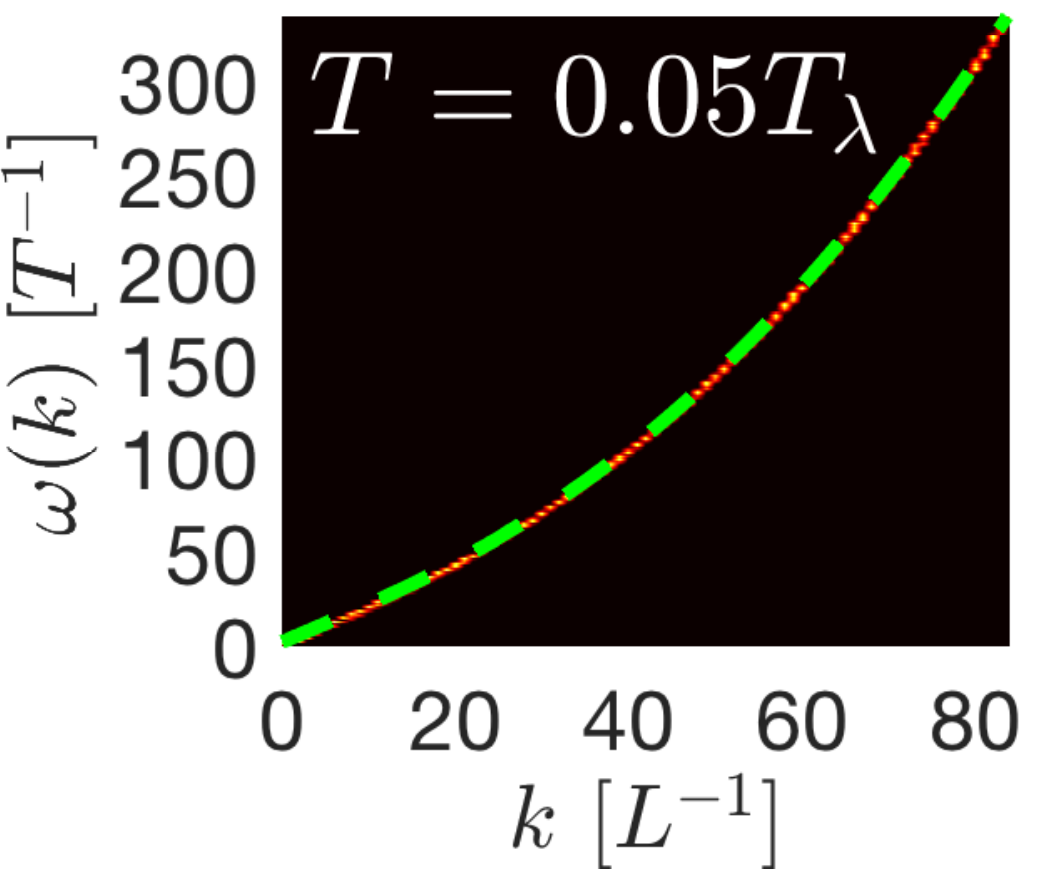}
    \includegraphics[width=0.49\linewidth]{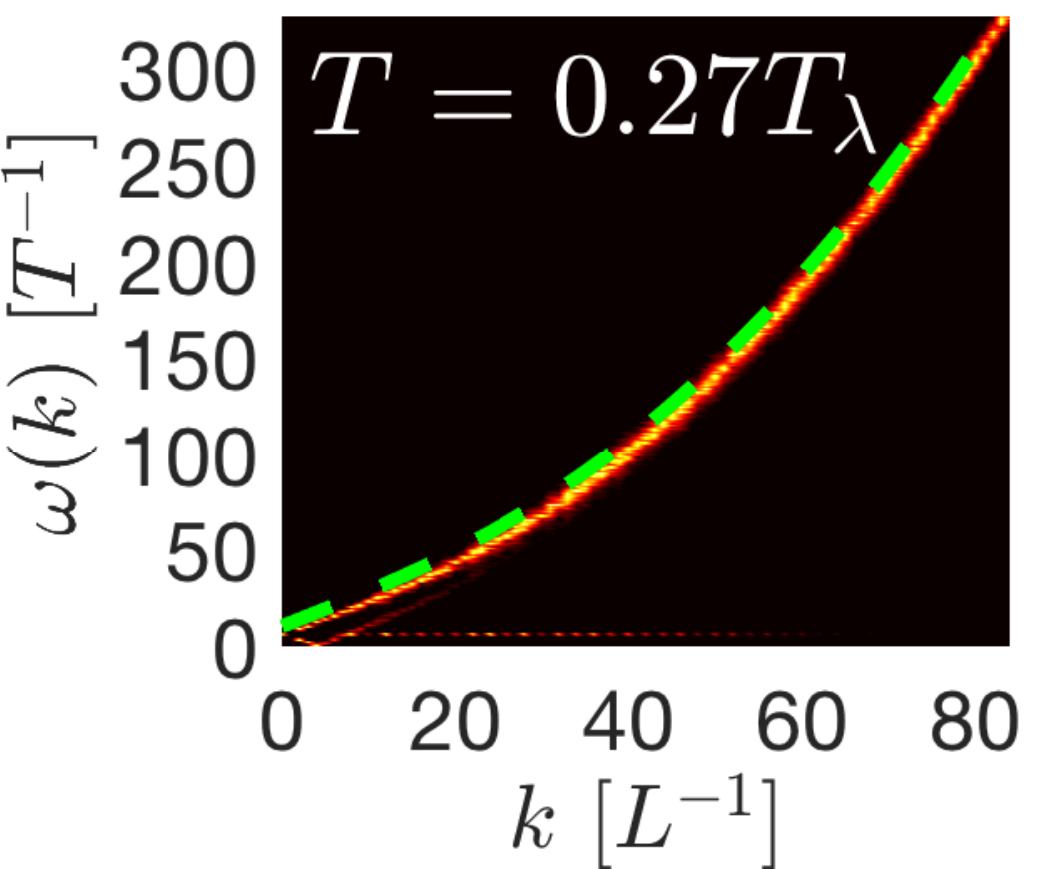}
    \includegraphics[width=0.49\linewidth]{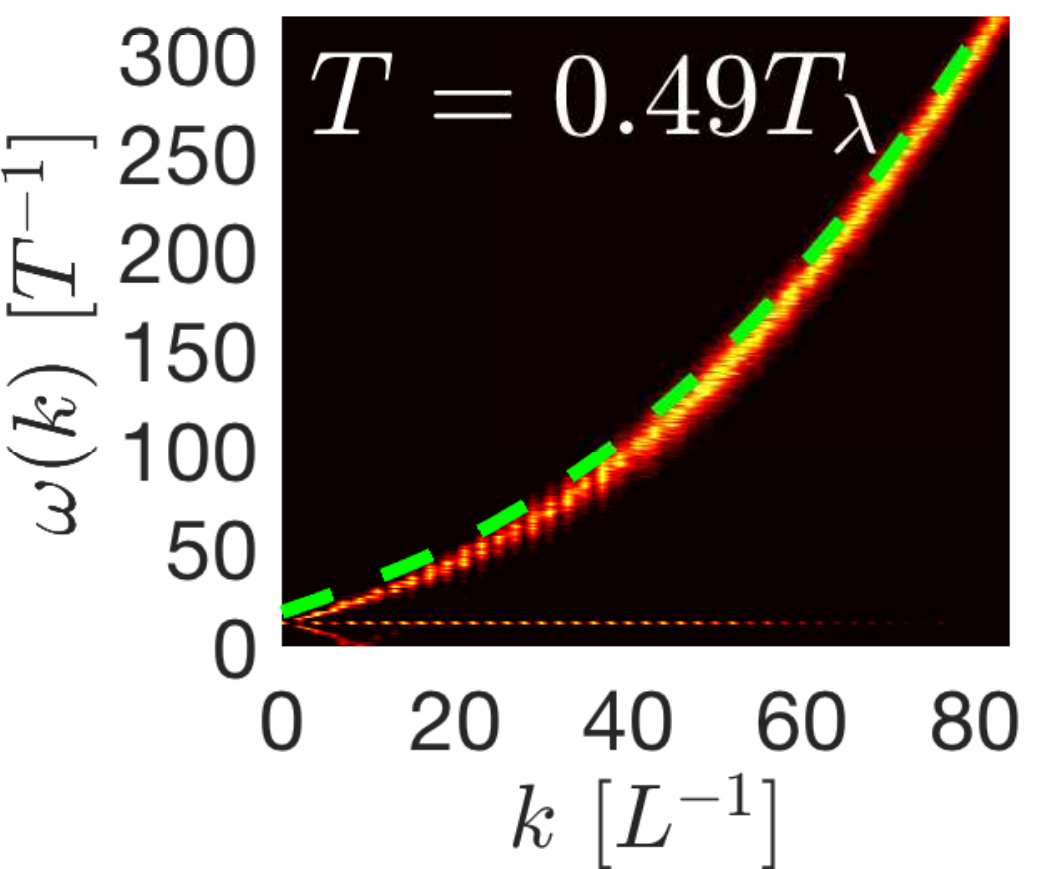}
    \includegraphics[width=0.49\linewidth]{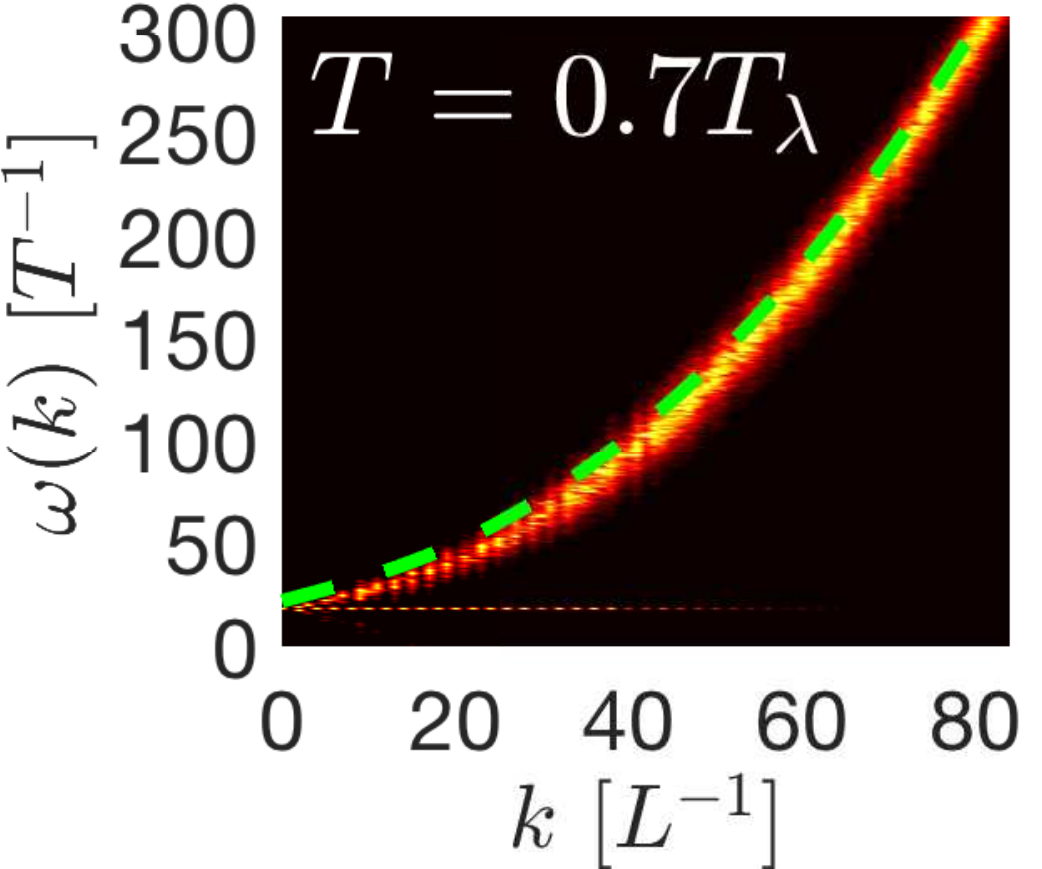}
    \includegraphics[width=0.49\linewidth]{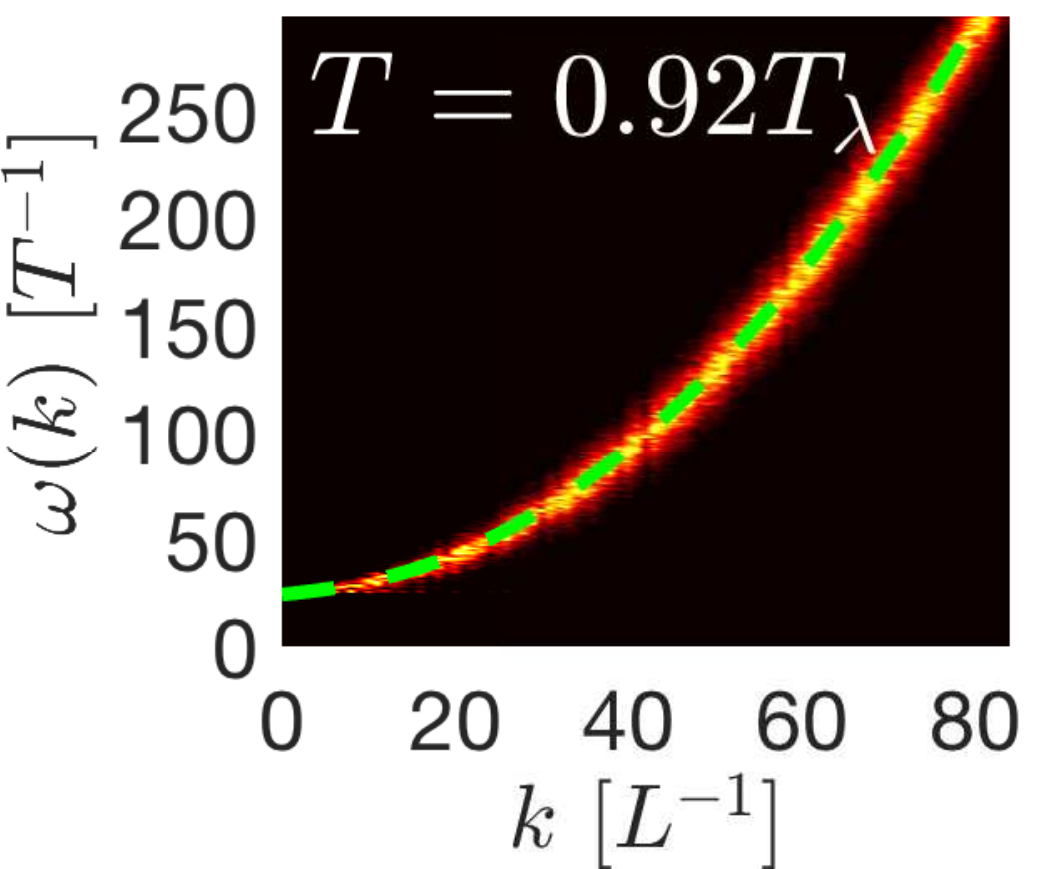}
    \includegraphics[width=0.49\linewidth]{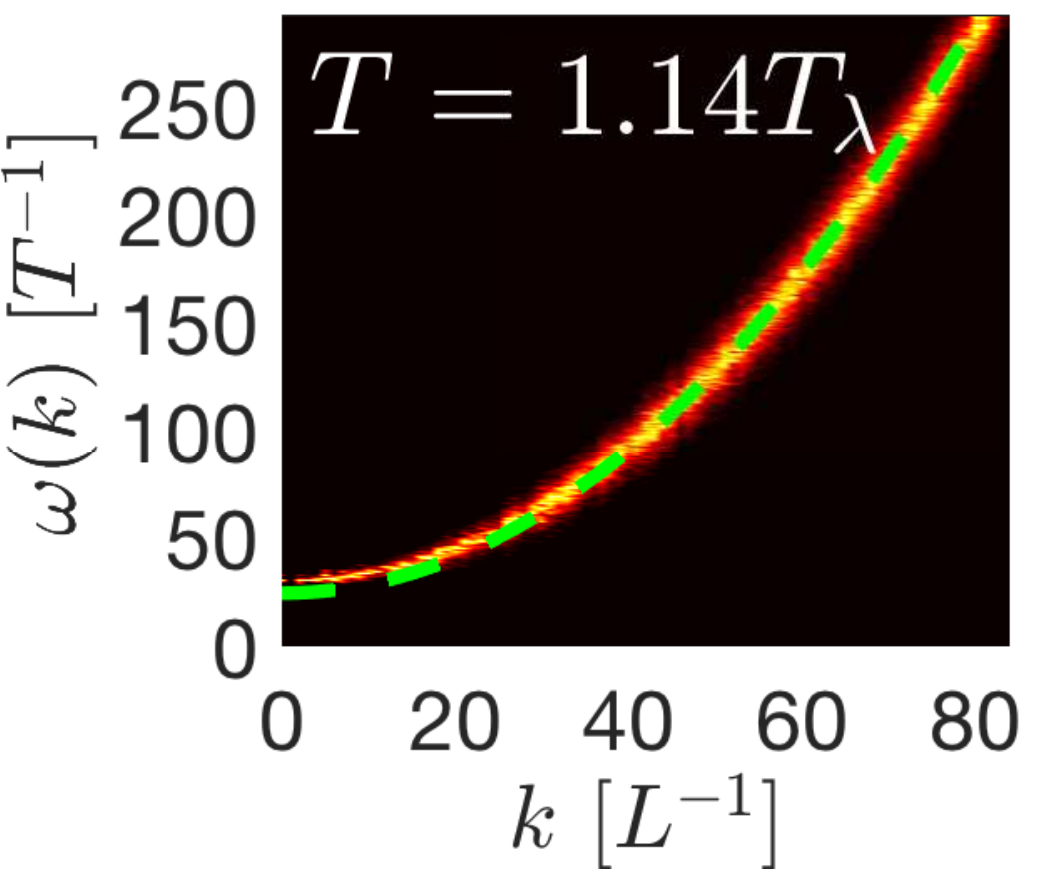}
    \caption{{\it (Color online)} Spatio-temporal spectra with $\xi
      k_{\rm max}=2.5$ and $N=256^3$, for different temperatures as
      indicated in each figure. The (green) dashed line indicates the
      theoretical Bogoliubov dispersion relation. Bright (red to
      yellow) areas indicate modes with large excitation.
	%Labels are the same as in Fig.~\ref{Fig:STS15}.
	}
    \label{Fig:STS25}
\end{figure}

The STS for runs with $N=256^3$ and $\xi k_{\rm max}=1.5$, and for
different temperatures, is shown in Fig.~\ref{Fig:STS15}. At very low
temperature ($T=0.05T_\lambda$) the non-linear interaction is very
weak, leading to exacts resonances in the periodic domain. The only
modes excited are those satisfying the Bogolioubov dispersion
relation given by Eq.~\eqref{eq:Bog} (indicated as a reference by the
dashed line), and modes with $\omega=0$ which correspond to the
condensate. As the temperature increases, non-linear interactions
become important and the dispersion relation broadens, as can be seen,
e.g., for $T=0.31T_\lambda$. Also note that as the temperature is
increased, the condensate (which in this figure appears as a straight
horizontal bright line) is shifted to higher frequencies, as it takes
place at energies $\hbar \omega = \mu$. The excitation of sound waves
around the Bogoliubov dispersion relation keeps broadening for larger
and larger temperatures up to $T_\lambda$, which is expected as
the broadening should be the strongest close to the transition. For
temperatures much larger than $T_\lambda$ the dispersion relation is
given by free particles, and we expect to recover the standard 4-wave
interaction. Also, note that for $T=1.08T_\lambda$ the condensate
disappears. Figure \ref{Fig:STS25} displays the same STS for runs with
$\xi\kmax=2.5$. Although the qualitative behavior is the same, note
that at a fixed temperature the spectral broadening is smaller,
consistent with the fact that for large $\xi\kmax$ the non-linear
interaction is expected to be weaker. 

\begin{figure}
    \centering
    \includegraphics[width=0.98\linewidth]{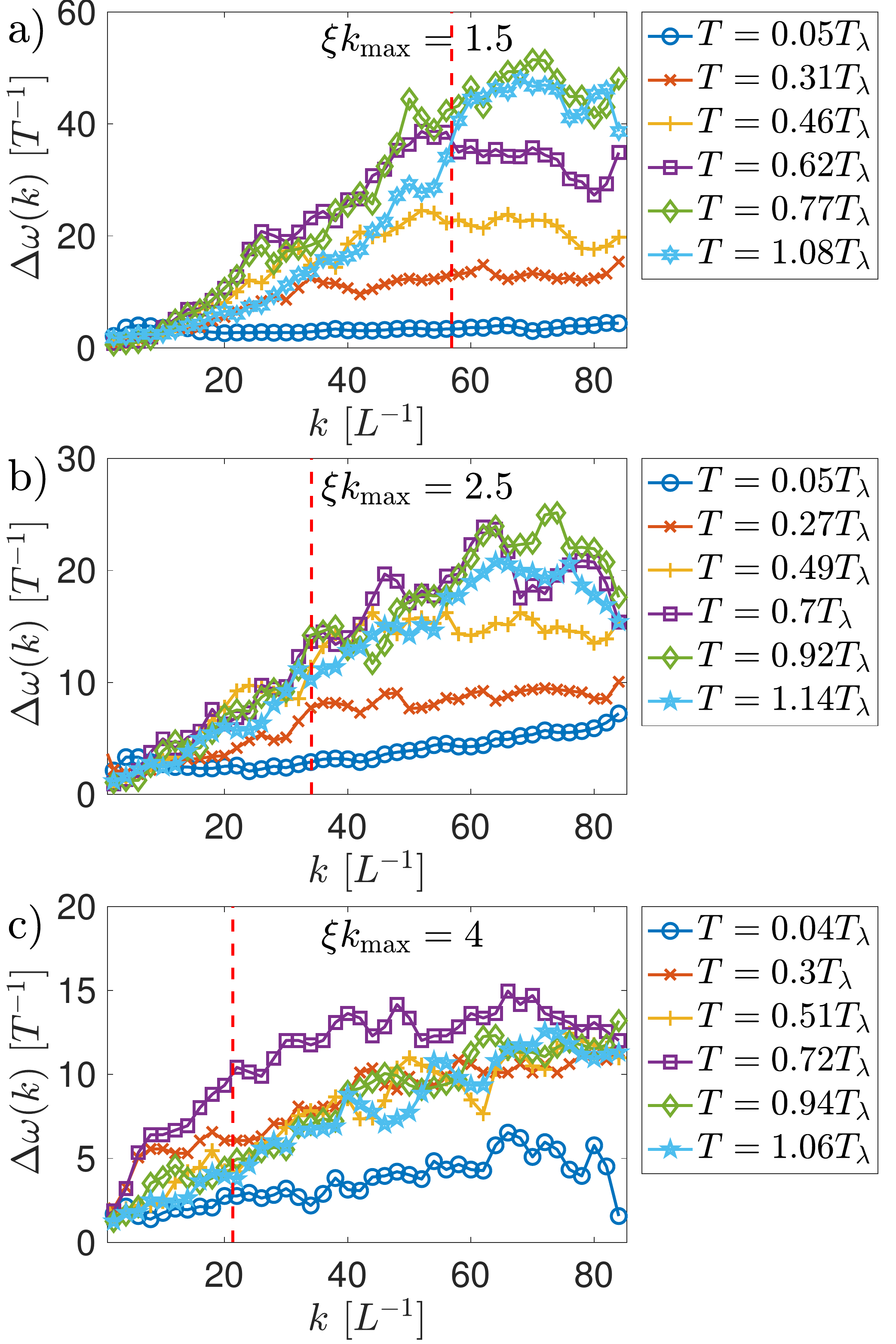}
    \caption{{\it (Color online)} Spectral broadening $\Delta \omega
      (k)$ of the dispersion relation, for different temperatures and
      values of $\xi\kmax$. The vertical dashed lines indicate the
      value of $1/\xi$ in each case: (a) as a function of $k$ for
      different temperatures and for $\xi\kmax =1.5$, (b) same for
      $\xi\kmax =2.5$, and (c) same for $\xi\kmax =4$.}
    \label{Fig:STS2}
\end{figure}

Based on these results, for a fixed $k$ we define the spectral
broadening $\Delta \omega (k)$ as the width for which the 
accumulation of spectral power around the dispersion relation goes to
half of its maximum amplitude. Note that $\Delta \omega(k)$ is
associated with the inverse of the time of non-linearly interacting
waves. The values of $\Delta \omega (k)$ extracted from the STS for
different temperatures and values of $\xi\kmax$, as a function of $k$,
can be seen in Fig.~\ref{Fig:STS2}. Note that: (1) For fixed
temperature, $\Delta \omega (k)$ increases with $k$, reaching its
maximum value for $k\sim 1/\xi$, and then growing slowly or 
remaining approximately constant. (2) For fixed $k$, 
$\Delta \omega (k)$ increases with temperature, reaching its maximum
value close to $T_\lambda$. And finally, (3) the role of the parameter
$\xi\kmax$ controlling the strength of non-linear interactions between
waves is confirmed by the values of  $\Delta \omega (k)$, as the
amplitude of this quantity is significantly reduced for increasing 
$\xi\kmax$.

%%%%%%%%%%%%%%%%%%%%%%%%%%%%%%%%%%%%%%%%%%%
%%%%%%%%%%%%%%%%%%%%%%%%%%%%%%%%%%%%%%%%%%%

%%%%%%%%%%%%%%%%%%%%%%%%%%%%%%%%%%%%%%%%%%%
%%%%%%%%%%%%%%%%%%%%%%%%%%%%%%%%%%%%%%%%%%%

%%%%%%%%%%%%%%%%%%%%%%%%%%%%%%%%%%%%%%%%%%%
%%%%%%%%%%%%%%%%%%%%%%%%%%%%%%%%%%%%%%%%%%%

\subsection{Mean free path and effective viscosity}
\label{sec:mfp}

\begin{figure}
    \centering
    \includegraphics[width=0.98\linewidth]{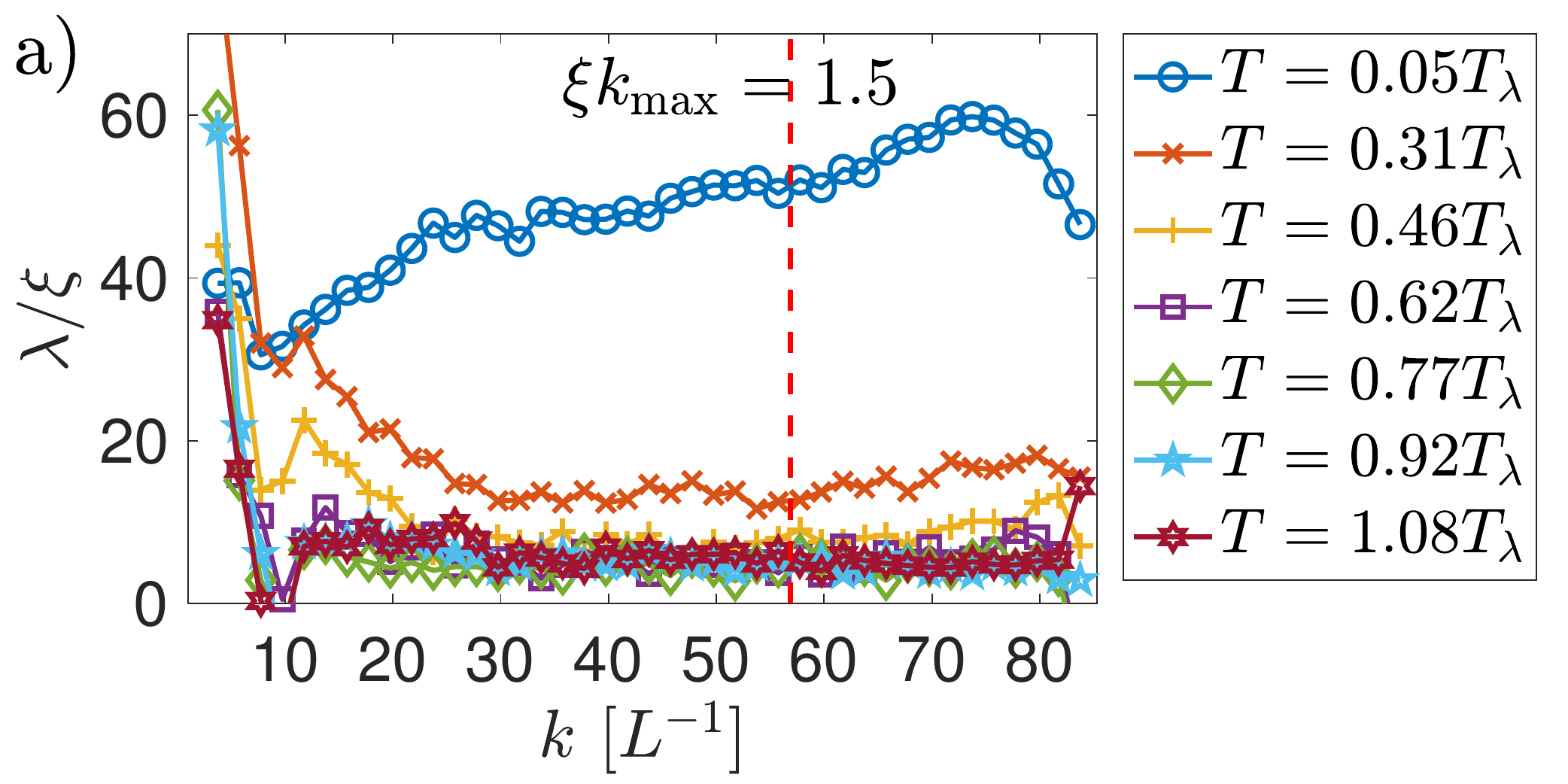}
    \includegraphics[width=0.98\linewidth]{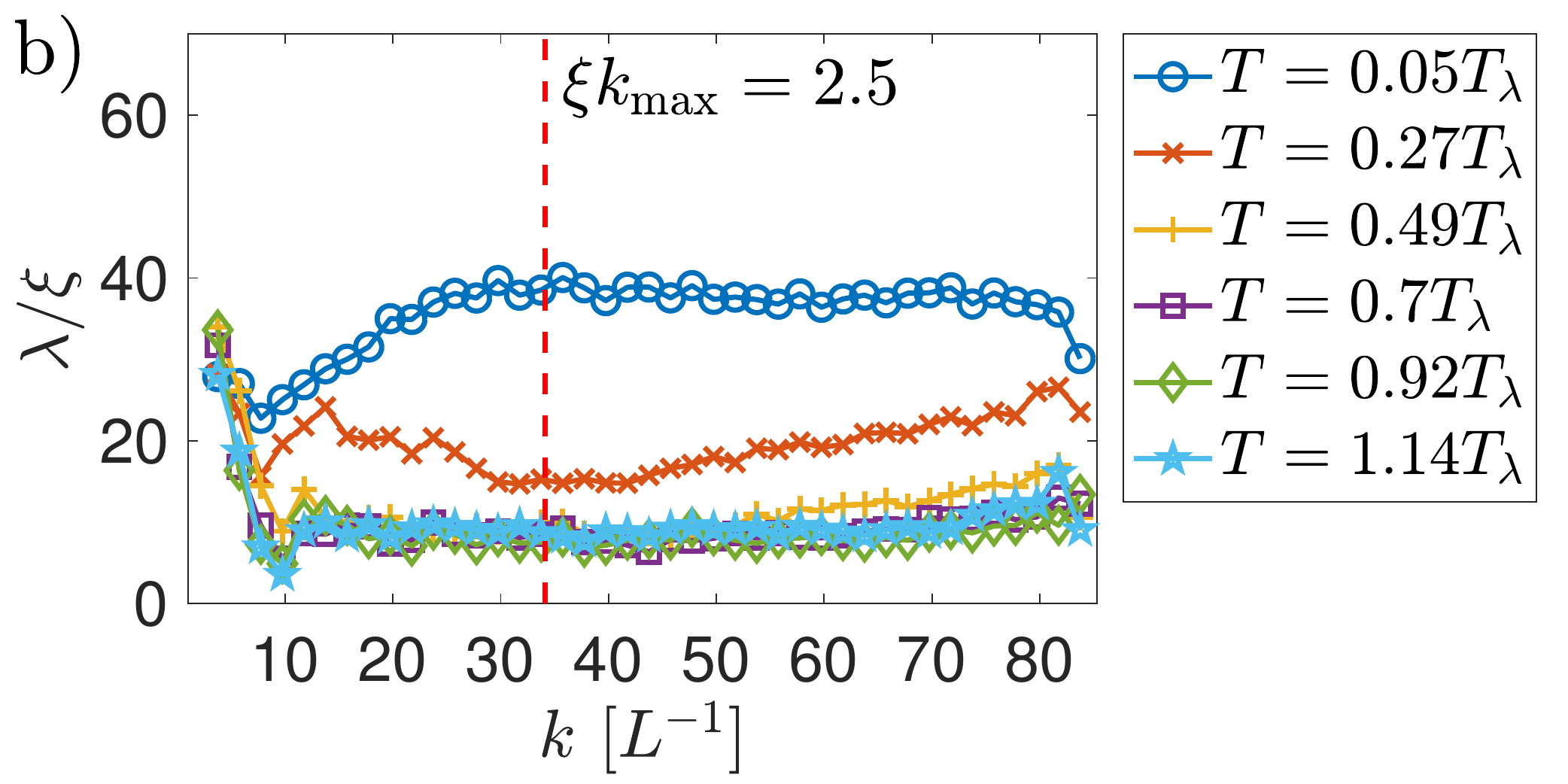}
    \includegraphics[width=0.98\linewidth]{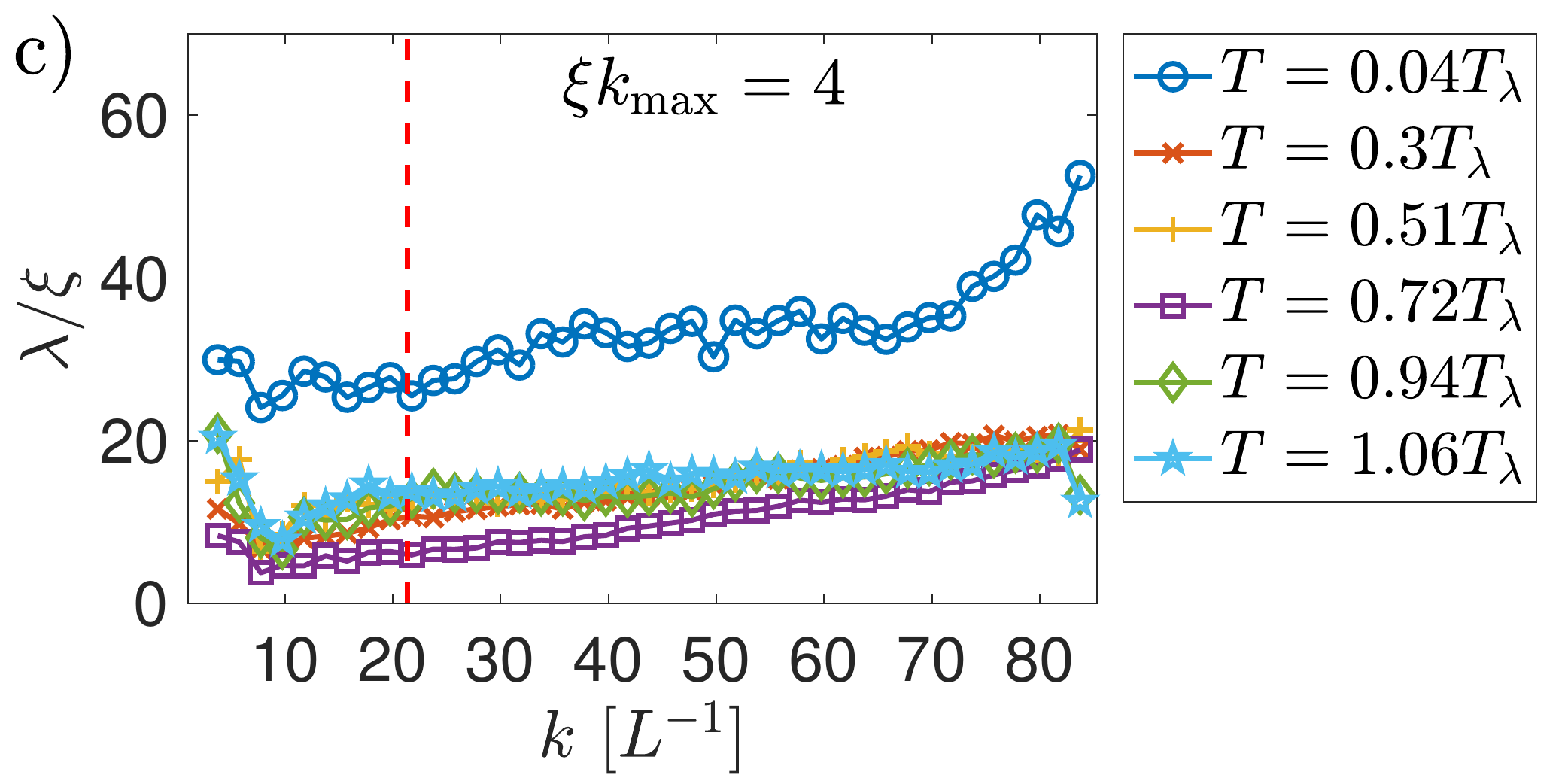}
    \caption{{\it (Color online)} Mean free-path as a function of
      $k$ for different temperatures in simulations with
      $\xi\kmax=1.5$ (a), $\xi\kmax=2.5$ (b) and $\xi\kmax=4$ (c). The
      vertical dashed line indicates the value of $1/\xi$.}
    \label{fig:Lambda}
\end{figure}

\begin{figure}[h]
    \centering
    \includegraphics[width=0.98\linewidth]{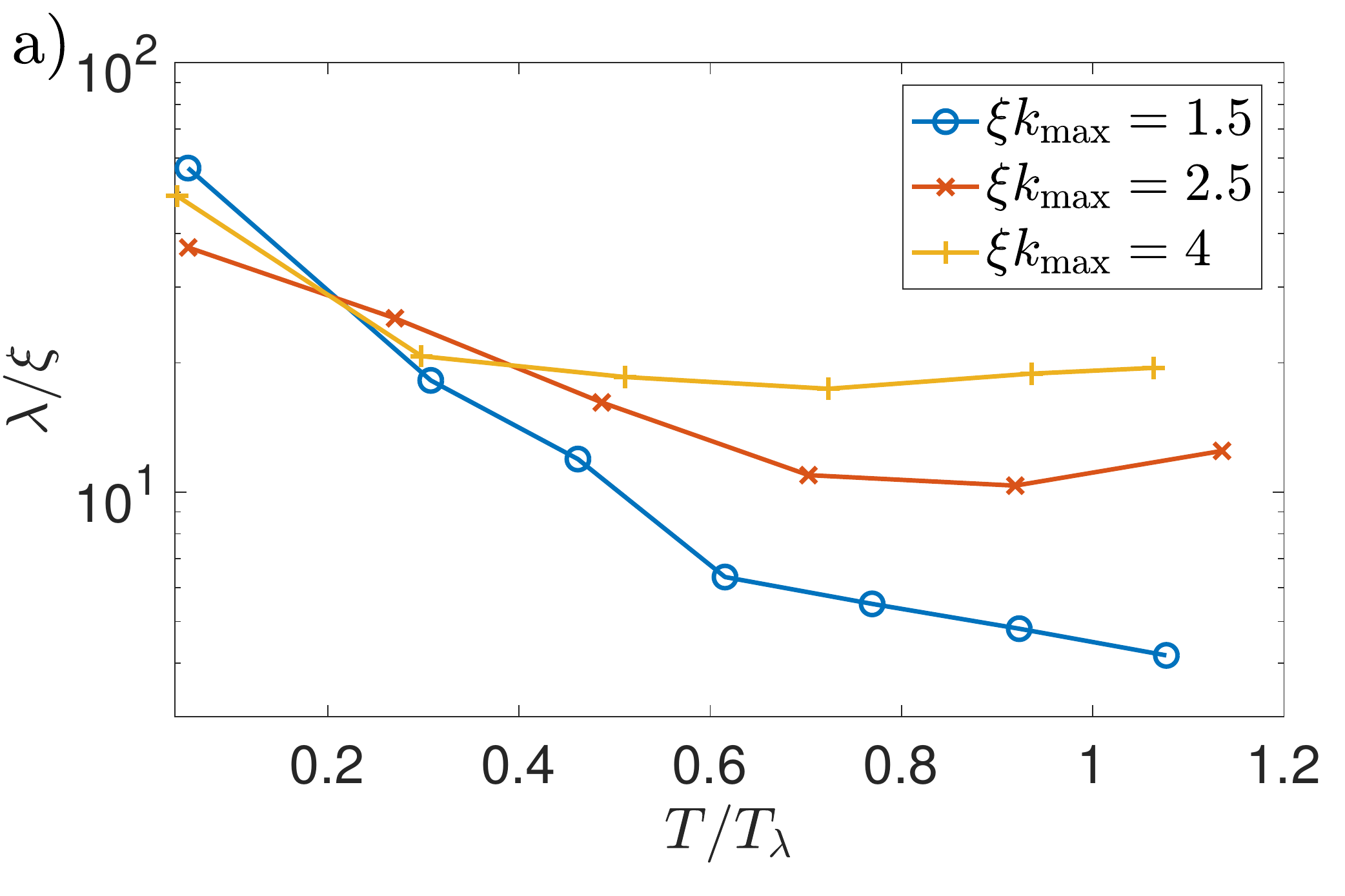}
    \includegraphics[width=0.98\linewidth]{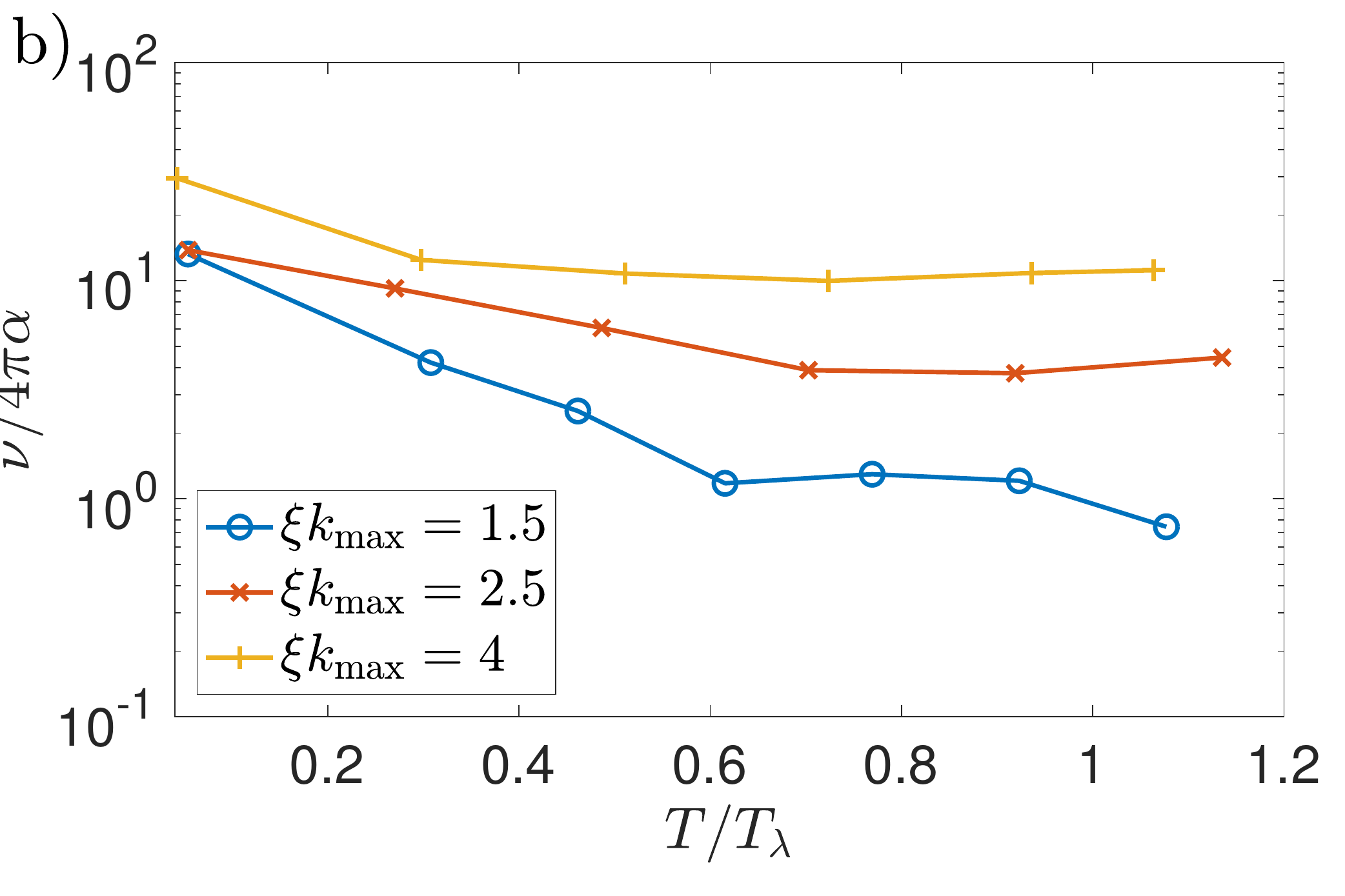}
    \caption{{\it (Color online)} (a) Mean free-path as a function of
      temperature $T$ for different values of  $\xi k_{\rm max}$. 
	(b) Effective viscosity (acting on the normal fluid),
        normalized by the quantum of circulation $4\pi\alpha=4\pi
        c\xi/\sqrt{2}$, as a function of temperature $T$ for different
	values of $\xi k_{\rm max}$.}
    \label{fig:Nu}
\end{figure}

We turn now to estimate the mean free path and the effective viscosity
for TGPE flows. As already mentioned, the time of non-linear wave
interactions is associated to the inverse of $\Delta\omega$; during
a time proportional to this quantity, waves travel without being
scattered by other waves \cite{Nazarenko}. As a result, the mean free
path can be constructed as 
\begin{equation}
    \lambda(k,T)\sim \frac{1}{\Delta\omega(k,T)}\frac{d\omega_B}{d k}.
\end{equation}
In Fig.~\ref{fig:Lambda} we show the mean-free path of the simulations
at different values of temperatures and $\xi \kmax$. The dispersion
relation at different temperatures was directly measured from the STS in
Fig.~\ref{Fig:STS25}. For wavenumbers larger than $k\approx 1/\xi$,
$\lambda(k,T)$ seems to saturate (or to fluctuate around a mean value
for a range of wavenumbers larger than $1/\xi$) in many cases.  We can
then define a mean-free path at each temperature by taking
$\lambda(T)=\lambda(k\approx k_{\max},T)$. The resulting plot of
$\lambda(T)$ is shown in Fig.~\ref{fig:Nu}(a) for different values of
$\xi\kmax$. As expected, it increases when $T$ goes to zero. Larger
values of $\xi\kmax$ reduce the non-linear interactions and thus
increase the mean-free path.

From the mean free path, we can estimate the effective viscosity by
writing
\begin{equation}
    \nu_{\rm eff}(T) \sim \lambda(T) \left.\frac{d\omega_B}{d
        k}\right|_{k\approx k_{\max}}
    =\frac{\left(\left.\frac{d\omega_B}{d
    k}\right|_{k\approx k_{\max}} \right)^2}{\Delta\omega(k\approx
        k_{\max},T)};
\end{equation}
note that we evaluate $\Delta\omega(T)$ and $d\omega_B/dk$ at 
$k=80\approx k_{\max}$. In Fig.~\ref{fig:Nu}(b) we show 
$\nu_{\rm eff}(T)$ normalized by the quantum of circulation 
$4\pi\alpha=4\pi c\xi\sqrt{2}$. For temperatures above 
$0.5T_\lambda$, both $\lambda$ and $\nu_{\rm eff}$ are relatively
constant.  For the runs with $\xi\kmax=2.5$ we have
\begin{eqnarray}
    \lambda(T) &\sim& 10 \xi, \\
    \nu_{\rm eff}(T) &\sim& 50 \alpha=50 c \xi
                            /\sqrt{2}. \label{eq:nuphy}
\end{eqnarray}
Equation (\ref{eq:nuphy}) gives a physical estimation of the scaling
of the effective viscosity in terms of the sound velocity $c$ and of
the healing length $\xi$. In the simulations, using $c=2U$ and 
$\xi = 2.5/k_{\max} = 2.5 \times 3L/N$, the effective viscosity then
becomes
\begin{equation}
    \nu_{\rm eff}(T) \sim L \, U \frac{500}{N} ,
\label{eq:nueff}
\end{equation}
where $U$ and $L$ are the unit velocity and length, and $N$ is the
linear spatial resolution. In dimensionless units, with $U=L=1$, the
Reynolds number can then be estimated as
\begin{equation}
  \mathrm{Re}^{\textrm{(TG)}} = \frac{C}{\nu_{\rm eff}} = \frac{CN}{500} ,
\label{eq:reeffTG}
\end{equation}
where $C$ is a prefactor of order unity (as $\nu_{\rm eff}$ is an
effective transport coefficient, we can only ascertain its value from
the mean free path up to a multiplicative constant).

The definition of the Reynolds number as $1/\nu_{\rm eff}$ is the
usual definition in simulations of the Taylor-Green flow
\cite{Brachet1983}. As in the next section (Sec.~\ref{sec:NS}) we will
compare GPE runs with simulations of the Navier-Stokes equations at
low Reynolds numbers, it will be convenient to also use a definition
of this number based on the dynamic r.m.s.~flow velocity 
\begin{equation}
U_0= \sqrt{2 E}
\label{eq:Urms}
\end{equation}
and the
flow correlation length 
\begin{equation}
L_0= 2\pi\,\frac{ \int{E(k)/k \, dk }}{ \int{E(k) \, dk }}
\label{eq:Lintegral}
\end{equation}
(i.e., the flow integral scale). Writing
$U_0$ and $L_0$ in units of $U$ and $L$, this Reynolds number is
\begin{equation}
  \mathrm{Re} = C \frac{U_0 L_0}{\nu_{\rm eff}} = C \frac{U_0 L_0
    N}{500}.
\label{eq:reeff}
\end{equation} 

It is worth pointing out that $\nu_{\rm eff}(T)$ depends on the value
of $\xi k_{\max}$, and that the strength of the nonlinear interactions
goes down with increasing $\xi k_{\max}$. Thus, in the simulations
with $\xi\kmax=1.5$ we have stronger turbulence. Moreover,   
the mean free path (see Fig.~\ref{fig:Nu}) in these runs is 
$\lambda(T)\sim 5\xi$, which gives a smaller 
$\nu_{\rm eff}(T) \sim 15\alpha$. This results, for $\xi\kmax=1.5$, in
a larger Reynolds number
\begin{equation}
  \mathrm{Re} = C \frac{U_0 L_0}{\nu_{\rm eff}} = C \frac{U_0 L_0
    N}{90} .
\label{eq:reeff2}
\end{equation} 
However, we cannot arbitrarily decrease $\xi k_{\max}$ to obtain
higher values of $Re$. At a fixed spatial resolution, $\xi k_{\max}$
must be larger than unity if we want to properly resolve the vortices
in simulations.

%%%%%%%%%%%%%%%%%%%%%%%%%%%%%%%%%%%%%%%%%%%
%%%%%%%%%%%%%%%%%%%%%%%%%%%%%%%%%%%%%%%%%%%

%%%%%%%%%%%%%%%%%%%%%%%%%%%%%%%%%%%%%%%%%%%
%%%%%%%%%%%%%%%%%%%%%%%%%%%%%%%%%%%%%%%%%%%

In a two-fluid framework, the viscosity in Eq.~(\ref{eq:nueff})
corresponds to a viscosity acting on the normal fluid (as it was
obtained from the thermalized component). This is consistent with
derivations of damping from stochastic equations for quasiclassical
fields (see, e.g., \cite{Gardiner00,Gardiner02}), where modes below an
energy cut-off are considered as the condensate, and modes above the
cut-off are considered as thermalized noise. In our case the
distinction is made using the STS, and from an extraction of 
the excitations lying in the vicinity of the dispersion relation of
the waves. Thus, at very high temperatures we basically have one fluid
with viscosity $\nu_{\rm eff}$. At intermediate temperatures, if the
mutual friction is large enough, the two fluids are then locked
together. Mutual friction in this case is estimated to be proportional
to $\rho_n/\rho_0$ (see \cite{Krstulovic11}). Thus, at intermediate
temperatures we can assume we have the fluids locked with an effective
mutual viscosity
\begin{equation}
    \nu_{\mathrm{eff}}' = \frac{\rho_n(T)}{\rho_0} \nu_{\rm eff}(T) .
\end{equation}

These results are consistent with the following interpretation. On the
one hand, at very large temperatures we have a very viscous flow. In
the next subsection this will be verified by performing Navier-Stokes
simulations using $\nu_{\mathrm{eff}}$ and comparing with runs of the
TGPE. This will further confirm that the effective Reynolds numbers of
flows in the TGPE are low even for the high resolution runs presented
in this work, and also allow us to estimate the prefactor $C$ in
Eq.~(\ref{eq:reeff}). Moreover, this is also consistent with previous
estimations based on the free decay of the incompressible kinetic
energy in \cite{Clark18}. In fact, based on the estimation in
Eq.~(\ref{eq:reeff}), to obtain a turbulent normal fluid described by
the TGPE near or above  $T_\lambda$, would require resolutions that
are not achievable even in the largest supercomputers available today
(thus, doing a classical turbulent Navier-Stokes simulation with the
TGPE can be very expensive!). On the other hand, at small temperatures
we have a problem that could be modeled with the Euler and Boltzmann
equations with small coupling, or more formally with a stochastic
equation for a quasiclassical field \cite{Gardiner02}, with the system
being close to GPE dynamics. Finally, in the middle region the system
probably behaves close to Euler and Navier-Stokes fluids with
coupling, and its modelling is left for future work.

%%%%%%%%%%%%%%%%%%%%%%%%%%%%%%%%%%%%%%%%%%%
\begin{figure*}
    \centering
\includegraphics[width=0.3\linewidth]{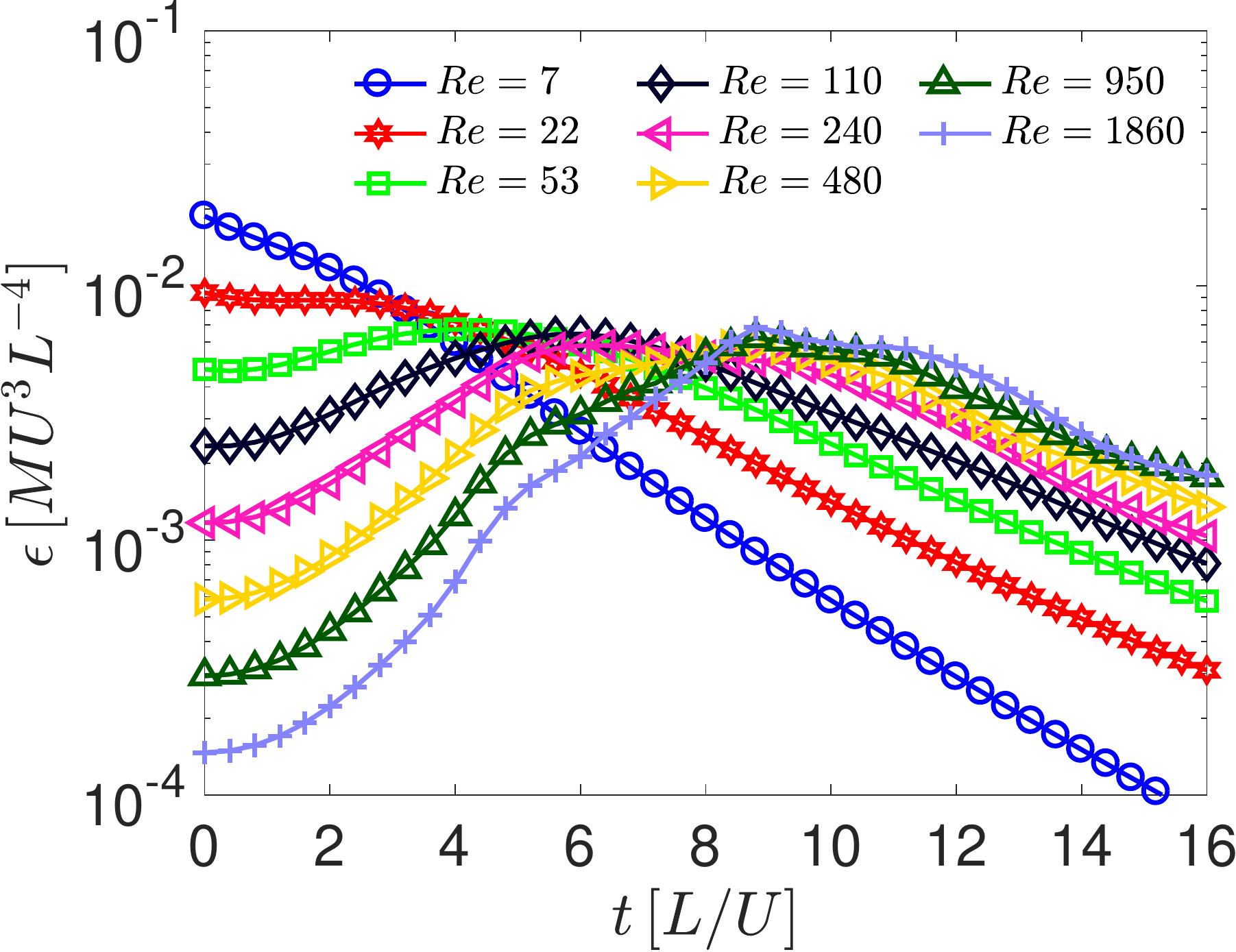}
\includegraphics[width=0.3\linewidth]{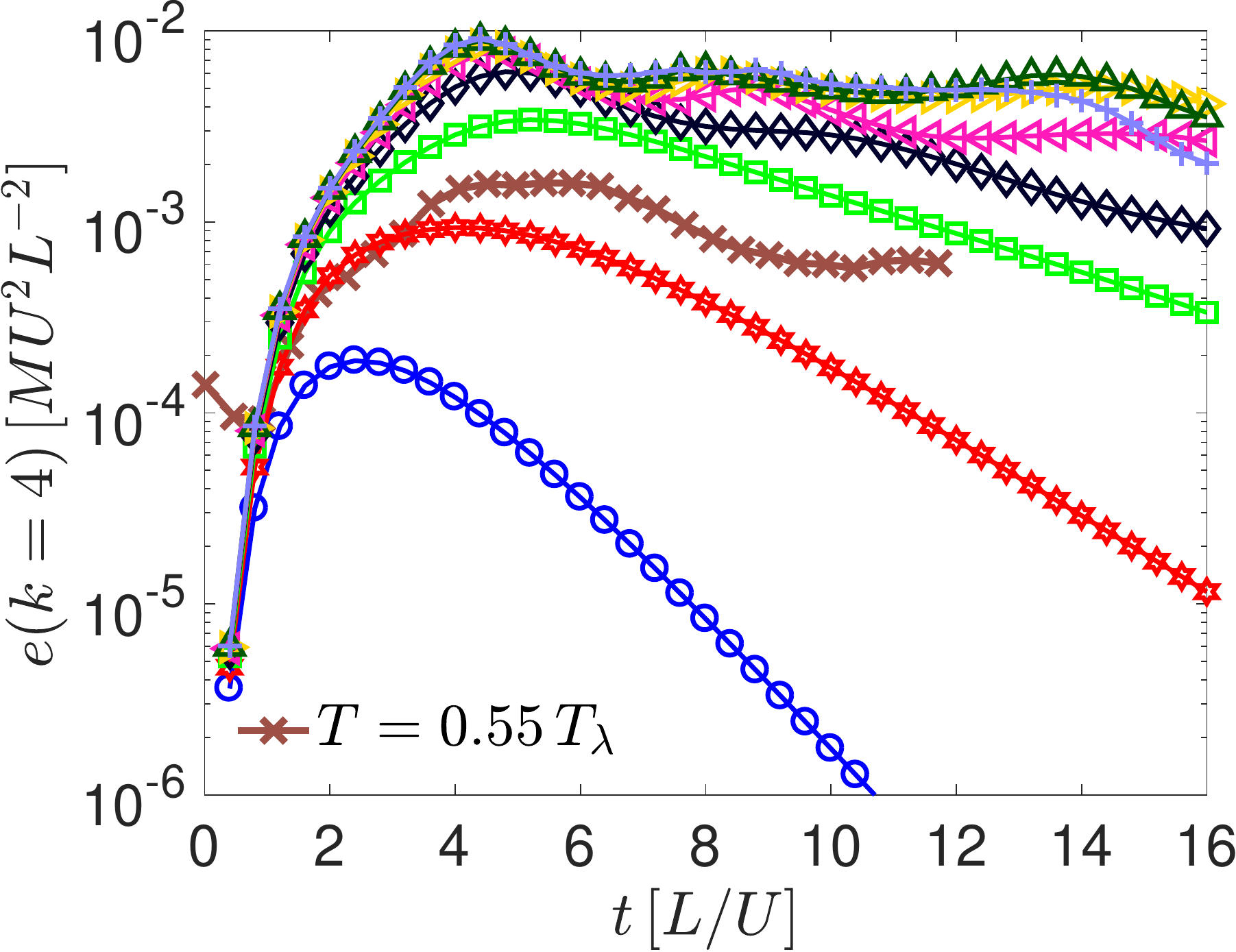}
\includegraphics[width=0.3\linewidth]{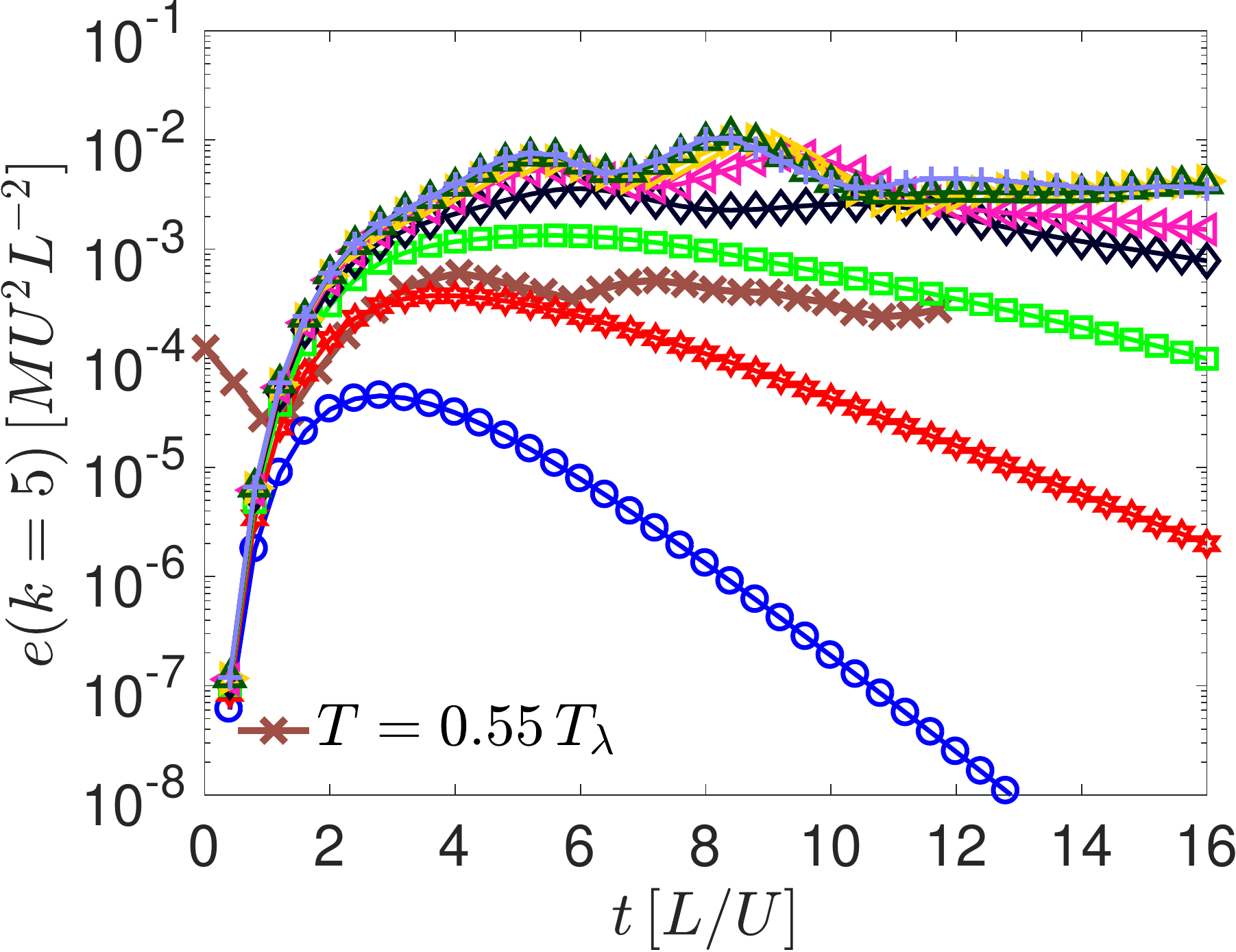}
\\
\includegraphics[width=0.3\linewidth]{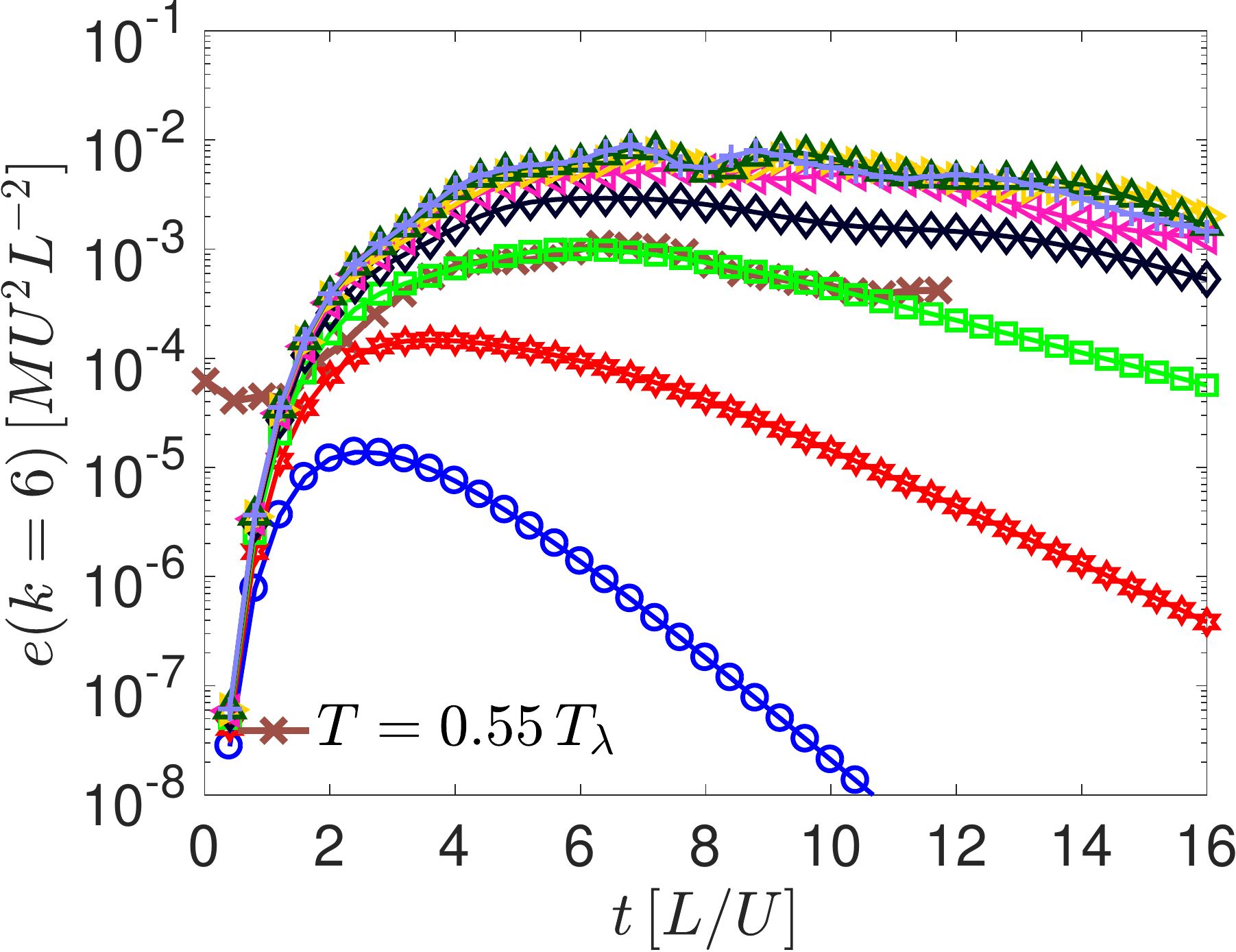}
\includegraphics[width=0.3\linewidth]{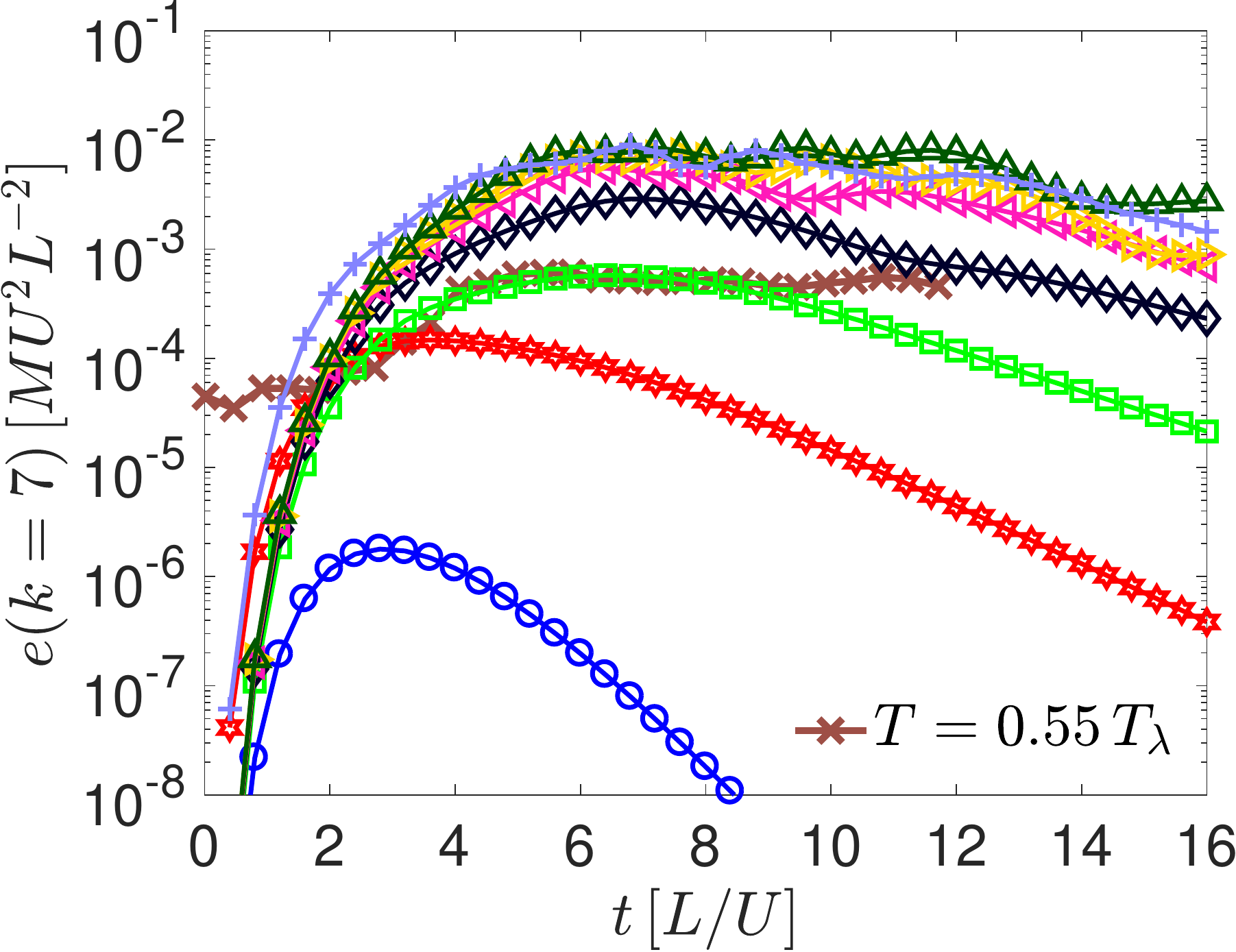}
\includegraphics[width=0.3\linewidth]{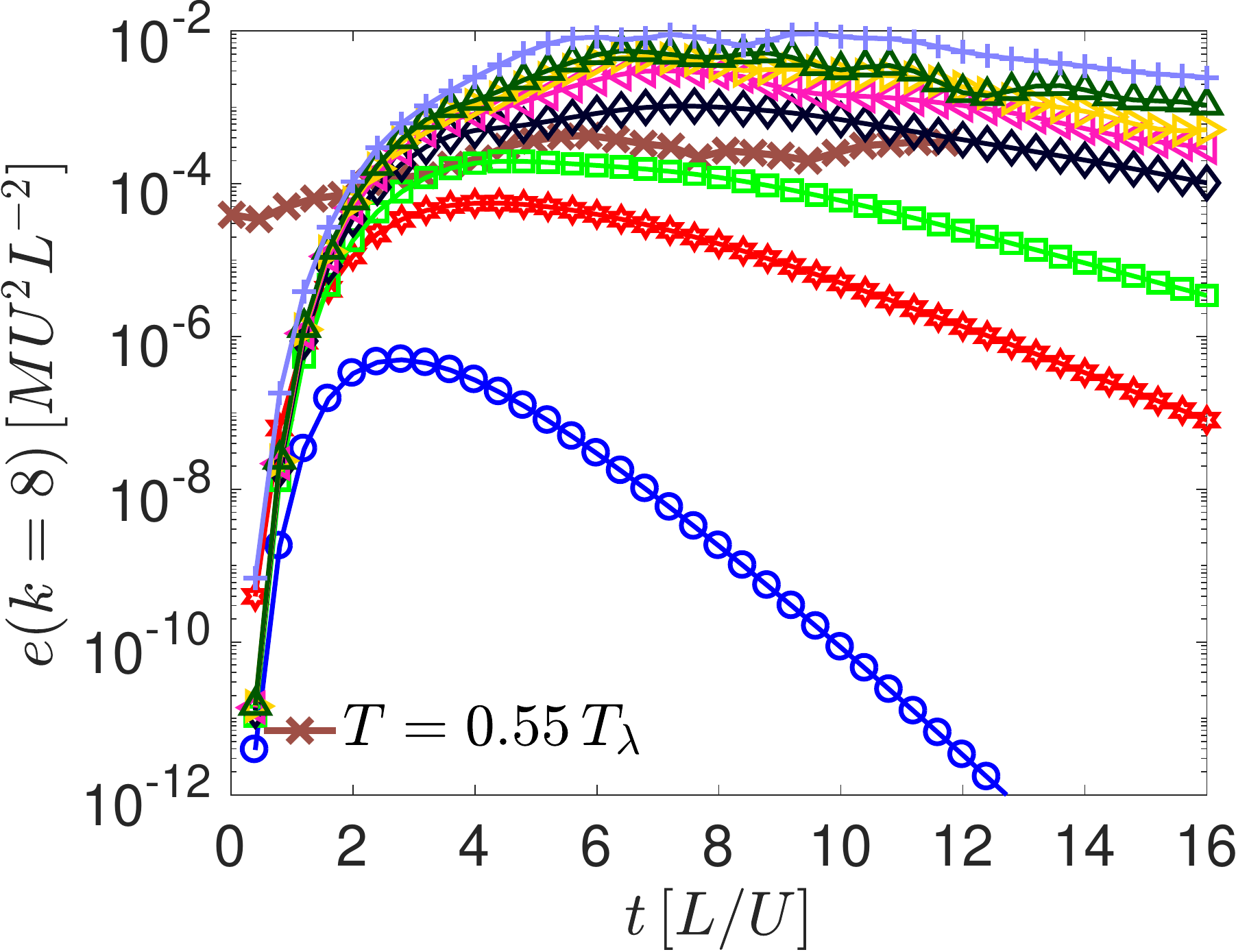}
\caption{{\it (Color online)} 
  Results from the DNS of low-Reynolds number Navier-Stokes TG runs
  at different Reynolds numbers (${\rm Re}=U_0L_0/\nu$, based on the
  integral scale $L_0$ and the flow r.m.s.~velocity $U_0$, see 
	Eqs.\eqref{eq:Urms} and~\eqref{eq:Lintegral}). (a) Plots
  of energy dissipation rate $\epsilon$ vs time. The maximum of  
  $\epsilon$ for the run with $Re=1860$ occurs at $t\approx 9\,L/U$.
  (b)-(f) Plots of the temporal evolution of the energy in different
  $k$-shells in Fourier space (i.e., $e(k,t)$ for fixed $k$ as a
  function of time) for $k=4$ to $8$ from the Navier-Stokes TG runs
  and the TGPE run at temperature $T=0.55\,T_{\lambda}$ (linear
  resolution $N=1024$ and $\xi k_{\rm max}=2.5$; for GPE $e(k,t)$
  corresponds to $e^i_{\textrm{kin}}(k,t)$). In all panels, the GPE
  run at $T=0.55\,T_{\lambda}$ is indicated by the (brown) curve with
  crosses, while all Navier-Stokes runs at different Reynolds numbers
  are labeled as indicated in panel (a). To vary the Reynolds numbers,
  the Navier-Stokes runs were performed 
  with kinematic viscosities $\nu=1/10$ and $\nu=1/40$ at linear spatial resolutions $N=64$; 
	$\nu=1/80$ and $\nu=1/160$ ($N=128$); $\nu=1/320$, $\nu=1/640$ and 
	$\nu=1/1280$ ($N=256$); $\nu=1/2560$ ($N=512$).
}
\label{Fig:NSvsGPE}
\end{figure*}
%%%%%%%%%%%%%%%%%%%%%%%%%%%%%%%%%%%%%%%%%%%

\subsection{Comparisons with Low-Reynolds runs}
\label{sec:NS}

To verify that the evolution of a finite temperature TG flow is
similar to a highly viscous classical flow, and to estimate the value
of the factor $C$ in Eq.~(\ref{eq:reeff}), we perform a set of
simulations of freely decaying ``classical'' TG flows obeying the
Navier-Stokes equations at different Reynolds numbers, 
$\textrm{Re}=U_0 L_0/\nu = 7$ (with a spatial resolution $N^3=64^3$),
22 (with $64^3$ resolution), 53 ($128^3$), 110 ($128^3$), 240
($256^3$), 480 ($256^3$), 950 ($256^3$), and 1860 ($512^3$). Here
$\nu$ is the kinematic viscosity, and the r.m.s.~flow velocity $U_0$
and the flow correlation length $L_0$ correspond to the averaged in
time values between $t=4$ and $10$, when turbulence has developed.
We compare these runs with the TGPE run at $T=0.55\,T_{\lambda}$ with
linear resolution $N=1024$ and $\xi k_{\rm max}=2.5$.

In Fig.~\ref{Fig:NSvsGPE}(a) we show the time evolution of the energy
dissipation rate, $\epsilon$, for these TG Navier-Stokes runs. We
observe that as we increase $\rm Re$, the time to achieve the maximum
energy dissipation rate also increases, after which turbulence
develops. To provide a detailed comparison between the runs, in
Fig.~\ref{Fig:NSvsGPE} (b)-(f) we show the temporal evolution of the
incompressible kinetic energy in Fourier shells $k=4\,L^{-1}$ to
$8\,L^{-1}$, respectively, for the above mentioned Navier-Stokes runs
and for the TGPE run.  We find it remarkable that the shell by shell
evolution of these two systems shows a considerable overlap for
Reynolds numbers in the range $Re=22$ to $53$; in other words, the
time evolution of the energy in each shell in the TGPE run is in
between these two Navier-Stokes runs. This is in reasonable
agreement, at least in terms of order of magnitude, with the
predictions given by Eq.~(\ref{eq:reeff}). Computing $L_0$ and $U_0$
in the TGPE run in the same time inverval (between $t=4$ and 10), we
obtain $\textrm{Re} = 5C$, suggesting $C\approx 7$.

\section{Conclusion}
\label{sec:Conclusion}

The results presented in this paper significantly extend our knowledge
of quantum turbulence in the Taylor-Green vortex. At zero temperature,
runs in three dimensions computed with linear spatial resolutions up
to $N=4096$ grid points allowed us to characterize the presence of a
Kolmogorov scaling range at scales larger than the intervortex
distance $\ell$, and to observe another scaling range at scales
smaller than $\ell$. The presence of tangled  substructures is
apparent in vortex line visualizations.

Then, using thermal equilibria and spatio-temporal spectra, we were
able to separate the condensed phase from the interacting waves, and
to estimate from the nonlinear-broadening the mean free-path and the
effective viscosity as a function of the temperature. The actual
(large) values of our estimated effective viscosity near the 
$\lambda$-transition, $\nu_{\rm eff} \sim 500/N$ (in dimensionless
units) for $\xi \kmax = 2.5$, and $\nu_{\rm eff} \sim 90/N$ for 
$\xi \kmax = 1.5$, correspond respectively to effective Reynolds
numbers $\mathrm{Re} \sim U_0L_0 N/500$ and 
$\mathrm{Re} \sim U_0L_0 N/90$ (up to prefactors of order unity),
where $N$ is the linear resolution of the simulation.

Finally, the comparison of finite temperature quantum turbulence using
linear resolutions of $N=1024$ grid points against low-Reynolds
Navier-Stokes numerical simulations further confirmed our estimations
of the effective viscosity $\nu_{\rm eff}$ based on the mean free-path
of the thermal excitations, and allowed us to get a first estimation
of the amplitude of the unknown prefactors.

It is well known (see, e.g., Ref.~\cite{Brachet1983}) that Kolmogorov
scaling becomes apparent in Navier-Stokes numerical simulations of 
the Taylor-Green flow with linear resolutions of $N=256$, for 
$\mathrm{Re}\approx 1600$. We can thus conclude that an equivalent
direct numerical simulation using the truncated Gross-Pitaevskii
equation performed at  $T \approx T_\lambda$ would need a resolution
of about $N\approx 10,000$ grid points in each spatial direction for 
$\xi \kmax = 1.5$, and of $N\approx 43,000$ for $\xi \kmax = 2.5$, 
to achieve a similar Reynolds number and a classical direct energy
cascade with similar scale separation for the normal fluid. These
resolutions are out of reach using present day computing resources. At
smaller values of $T$ this situation changes drastically, as the
mutual friction between the fluid and the superfluid depends on the
density of the normal fluid as $\rho_n/\rho_0$ \cite{Krstulovic11}.

Looking back at the estimates of effective viscosity, it can be seen
that the high value of $\nu_{\rm eff}$ traces back to the high value of
$(d\omega_B/d k)^2/\Delta\omega$ at $k=k_{\rm max}$ for the
Gross-Pitaevskii equation. This brings into mind the possibility of
modifying the Bogoliubov dispersion relation through modifications in
Gross-Pitaevskii, and therefore changing the value of
$(d\omega_B/d k)^2$ at high wavenumbers. It is well known that, by
changing the cubic term in the Gross-Pitaevskii equation into a
non-local term of the form 
$\psi \int |\psi({\bf x'})|^2 V(|{\bf x- x'}|)\, d{\bf x'} $, 
the first term in the Bogoliubov dispersion relation can be changed to
a term involving a potential
$\widetilde V(k)=\int V(r) \exp[-i {\bf k}\cdot {\bf r}]\, d{\bf r}$. 
In this way, it is possible to ``adjust'' the dispersion relation,
see, e.g., Eqs.~(3) and (4) of Ref.~\cite{Berloff14}. This, besides 
allowing the modeling of rotons in superfluid $^4$He at low
temperatues, can also result in a decrease of the effective viscosity
at temperatures close to the $\lambda$-transition. The impact of these
changes in $\nu_{\rm eff}$ is left for future work.

\begin{acknowledgements}
    The authors acknowledge financial support from ECOS-Sud grant
    No.~A13E01, and computing hours in the IDRIS supercomputer 
    granted by Project IDRIS 100591 (DARI x20152a7493). Computations
    were also carried out on the M\'esocentre SIGAMM hosted at the
    Observatoire de la C\^ote d'Azur P.C.dL. acknowledges funding from
    the European Research Council under the European Community's
    Seventh Framework Program, ERC Grant Agreement No.~339032. PDM
    acknowledges funding from grant PICT No.~2015-3530, and useful
    discussions with E.~Calzetta.
\end{acknowledgements}

%\bibliographystyle{apsrev}
%\bibliography{bibli}

\end{document}